\documentclass[aps,twocolumn,amsmath,amssymb,notitlepage,superscriptaddress]{revtex4-1}
\usepackage{graphicx}
\usepackage{dcolumn}
\usepackage{bm}
\usepackage{upgreek}
\usepackage{siunitx}
\usepackage{colonequals}

\usepackage[usenames,dvipsnames]{xcolor}

\usepackage{enumitem}

\usepackage{xr-hyper}
\definecolor{darkblue}{rgb}{0,0,0.6}
\definecolor{darkred}{rgb}{0.6,0,0}
\usepackage[colorlinks=true,urlcolor=darkblue, citecolor=darkblue, linkcolor=black, hyperfootnotes=false]{hyperref}

\usepackage{mathrsfs}

\usepackage{tikz}
\usepackage[kerning=true]{microtype}
\usetikzlibrary{patterns}

\newcommand*\circled[1]{\tikz[baseline=(char.base)]{
            \node[shape=circle,draw,inner sep=1.pt,minimum size=3ex] (char) {#1};}}

\newcommand{\ueta}{\underline{\eta}}

\newcommand{\dd}{\mathrm{d}}

\DeclareMathOperator{\erf}{erf}
\DeclareMathOperator{\erfc}{erfc}
\DeclareMathOperator{\sign}{sign}
\def \equi#1{\mathrel{\mathop{\kern 0pt\sim}\limits_{#1}}} 

\def\Xint#1{\mathchoice
  {\XXint\displaystyle\textstyle{#1}}%
  {\XXint\textstyle\scriptstyle{#1}}%
  {\XXint\scriptstyle\scriptscriptstyle{#1}}%
  {\XXint\scriptscriptstyle\scriptscriptstyle{#1}}%
  \!\int}
\def\XXint#1#2#3{{\setbox0=\hbox{$#1{#2#3}{\int}$}
    \vcenter{\hbox{$#2#3$}}\kern-.5\wd0}}
\def\intVp{\Xint-}

\newcommand{\Df}{\mathcal{D}}

\def\e{\mathrm{e}}
\def\I{\mathrm{i}}
\def\O{\mathcal{O}}

\newcommand{\abs}[1]{\ensuremath{\left| #1 \right|}}

\newcommand{\moy}[1]{\ensuremath{\left\langle #1 \right\rangle}}

\def\rhs{\circled{?}}

\begin{document}

\title{Exact spatial correlations in single-file diffusion}

\author{Aur\'elien Grabsch}
\affiliation{Sorbonne Universit\'e, CNRS, Laboratoire de Physique Th\'eorique de la Mati\`ere Condens\'ee (LPTMC), 4 Place Jussieu, 75005 Paris, France}

\author{Pierre Rizkallah}
\affiliation{Sorbonne Universit\'e, CNRS, Laboratoire de Physico-Chimie des \'Electrolytes et Nanosyst\`emes Interfaciaux (PHENIX), 4 Place Jussieu, 75005 Paris, France}

\author{Alexis Poncet}
\affiliation{Univ. Lyon, ENS de Lyon, Univ. Claude Bernard, CNRS, Laboratoire de Physique, F-69342, Lyon, France}

\author{Pierre Illien}
\affiliation{Sorbonne Universit\'e, CNRS, Laboratoire de Physico-Chimie des \'Electrolytes et Nanosyst\`emes Interfaciaux (PHENIX), 4 Place Jussieu, 75005 Paris, France}

\author{Olivier Bénichou}
\affiliation{Sorbonne Universit\'e, CNRS, Laboratoire de Physique Th\'eorique de la Mati\`ere Condens\'ee (LPTMC), 4 Place Jussieu, 75005 Paris, France}

\date{\today}

\begin{abstract}
Single-file diffusion refers to the motion of diffusive particles in narrow channels, so that they cannot bypass each other. This constraint leads to the subdiffusion of a tagged particle, called the tracer. This anomalous behaviour results from the strong correlations that arise in this geometry between the tracer and the surrounding bath particles. Despite their importance, these bath-tracer correlations have long remained elusive, because their determination is a complex many-body problem. Recently, we have shown that, for several paradigmatic models of single-file diffusion such as the Simple Exclusion Process, these bath-tracer correlations obey a simple exact closed equation. In this paper, we provide the full derivation of this equation, as well as an extension to another model of single-file transport: the double exclusion process. We also make the connection between our results and the ones obtained very recently by several other groups, and which rely on the exact solution of different models obtained by the inverse scattering method.
\end{abstract}

\maketitle

\tableofcontents

\hypersetup{linkcolor=darkred}

\section{Introduction}

Single-file diffusion refers to the dynamics of diffusive particles in narrow channels with the constraint that they cannot bypass each other. It is a fundamental model for the subdiffusion of a tagged particle, a tracer, in a confined environment~\cite{Levitt:1973,Arratia:1983}. Indeed, the constraint that the particles must remain in the same order at all times leads to a subdiffusive behaviour of the position $X_t$ of the tracer at time $t$, $\moy{X_t^2} \propto \sqrt{t}$~\cite{Harris:1965}, in contrast with the normal diffusion of an isolated particle $\moy{X_t^2} \propto t$. This prediction has been observed experimentally, at different scales, from the motion of molecules in zeolites to the dynamics of colloids in narrow trenches~\cite{Hahn:1996,Wei:2000,Lin:2005}.

A central model in the study of single-file diffusion is the Symmetric Exclusion Process (SEP). It is a model of particles on a one-dimensional infinite lattice, which perform symmetric random walks in continuous time, with the constraint that each site can host at most one particle, corresponding to a hard-core repulsion between the particles. In this article, we will consider that, at $t=0$, each site is filled independently with probability $\rho$ (annealed initial condition). Beyond modelling single-file diffusion, the SEP has become a paradigmatic model of statistical physics, which has been the focus of numerous works, both in the physical and mathematical literature (see for instance Refs.~\cite{Levitt:1973,Arratia:1983,Spitzer:1970,Derrida:2009}). An important progress has been achieved a few years ago with the computation of the large deviation function of the position $X_t$ of a tracer~\cite{Imamura:2017,Imamura:2021}, which fully characterises the distribution of $X_t$ in the large time limit. This function gives access to the large time behaviour of the cumulants of the position of the tracer, which all display an anomalous behaviour in $\sqrt{t}$~\cite{Imamura:2017,Krapivsky:2015a}. Similarly, the cumulants of an other observable, the time integrated current through the origin $Q_t$, have been computed and all display the same anomalous scaling in $\sqrt{t}$.

These anomalous behaviours in the SEP come from the strong spatial correlations which arise in the single-file geometry, due to the constraint that the particles cannot bypass each other. These correlations are thus fundamental quantities to analyse single-file diffusion. Although this fact is well recognised, the determination of these bath-tracer correlations has long remained an open question, because they obey an infinite hierarchy of equations, as usual in many-body problems. These correlations have first been determined in the high and low density limits of the SEP~\cite{Poncet:2021}. Recently, we have solved this problem by finding a strikingly simple exact closed equation satisfied by these bath-tracer correlations, at any density~\cite{Grabsch:2022}. We further argued that this equation plays a central role in single-file diffusion by showing that it also holds in out-of-equilibrium situations, and applies to other observables like currents and other models of single-file systems than the SEP~\cite{Grabsch:2022}.

\begin{figure}
    \centering
    \includegraphics[width=\columnwidth]{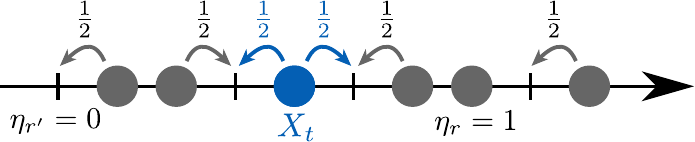}
    \caption{\textbf{The Symmetric Exclusion Process (SEP)} with a tracer (in blue). Each particle can hop, in either direction, with rate $\frac{1}{2}$, only if the neighbouring site in that direction is empty. We denote $X_t$ the position of the tracer at time $t$, and $\eta_r = 0$ or $1$ the occupation of the site $r$ by the bath particles (in gray).}
    \label{fig:SEP_def}
\end{figure}

In this article, we provide the full derivation of this key equation for the bath-tracer correlation profiles. We also make the connection between our results and the ones obtained recently by other groups~\cite{Bettelheim:2022,Bettelheim:2022a,Mallick:2022,Krajenbrink:2022}, which rely on a different approach based on the inverse scattering technique~\cite{Ablowitz:1981}, and show, in fact, the very same equation appears also in these papers, but in a different form.
Finally, we determine the bath-tracer correlations for another model of single-file diffusion, the double exclusion process, using a mapping between this model and the SEP~\cite{Rizkallah:2022}, giving an extra example which further confirms the ubiquity of this equation.

The paper is organised as follows: in Section~\ref{sec:DefObs} we introduce the notations and the observables. In Section~\ref{sec:SEPhydroGen} we derive the microscopic equations for the SEP and then deduce the hydrodynamic version of these equations, valid in the limit of large time and large distances. We then discuss limiting cases in Section~\ref{sec:LimCases} before presenting in Section~\ref{sec:CloseEqSEPGen} the approach that led us to the closed equation for the bath-tracer correlations in the SEP, and the computation of these correlations from this equation. We then discuss extensions of this equation to other observables and out-of-equilibrium situations in Section~\ref{sec:ExtOtherSit} and to other models than the SEP in Section~\ref{sec:ExtOtherModels}. We finally show that the same equation arises in various contexts studied recently using the inverse scattering method in Section~\ref{sec:CompInvScat}.

\section{Definitions and observables}
\label{sec:DefObs}

We consider the SEP on a 1D infinite lattice. Initially each site is filled independently with probability $\rho \in [0,1]$, which is the mean density of the system, except the origin which is always occupied by a particle, which we call the tracer. We denote $X_t$ the position of the tracer at time $t$, with $X_0=0$. The other particles are described by a set of occupation numbers $\{ \eta_i(t) \}_{i \in \mathbb{Z}}$, with $\eta_i(t) = 1$ if site $i$ is occupied at time $t$ and $0$ otherwise. We use the convention that the tracer is not a bath particle, so $\eta_{X_t}(t) = 0$ since the site occupied by the tracer cannot host a bath particle due to the exclusion rule.

The statistical properties of the position of the tracer are encoded in the cumulant generating function
\begin{equation}
    \label{eq:def_psi}
    \psi(\lambda, t) \equiv \ln \left\langle \e^{\lambda X_t} \right\rangle
    \:,
\end{equation}
whose expansion in powers of $\lambda$ give the cumulants $\kappa_n(t)$,
\begin{equation}
    \psi(\lambda, t) = \sum_{n=1}^\infty \frac{\lambda^n}{n!} \kappa_n(t)
    \:.
\end{equation}
To quantify the bath-tracer correlations, we consider the joint cumulant generating function of the position $X_t$ of the tracer and of the the occupation $\eta_{X_t+r}(t)$ of the site located at a distance $r$ from the tracer. Since the occupation numbers $\eta_{X_t+r}$ can only take values $0$ and $1$, we can write this joint cumulant generating function as
\begin{equation}
    \ln \moy{\e^{\lambda X_t + \chi \eta_{X_t+r}}}
    = \psi(\lambda,t) + \ln \left( 1 + (\e^{\chi}-1) w_r(\lambda,t) \right)
    \:,
\end{equation}
where we have introduced
\begin{equation}
    \label{eq:def_wr}
    w_r(\lambda, t) \equiv 
    \frac{\langle \eta_{X_t+r}(t) \: \e^{\lambda X_t}\rangle}{\langle \e^{\lambda X_t}\rangle}
    \:.
\end{equation}
The functions $\psi$ and $w_r(\lambda,t)$ fully characterise the joint cumulant generating function of $(X_t, \eta_{X_t+r}(t))$. We call $w_r(t)$ the generalised density profiles generating function, since its expansion in powers of $\lambda$ yields all the joint cumulants $\moy{X_t^n \eta_{X_t+r}(t)}_c$ between $X_t$ and $\eta_{X_t+r}$,
\begin{equation}
    w_r(\lambda, t) = 
    \sum_{n=0}^\infty \frac{\lambda^n}{n!} \moy{X_t^n \eta_{X_t+r}(t)}_c
    \:,
\end{equation}
which encode all the bath-tracer correlation functions. For instance, for $n=1$, $\moy{X_t \eta_{X_t+r}(t)}_c$ reads
\begin{equation}
    \label{eq:Phi1Cov}
    \moy{X_t \eta_{X_t+r}} - \moy{X_t} \moy{\eta_{X_t+r}}
    = \mathrm{Cov}(X_t , \eta_{X_t+r})
    \:,
\end{equation}
which is the covariance between the position of the tracer and the density of bath particles at a distance $r$ from it.

We will show that, in the large time and large distance limit, the density profiles $w_r$ can be determined from a simple closed integral equation.

\section{Hydrodynamic limit of the SEP from the master equation}
\label{sec:SEPhydroGen}

\subsection{Microscopic equations}
\label{sec:MicroEq}

Our starting point to derive the hydrodynamic equations for the profiles $w_r$ in the SEP is the master equation, which describes the time evolution of the probability $P_t(X, \ueta)$ to observe the configuration $\{ X, \ueta \}$ at time $t$. It verifies
\begin{multline}
    \label{eq:master_eq}
    \partial_t P_t(X, \ueta) 
    =~ \frac{1}{2}\sum_{r\neq X, X-1} \left[P_t(X, \ueta^{r,+})-P_t(X, \ueta)\right] 
    \\
    + \frac{1}{2} \sum_{\mu=\pm 1} 
    \left\{ (1 - \eta_X) P_t(X-\mu, \ueta) 
    - (1-\eta_{X+\mu}) P_t(X, \ueta) \right\} 
    \:,
\end{multline}
where $\ueta^{r,+}$ is the configuration $\ueta$ in which the occupations of sites $r$ and $r+1$ are exchanged. In~\eqref{eq:master_eq}, the first sum corresponds to the jumps of the bath particles, while the second one describes the displacement of the tracer.

From this equation, we compute the time evolution of any observable expressed in terms of the position $X$ and the occupations $\ueta$. For instance, taking the time derivative of the cumulant generating function~\eqref{eq:def_psi}, we obtain
\begin{align}
    \label{eq:evolPsi}
    \nonumber
    \frac{\dd \psi}{\dd t}
    &= \frac{1}{\moy{\e^{\lambda X_t}}} \sum_{X} \sum_{\ueta} 
    \e^{\lambda X} \partial_t P_t(X,\ueta)
    \\
    &= \frac{1}{2}\left[ (\e^\lambda-1)(1-w_1) + (\e^{-\lambda}-1)(1-w_{-1}) \right]
    \:.
\end{align}
For the GDP-generating function, we must treat separately the cases $r = \pm 1$ and $r \neq \pm 1$ due to the presence of the tracer on a neighbouring site in the former case. Combining the definition~\eqref{eq:def_wr} with the master equation~\eqref{eq:master_eq}, we obtain for $r \neq \pm 1$,
\begin{multline}
    \label{eq:gdp_bulk_discr0}
    \partial_t w_r 
    = \frac{1}{2} \Delta w_r
    \\
    + \frac{1}{2} \sum_{\mu=\pm 1} 
    \left( \e^{\mu\lambda} C_{\mu, r+\mu} - C_{\mu, r} - (\e^{\mu\lambda}-1)(1-w_\mu) w_r \right)
    \:,
\end{multline}
where we have denoted {$\Delta u_r = u_{r+1} -2 u_r + u_{r-1}$} the discrete Laplacian and introduced the higher order correlation functions
\begin{equation}
    C_{\mu,r} \equiv
    \frac{
    \moy{\eta_{X_t+r}(1-\eta_{X_t+\mu}) \e^{\lambda X_t}}}
    {\moy{\e^{\lambda X_t}}}
    \:.
\end{equation}
Similarly, for $r = \nu = \pm 1$, we obtain
\begin{multline}
    \label{eq:gdp_bound_discr0}
    \partial_t w_{\nu} 
    = \frac{1}{2} \nabla_{\nu} w_{\nu}
    + \e^{\mu\lambda} C_{\nu, 2\nu} - C_{-\nu, \nu}
    \\
     - (\e^{-\nu\lambda}-1)(1-w_{-\nu}) w_\nu
    \:,
\end{multline}
where we have denoted $\nabla_\mu u_r = u_{r+\mu} - u_r$ the discrete gradients. The equations~(\ref{eq:gdp_bulk_discr0},\ref{eq:gdp_bound_discr0}) are not closed: they involve the higher order correlations $C_{\mu,r}$. In fact, we are facing an infinite hierarchy of equations. This is standard for many-body systems, and it constitutes the main obstacle to get exact analytical results. Before tackling this issue, let us first rewrite these equations in a more convenient form. We introduce the ``modified centered correlations'',
\begin{equation}
\label{eq:def_f}
    f_{\mu, r}(\lambda, t) \equiv \displaystyle
    C_{\mu,r} - 
    \begin{cases}
        (1-w_\mu) w_{r-\mu} & \text{ if } \mu r > 0 \:, \\
        (1-w_\mu) w_r  & \text{ if } \mu r < 0 \:,
    \end{cases}
\end{equation}
which allow us to rewrite the equations~(\ref{eq:gdp_bulk_discr0},\ref{eq:gdp_bound_discr0}) for the time evolution of the GDP-generating function in the compact form
\begin{equation}
    \label{eq:gdp_bulk_discr}
    \partial_t w_r 
    = \frac{1}{2} \Delta w_r - B_\nu \nabla_{-\nu}w_r 
    + \frac{1}{2} \sum_{\mu=\pm 1} 
    \left( \e^{\mu\lambda} f_{\mu, r+\mu} - f_{\mu, r} \right)
    \:,
\end{equation}
for $r \neq \pm 1$, and 
\begin{equation}
    \label{eq:gdp_bound_discr}
    \partial_t w_{\pm 1} 
    = \frac{1}{2} \nabla_\pm w_{\pm 1} + B_{\pm} w_{\pm 1} 
    + \frac{1}{2} \left( \e^{\pm\lambda} f_{\pm 1, \pm 2} - f_{\mp 1, \pm 1} \right)
    \:.
\end{equation}
We have denoted $\nu$ the sign of $r$, and
\begin{equation}
    \label{eq:DefBnu}
    B_{\pm} = \frac{\partial_t \psi}{\e^{\pm\lambda} - 1}
    \:.
\end{equation}
We have introduced the functions $f_{\mu,r}$ because of the scaling with space and time of this function, discussed below.

Finally, the GDP-generating function verifies
\begin{equation}
    \label{eq:CondInf_wr_discr}
    \lim_{r\to\pm\infty}w_r = \rho
    \:,
\end{equation}
because the density of bath particles far away from the tracer is not affected by the displacement of the tracer. Having derived the microscopic equations from the master equation of the SEP, we now turn to the hydrodynamic description of these quantities.

\subsection{Hydrodynamic limit at large times}
\label{sec:HydroLim}

\begin{figure}
    \centering
    \includegraphics[width=0.5\textwidth]{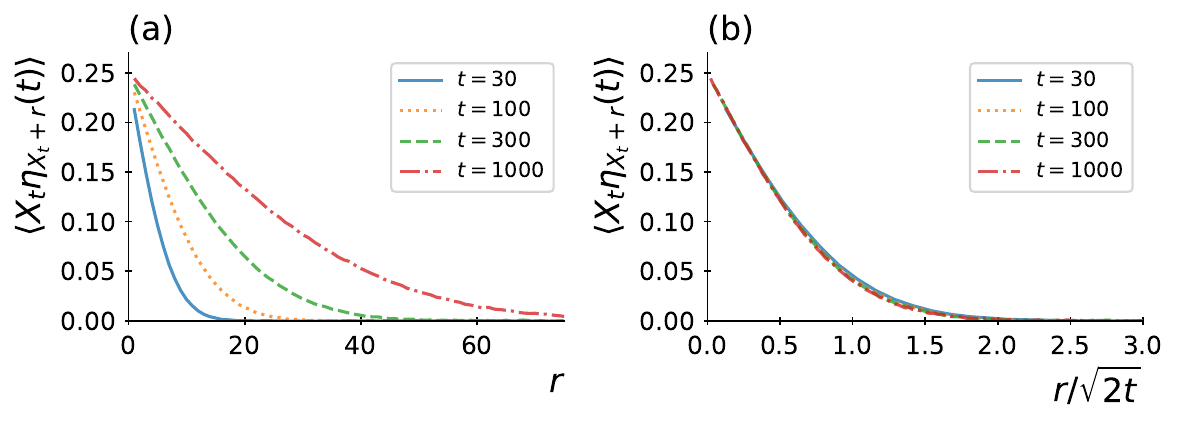}
    \caption{
First order in $\lambda$ of $w_r(\lambda,t) = \rho + \lambda \moy{\eta_{X_t+r}(t) X_t} + \O(\lambda^2)$ obtained by numerical simulations of the SEP (see Appendix~\ref{sec:AppNumSim}) performed on a lattice with $5000$ sites and $2500$ particles, at different times $t=30$, $t=100$, $t=300$ and $t=1000$. \textbf{(a)} $w_r(\lambda,t)$ represented as a function of $r$. We see that the correlations spread with time. \textbf{(b)} $w_r(\lambda,t)$ as a function of the scaling variable $r/\sqrt{2t}$. The collapse of the curves supports the scaling form~\eqref{eq:scaleing_wr}. The same collapse holds for higher orders in $\lambda$ of $w_r$, and for the functions $f_{\mu,r}$~\eqref{eq:scaling_f}.}
    \label{fig:SEP_scaling}
\end{figure}

At large time $t$, all the cumulants of the tracer present an anomalous $\sqrt{t}$ scaling~\cite{Imamura:2017,Imamura:2021}, leading us to define
\begin{equation}
\label{eq:scaling_psi}
    \hat\psi(\lambda) = 
    \lim_{t \to \infty} \frac{\psi(\lambda,t)}{\sqrt{2t}}
    \:,
    \quad
    \hat{\psi}(\lambda) = \sum_{n \geq 1} \hat\kappa_n \frac{\lambda^n}{n!}
    \:,
\end{equation}
where $\hat\kappa_n$ is the $n^{\mathrm{th}}$ cumulant of the tracer's position (rescaled by $\sqrt{2t}$). We also need the large $t$ scaling of the GDP-generating function~\eqref{eq:def_wr}. Relying on observations based on numerical simulations (see Fig.~\ref{fig:SEP_scaling}), we found that they satisfy a diffusive scaling
\begin{equation} 
    \label{eq:scaleing_wr}
    w_r(\lambda, t) \equi{t\to\infty}  \Phi\left(v = \frac{r}{\sqrt{2t}}, \lambda\right)
    \:,
    \quad
    \Phi(v, \lambda) = \sum_{n \geq 1} \Phi_n(v) \frac{\lambda^n}{n!}
    \:.
\end{equation}
In addition, we will see below that this scaling form is compatible with the known scaling of the cumulant generating function~\eqref{eq:scaling_psi}. Finally, we also need the scaling of the higher order correlation functions $f_{\mu,r}$~\eqref{eq:def_f} which appear in the time evolution of the GDP-generating functions~(\ref{eq:gdp_bulk_discr},\ref{eq:gdp_bound_discr}). Relying again on numerical observations, we infer the scaling form
\begin{multline} \label{eq:scaling_f}
    f_{\mu, r}(\lambda, t) 
    = \frac{1}{\sqrt{2t}} F_\mu\left(v=\frac{r}{\sqrt{2t}}, \lambda\right)
    \\
    + \frac{1}{2t} G_\mu\left(v=\frac{r}{\sqrt{2t}}, \lambda\right) + \mathcal{O}(t^{-3/2})
    \:.
\end{multline}
As we will see below, the leading order $F_\mu$ of these correlations will not be sufficient to write the long-time limit of the bulk equation~\eqref{eq:gdp_bulk_discr}, we will also need the corrections $G_\mu$. In the following, we will drop the dependence on $\lambda$ of $\Phi$, $F_\mu$ and $G_\mu$ for simplicity.

Having written the scalings at large time of all the functions of interest, we can now turn to the derivation of the hydrodynamic equations for the SEP. We first straightforwardly obtain the boundary conditions at $\pm \infty$ for $\Phi$ from~\eqref{eq:CondInf_wr_discr}, which read
\begin{equation}
    \label{eq:CondInf_Phi}
    \lim_{v \to \pm \infty} \Phi(v) = \rho
    \:.
\end{equation}
Next, we consider the equation for the cumulant generating function $\psi$~\eqref{eq:evolPsi}. Since $\psi$ has the anomalous scaling~\eqref{eq:scaling_psi}, the l.h.s. of Eq.~\eqref{eq:evolPsi} behaves as $1/\sqrt{t}$. This implies that the r.h.s. must also vanish at order $0$ in $t$, yielding
\begin{equation}
    (\e^\lambda -1) (1-\Phi(0^+)) + 
    (\e^{-\lambda} -1) (1-\Phi(0^-))
    = 0 \:,
\end{equation}
which we can rewrite as
\begin{equation}
\label{eq:cancel_veloc}
    \frac{1-\Phi(0^-)}{1-\Phi(0^+)}
    = \e^{\lambda}
    \:.
\end{equation}
This yields a first equation, valid in the hydrodynamic limit, relating the values of the scaling function $\Phi$ on both sides of the tracer. Note that corrections to the r.h.s. of Eq.~\eqref{eq:evolPsi} must be of order $1/\sqrt{t}$ because $\psi \propto \sqrt{t}$, which is compatible with the scaling form~\eqref{eq:scaleing_wr} of $w_r$.

We now examine the bulk equation~\eqref{eq:gdp_bulk_discr}, describing the time evolution of $w_r$. Considering $r$ of order $\sqrt{t}$, we get for the l.h.s. $\partial_t w_r = v \Phi'(v)/(2t)$ with $v = r/\sqrt{2t}$. On the other hand, for the r.h.s. we first get a contribution at order $t^{-1/2}$ which must therefore vanish, and thus leads to
\begin{equation}
\label{eq:cancel_F}
    (\e^\lambda - 1) F_1(v) + (\e^{-\lambda} - 1) F_{-1}(v) = 0
    \:.
\end{equation}
Writing Eq.~\eqref{eq:gdp_bulk_discr} at order $t^{-1}$, we get
\begin{equation}
    \label{eq:bulk_hydro_0}
    \Phi''(v) + 2 \left(v + \frac{\nu\hat{\psi}}{\e^{\nu \lambda} - 1} \right) \Phi'(v) + C(v) = 0
    \:,
\end{equation}
where $\nu = \sign(v)$ and we have defined
\begin{equation}
    \label{eq:exprCv}
    C(v) = (\e^\lambda - 1)F_1'(v) + 
    \sum_{\mu=\pm 1} (\e^{\mu\lambda}-1) G_\mu(v)
    \:.
\end{equation}
Let us turn to Eq.~\eqref{eq:gdp_bound_discr} for the time evolution of $w_{\pm 1}$. Due to the scaling form~\eqref{eq:scaleing_wr}, the l.h.s. behaves as $t^{-3/2}$. However, the r.h.s. is of order $t^{-1/2}$, which yields
\begin{equation}
    0 = \frac{1}{2} \Phi'(0^\pm) + \frac{\hat\psi}{\e^{\pm \lambda}-1} \Phi(0^\pm)
    + \frac{1}{2} \left( \e^{\pm \lambda} F_{\pm 1}(0^\pm) - F_{\mp 1}(0^\pm) \right)
    \:.
\end{equation}
Using the relation between $F_1$ and $F_{-1}$~\eqref{eq:cancel_F} derived previously, we get that the last term of the previous equation vanishes, so that we obtain
\begin{equation}
\label{eq:Bound_Phi}
    \Phi'(0^\pm) \pm \frac{2\hat\psi}{\e^{\pm \lambda}-1} \Phi(0^\pm) = 0
    \:.
\end{equation}
We stress that, although the microscopic equation~\eqref{eq:gdp_bound_discr} involves the higher order correlations $f_{\mu,r}$, its hydrodynamic version~\eqref{eq:Bound_Phi} is closed: it involves only the functions $\hat\psi$ and $\Phi$ of interest. Therefore, in this limit, only the bulk equation~\eqref{eq:bulk_hydro_0} is not closed since it involves the unknown functions $F_\mu$ and $G_\mu$ which describe higher order correlation functions.
Our main goal now is to find a closed bulk equation, that would allow to break the infinite hierarchy and compute the profile $\Phi$.

\section{Limiting cases}
\label{sec:LimCases}

In this Section, we consider limiting cases in which the bulk equation~(\ref{eq:bulk_hydro_0}) can be closed more easily than in the general case, and can thus be solved for the scaling function $\Phi$ of the GDP-generating function.

The results presented have been previously published in Ref.~\cite{Poncet:2021}. We reproduce them here for completeness, as they form the starting point of our discussion in Section~\ref{sec:CloseEqSEP}, and provide details on their derivation.

\subsection{First order in $\lambda$}

At lowest order in $\lambda$, the correlations $f_{\mu,r}$~(\ref{eq:def_f}) take the form
\begin{equation}
  f_{\mu,r} = \moy{\eta_{X_t+r}(1-\eta_{X_t+\mu})} -
  \rho (1-\rho)
  + \O(\lambda)
  \:,
\end{equation}
where we have used that $\moy{\eta_r} = \rho$. Since, at equilibrium, the occupation numbers of the SEP are independent~\cite{Spitzer:1974}, we thus have that $f_{\mu, r} = \O(\lambda)$. In the long time limit, this implies for the scaling form~(\ref{eq:scaling_f}) that $F_\mu(v) = \O(\lambda)$ and $G_\mu(v) = \O(\lambda)$. Consequently, we find that the unknown function $C(v) = \O(\lambda^2)$ due to the definition~(\ref{eq:exprCv}). This implies that the bulk equation for $\Phi_1$ is closed, and can be written as
\begin{equation}
\label{eq:EqOrder1}
  \Phi_1''(v) + 2 v \Phi_1'(v) = 0
  \:.
\end{equation}
The solution on the two domains $\mathbb{R}^\pm$ is $\Phi_1(v) = a_{\pm} \erfc(\pm v)$ for $v \gtrless 0$, where we have used the condition $\Phi_1(\pm \infty) = 0$ deduced from~(\ref{eq:CondInf_Phi}). We determine the integration constants $a_\pm$ by using the boundary conditions~(\ref{eq:Bound_Phi}) at first order in $\lambda$,
\begin{equation}
  \Phi_1'(v) + \hat\kappa_2 \rho = 0
  \Rightarrow a_\pm = \pm \frac{\sqrt{\pi}}{2} \hat\kappa_2 \rho
  \:.
\end{equation}
The second cumulant $\hat\kappa_2$ is determined by writing Eq.~(\ref{eq:cancel_veloc}) at first order in $\lambda$, from which we recover the well-known expression~\cite{Arratia:1983}
\begin{equation}
  \label{eq:kappa2_SEP}
  \hat\kappa_2 = \frac{1-\rho}{\rho \sqrt{\pi}}
  \:,
\end{equation}
and therefore
\begin{equation}
  \label{eq:Phi1_SEP}
  \Phi_1(v) = \sign(v) \frac{1-\rho}{2} \erfc(\abs{v})
  \:.
\end{equation}
This expression, obtained in~\cite{Poncet:2021}, gives the large time behaviour of the correlation function $\moy{X_t \eta_{X_t+r}} - \moy{\eta_{X_t+r}} \moy{X_t} \equiv \mathrm{Cov}(X_t,\eta_{X_t+r})$, with $v=r/\sqrt{2t}$, in the SEP, at arbitrary density $\rho$.

\subsection{Dense limit}

We now consider the high density limit $\rho\to 1$. In this case, both the cumulants and the shifted GDP-generating function $\Phi(v)-\rho$ scale as $1-\rho$ (see for instance~(\ref{eq:kappa2_SEP}) and~(\ref{eq:Phi1_SEP}) for the scaling at lowest order in $\lambda$). This leads us to define
\begin{equation}
  \label{eq:defScalHighDens}
  \check\Phi(v) = \lim_{\rho\to 1} \frac{\Phi(v) - \rho}{1-\rho}
  \:.
\end{equation}
The unknown function $C(v)$~(\ref{eq:exprCv}) is defined from higher order correlation functions, which involve product of occupation numbers at different sites. It is thus of order $(1-\rho)^2$ and is subleading in the bulk equation~(\ref{eq:bulk_hydro_0}) in this limit, which becomes
\begin{equation}
\label{eq:bulkHighDens}
    \check \Phi''(v) + 2v \check \Phi'(v) = 0.
\end{equation}
Solving this equation on both domains $v>0$ and $v<0$, combined with
the condition $\check\Phi(\pm \infty) = 0$ deduced
from~(\ref{eq:CondInf_Phi},\ref{eq:defScalHighDens}), we obtain
$\check\Phi(v) = A_\pm \erfc(\pm v)$ for $v \gtrless 0$. The
integration constants $A_\pm$ are then determined by the boundary
conditions at zero~(\ref{eq:Bound_Phi}) in the limit $\rho \to 1$,
\begin{equation}
  \check\Phi'(0^\pm) = \frac{\mp 2 \hat{\psi}}{\e^{\pm \lambda}-1}
  \Rightarrow
  A_\pm = \sqrt{\pi} \frac{\hat{\psi}}{\e^{\pm \lambda}-1}
  \:.
\end{equation}
Finally, using the last condition~(\ref{eq:cancel_veloc}), we determine $\hat{\psi}$ in terms of $\lambda$ in the dense limit:
\begin{equation}
  \hat{\psi}(\lambda) = \frac{2(1-\rho)}{\sqrt{\pi}} \sinh^2 \left( \frac{\lambda}{2} \right)
  \:,
\end{equation}
which coincides with the result obtained from a different approach in~\cite{Illien:2013}. Additionally, we get the profile~\cite{Poncet:2021}
\begin{equation}
    \label{eq:PhiHighDens}
  \Phi(v) = \rho + (1-\rho) \frac{1-\e^{\mp \lambda}}{2} \erfc(\pm v)
  \:,
\end{equation}
which quantify all correlation functions between the tracer and the bath of surrounding particles $\moy{X_t^n \eta_{X_t+r}}$ in the high density regime, for large times.

Note that, in the dense limit, the microscopic equations~(\ref{eq:gdp_bulk_discr},\ref{eq:gdp_bound_discr},\ref{eq:DefBnu}) can actually be solved at all times~\cite{Poncet:2021}. This leads to an exact time dependent solution for both the profiles and the generating function~\cite{Poncet:2022}.

\subsection{Dilute limit}

Despite the particle-hole symmetry of the SEP, the low density and the
high density limit are not equivalent, because we explicitly broke this
symmetry by following the dynamics of a particle (the tracer). In
order to properly define the low density limit, one should rescale the
distances by the mean distance between the particles ($1/\rho$) and
the time by the mean time between the interaction of two particles
($1/\rho^2$). We thus take $\rho \to 0$ with $z = \rho r$ and
$\tau = \rho^2 t$ fixed. We still have the same scaling variable
$v = r/\sqrt{2t} = z/\sqrt{2\tau}$.

Since the distances scale as $1/\rho$, $\lambda$ scales as $\rho$ so that the product $\lambda X_t$ remains of order $1$. This implies that the boundary conditions at zero~\eqref{eq:Bound_Phi} become
\begin{equation}
    \label{eq:Bound_Phi_lowDens}
    \Phi'(0^\pm) + \frac{\hat{\psi}}{\lambda} \Phi(0^\pm) = 0
    \:,
\end{equation}
and the cancellation of the velocity~\eqref{eq:cancel_veloc} yields
\begin{equation}
    \label{eq:cancel_veloc_lowDens}
    \Phi(0^+) - \Phi(0^-) = \lambda 
    \:.
\end{equation}
In the dilute limit, the unknown function $C(v)$ which encodes the higher order correlations is not negligible, and the bulk equation~\eqref{eq:bulk_hydro_0} is therefore not closed. In principle this requires to determine these higher order correlation functions in order to obtain $C(v)$. Nevertheless, in~\cite{Poncet:2021}, a closure relation was found, which allows to determine $C(v)$ and thus yields a closed bulk equation,
\begin{equation}
\label{eq:decoup_1_low_c}
    \Phi''(v) + 2(v+\xi) \Phi'(v) = 0
    \:,
    \quad
    \xi = \frac{\dd \hat\psi}{\dd \lambda}
    \:.
\end{equation}
Solving this equation on $v>0$ and $v<0$ with the condition $\Phi(\pm \infty) = \rho$, we get $\Phi(v) = \rho + A_{\pm} \erfc(\pm (v+\xi))$ for $v \gtrless 0$. The integration constants $A_\pm$ are determined by the boundary conditions at zero~\eqref{eq:Bound_Phi_lowDens},
\begin{equation}
    \label{eq:CoefsSolLowDens}
    A_\pm = 
    \frac{\rho \hat{\psi}}
    {\pm \lambda \frac{\e^{-\xi^2}}{\sqrt{\pi}} - \hat{\psi} \erfc(\pm \xi)}
    \:.
\end{equation}
Combined with the relation~\eqref{eq:cancel_veloc_lowDens}, this gives an implicit equation for the cumulant generating function (because $\xi = \frac{\dd \hat\psi}{\dd \lambda}$),
\begin{equation}
    \label{eq:EqCumulLowDens}
    \frac{\rho \hat{\psi} \erfc(\xi)}
    {\lambda \frac{\e^{-\xi^2}}{\sqrt{\pi}} - \hat{\psi} \erfc(\xi)}
    +\frac{\rho \hat{\psi} \erfc(-\xi)}
    {\lambda \frac{\e^{-\xi^2}}{\sqrt{\pi}} + \hat{\psi} \erfc(-\xi)}
    = \lambda 
    \:.
\end{equation}
Expanding in powers of $\lambda$ using~\eqref{eq:scaling_psi},
\begin{equation}
    \hat{\psi}(\lambda) = \sum_{n \geq 1} \hat\kappa_n \frac{\lambda^n}{n!}
    \:,
    \quad
    \xi = \frac{\dd \hat{\psi}}{\dd \lambda}
    = \sum_{n \geq 0} \hat\kappa_{n+1} \frac{\lambda^n}{n!}
    \:,
\end{equation}
we obtain from~\eqref{eq:EqCumulLowDens} the cumulants (the odd cumulants vanish by symmetry)
\begin{equation}
    \hat\kappa_2 = \frac{1}{\rho \sqrt{\pi}} 
    \:,
    \quad
    \hat\kappa_4 = \frac{3(4-\pi)}{\rho^3 \pi^{3/2}}
    \:,
    \quad
    \cdots
\end{equation}
which coincides with those obtained previously~\cite{Sadhu:2015}. Furthermore, we obtain the profiles $\Phi_n$~\eqref{eq:scaleing_wr} which encore the correlations between the displacement of the tracer and the density of bath particles. The exact expressions can be obtained by expanding~\eqref{eq:EqCumulLowDens} in powers of $\lambda$ and plugging the result into the solution of~\eqref{eq:decoup_1_low_c}. For instance the first profiles are found to be given by~\cite{Poncet:2021},
\begin{align}
    \Phi_0(v) &= \rho \:,
    \\
    \Phi_1(v) &= \frac{1}{2} \erfc(v)
    \:,
    \\
    \Phi_2(v) &= \frac{1}{2 \rho} \erfc(v) - \frac{2}{\pi \rho} \e^{-v^2}
    \:.
\end{align}
These expressions are more complex than those obtained in the high density regime, see Eq.~\eqref{eq:PhiHighDens}. Nevertheless they remain explicit at all orders in terms of the error function and its derivatives.

\section{SEP with a tracer at arbitrary density}
\label{sec:CloseEqSEPGen}

The main goal of this Section is to derive the closed equation verified by the profile $\Phi$ in the SEP, Eq.~\eqref{eq:LinEqOmegaSummary} below, which we have announced in Ref.~\cite{Grabsch:2022}. Here we present details of the approach that led us to obtain this equation, and discuss its consequences, as well as its resolution in Fourier space. We additionally provide a numerical scheme to directly compute this profile in real space.

\subsection{A closed integral equation}
\label{sec:CloseEqSEP}

The bulk equation~\eqref{eq:bulk_hydro_0}, obtained from the microscopic equation is valid at arbitrary density. But this equation is not closed: it involves the unknown function $C(v)$ which contains the higher order correlations. Our aim is to construct an equation for the profile $\Phi$ which is closed. 

From the closed bulk equations we obtained at high density~\eqref{eq:bulkHighDens} and low density~\eqref{eq:decoup_1_low_c}, we guess that the general equation at arbitrary density takes the form
\begin{equation}
\label{eq:expectedEqPhi}
  \Phi''(v) + 2(v+\xi) \Phi'(v) =
  \rhs
  \:,
\end{equation}
with $\rhs$ a right hand side to be determined. For this new equation to be closed, $\rhs$ should be expressed in terms of the function $\Phi(v)$ only. From our previous results~\cite{Poncet:2021} recalled in Section~\ref{sec:LimCases}, $\rhs$ vanishes in both limits $\rho \to 0$ and $\rho \to 1$, as well as at first order in $\lambda$ (because of~\eqref{eq:EqOrder1}). For this latter reason, when expanding this equation in powers of $\lambda$, we expect that, at order $n$, the r.h.s. $\rhs$ will act as a source term in the equation for $\Phi_n$, by involving only the profiles $\Phi_m$ at lower orders $m < n$. Furthermore Eq.~\eqref{eq:expectedEqPhi}, combined with the boundary conditions~(\ref{eq:CondInf_Phi},\ref{eq:cancel_veloc},\ref{eq:Bound_Phi}), has to reproduce the known cumulants of the position of the tracer~\cite{Imamura:2017,Imamura:2021}.

Finally, we have some constraints on how the different parameters ($\lambda$, $\hat{\psi}$, $\xi$ and $\rho$) should appear in $\rhs$:
\begin{enumerate}[label={(C\arabic*)}]
    \item \label{Constr:lamb} $\lambda$ is explicitly involved in the hydrodynamic equations of Section~\ref{sec:HydroLim} only through expressions of the form $\e^{\pm \lambda}-1$, so we expect that only these combinations appear.
    \item \label{Constr:params} In the low density equation~\eqref{eq:decoup_1_low_c}, $\hat{\psi}$ does not appear explicitly, only its derivative $\xi = \frac{\dd \hat{\psi}}{\dd \lambda}$ is involved, so we expect the same to happen at arbitrary density. Similarly, we do not expect other parameters, such as the density $\rho$, to appear explicitly in the equation.
    \item \label{Constr:time} The equation we write should have a ``proper scaling'' with time. Indeed, the bulk equation~\eqref{eq:bulk_hydro_0} (which we aim to replace) is obtained by expanding at order $1/t$ the microscopic equation~\eqref{eq:gdp_bulk_discr0}, so it should be the same for this new equation. For instance, the functions $\Phi$, $\Phi'$ and $\Phi''$ originate respectively from terms of orders $t^0$, $t^{-1/2}$ and  and $t^{-1}$ in the microscopic equations. Similarly, the scaling variable $v = r/\sqrt{2t}$ originates from a $t^{-1/2}$ term, and the same argument holds for $\xi$. One can thus check that, with these scalings, the l.h.s. of Eq.~\eqref{eq:expectedEqPhi} indeed originates from the scaling $1/t$. The same should hold for the r.h.s. $\rhs$.
\end{enumerate}

As a starting point to infer the structure of $\rhs$, we compute its lowest order in $\lambda$. To this end, we have computed $\Phi$ explicitly using the formalism of Macroscopic Fluctuation Theory (MFT, see Appendix~\ref{sec:AppendixMFT}) up to order $3$ in $\lambda$ included.
In particular, we recover the first order $\Phi_1$, given by~\eqref{eq:Phi1_SEP}, and the following orders are given by~\eqref{eq:PhiMFTflat}.
Plugging the expression of $\Phi$~\eqref{eq:PhiMFTflat} into the l.h.s. of~\eqref{eq:expectedEqPhi}, we get
\begin{multline}
    \label{eq:PhiFlatOrder3}
  \rhs =
  \frac{2(1-\rho)^2}{\rho}
  \Bigg(\frac{v}{4 \sqrt{2\pi}}
    \e^{-\frac{v^2}{2}} \erfc \left( \frac{\abs{v}}{\sqrt{2}} \right)
    \\
    - \sign(v) \frac{\e^{-v^2}}{4\pi} \Bigg) \lambda^3
  + \O(\lambda^4)
  \:.
\end{multline}
The goal is now to rewrite this expression in terms of the lowest orders of $\Phi$, as guessed above. In particular, we aim to rewrite the above expression in terms of $\Phi_1$, given by~\eqref{eq:Phi1_SEP}. This cannot be done straightforwardly, as the arguments of the error functions do not match between~\eqref{eq:PhiFlatOrder3} and~\eqref{eq:Phi1_SEP}. We need to find transformation which, acting on $\Phi_1$~\eqref{eq:Phi1_SEP}, changes the argument of the error function by $\sqrt{2}$. This can be done by using that
\begin{equation}
    \int_0^\infty \dd z \: \Phi_1'(v+z) \Phi_1'(-z)
    = \frac{(1-\rho)^2}{\sqrt{2\pi}} \e^{-\frac{v^2}{2}} \erfc \left( \frac{v}{\sqrt{2}} \right)
    \:,
\end{equation}
we can rewrite the r.h.s. of~\eqref{eq:PhiFlatOrder3} as
\begin{equation}
\label{eq:FirstGuessRHSflat}
 \rhs
  =
  \left\lbrace
  \begin{array}{ll}
  \displaystyle
    -\frac{\lambda}{\rho}
     \int_0^\infty \dd z \:
     \Phi''(-z)
     \Phi'(v+z)
  + \O(\lambda^4) & \text{for } v >0 \:, \\[0.4cm]
  \displaystyle
    -\frac{\lambda}{\rho}
     \int_0^\infty \dd z \:
     \Phi''(z)
     \Phi'(v-z)
  + \O(\lambda^4) & \text{for } v < 0 \:. 
  \end{array}
  \right.
\end{equation}
The constraint~\ref{Constr:params} imposes to rewrite $\rho$ in terms of $\Phi$. Since $\Phi(v) = \rho + \O(\lambda)$, we can for instance replace $\rho$ by $\Phi(0^\pm)$.
This intuition is confirmed by considering the case an in initial step initial condition (corresponding to an initial mean density $\rho_-$ for $x<0$ and $\rho_+$ for $x>0$). In this case, we also derive in Appendix~\ref{sec:AppendixMFT} the profiles at first orders,
\begin{equation}
\label{eq:PhiOrder0Step0}
    \Phi_0(v) = \frac{\rho_+}{2} \erfc(-v-\hat\kappa_1)
    + \frac{\rho_-}{2}\erfc(v+\hat\kappa_1)
    \:,
\end{equation}
\begin{multline}
    \Phi_1(v) =
    \hat\kappa_2 \Phi_0'(v)
    + \frac{1}{\Phi_0(0)} \Phi_0(v) (1-\Phi_0(v))
    \\
    - \frac{\rho_+(1-\rho_+)}{2 \Phi_0(0)} \erfc(-v-\hat\kappa_1)
    \\
    - \frac{(\rho_+-\rho_-)^2}{\Phi_0(0)} \int_0^\infty \frac{\dd z}{2 \sqrt{2\pi}} 
    \e^{- \frac{1}{2}(v+2\hat\kappa_1 + z)^2} \erfc \left( \frac{\abs{v-z}}{\sqrt{2}} \right)
    \:,
\end{multline}
where $\hat{\kappa}_1$ and $\hat{\kappa}_2$ can be determined explicitly, but their expressions are not required here. Plugging these expressions into the l.h.s. of~\eqref{eq:expectedEqPhi}, we obtain,
\begin{multline}
    \label{eq:rhsForStepOrder1}
  \rhs
  =
  - \sign(v) \frac{\Phi_0'(v)\Phi_0'(0)}{\Phi_0(0)} \lambda
  \\
  - \lambda \frac{(\rho_+-\rho_-)^2}{\Phi_0(0)}
  \partial_v \left(
    \frac{1}{2 \sqrt{2\pi}} \e^{-\frac{1}{2}(v+2\hat{\kappa}_1)^2}
    \erfc \left(
      \frac{\abs{v}}{\sqrt{2}}
    \right)
  \right)
  \\
  -
  \lambda \frac{(\rho_+-\rho_-)^2}{\Phi_0(0)}
  \frac{\hat{\kappa}_1}{\sqrt{2\pi}} \e^{-\frac{1}{2}(v+2\hat{\kappa}_1)^2}
    \erfc \left(
      \frac{\abs{v}}{\sqrt{2}}
    \right)
  + \O(\lambda^2)
  \:.
\end{multline}
Notice that, from the expression of the profile $\Phi_0$ at the lowest order~\eqref{eq:PhiOrder0Step0}, we have, for $v>0$,
\begin{multline}
    \int_0^\infty \dd z \:
    \Phi_0'(v+z)\Phi_0'(-z)
  =
  \\
   \frac{(\rho_+-\rho_-)^2}{2\sqrt{2\pi}}
  \e^{-\frac{(v+2\hat{\kappa}_1)^2}{2}}
  \erfc\left( \frac{v}{\sqrt{2}} \right)
  \:,
\end{multline}
and a similar expression holds for $v<0$. This allows us to rewrite the r.h.s. of~\eqref{eq:rhsForStepOrder1} as (after integration by parts, and using that $\xi = \hat\kappa_1 + \O(\lambda^2)$),
\begin{multline}
\label{eq:RHSOrder1Step}
  \rhs
  =
  -\frac{\lambda}{\Phi_0(0)}
     \int_0^\infty \dd z \:
     \Phi''(-z)
     \Phi'(v+z)
     \\
- 2 \xi \frac{\lambda}{\Phi_0(0)}
     \int_0^\infty \dd z \:
     \Phi'(-z)
     \Phi'(v+z)
  + \O(\lambda^2)
  \:,
\end{multline}
for $v>0$.
This equation reduces to~\eqref{eq:FirstGuessRHSflat} in the case of a uniform density $\rho$, since in that case $\xi = \O(\lambda)$ and $\Phi_0(0) = \rho + \O(\lambda)$.
Using now the constraint~\ref{Constr:lamb}, we can rewrite $\lambda/\Phi_0(0)$ as $(\e^{\lambda}-1)/\Phi(0^+)$ or $(\e^{-\lambda}-1)/\Phi(0^-)$ at lowest order in $\lambda$. Therefore, it leads us to propose an equation of the form
\begin{multline}
\label{eq:guessBulk0}
  \Phi''(v) + 2(v+\xi) \Phi'(v)
  =
  \frac{\e^{-\lambda}-1}{\Phi(0^-)}
     \int_0^\infty \dd z \:
     \Phi''(-z)
     \Phi'(v+z)
     \\
     + 2 \xi 
     \frac{\e^{-\lambda}-1}{\Phi(0^-)}
     \int_0^\infty \dd z \:
     \Phi'(-z)
     \Phi'(v+z)
     + \O(\lambda^4)
    \:,
\end{multline}
for $v>0$. Note that this equation satisfies the constraint~\ref{Constr:time}. Using this equation in the case of the uniform density, combined with the boundary conditions~(\ref{eq:cancel_veloc},\ref{eq:Bound_Phi}), we can compute the cumulants predicted by this equation. The first two cumulants $\hat\kappa_2$ and $\hat\kappa_4$ coincide with the ones computed previously~\cite{Arratia:1983,Krapivsky:2015a,Imamura:2017}, but not the following one $\hat\kappa_6$. This means that~\eqref{eq:guessBulk0} is not the desired equation for $\Phi$.

Having introduced convolutions when going from order $2$ to order $3$ in $\lambda$, we can try to add double convolutions to go beyond these orders, such as
\begin{equation}
    \int_0^\infty \dd z \int_0^\infty \dd z' \: \Phi'(z+z'+v) \Phi'(-z) \Phi'(-z')
    \:,
\end{equation}
or
\begin{equation}
    \int_0^\infty \dd z \int_0^\infty \dd z' \: \Phi'(z+v) \Phi'(-z-z') \Phi'(z')
    \:.
\end{equation}
Note that these are the only double convolutions that respect two features found in~\eqref{eq:guessBulk0}: (i) the sum of the arguments of the $\Phi'$s is $v$, and (ii) the arguments of $\Phi'$ do not change sign inside the integration domains.
One could then expect to find different combinations of these terms, with different prefactors $(\e^{\pm \lambda}-1)/\Phi(0^\pm)$. After trying several of these combinations,
we found that the equation (valid for $v>0$)
\begin{widetext}
\begin{multline}
  \label{eq:EqPhiOrder6p}
   \Phi''(v) + 2(v+\xi) \Phi'(v)
  =
   \frac{\e^{-\lambda}-1}{\Phi(0^-)}
    \int_{0}^\infty \dd z \:
    \Phi''(-z) \Phi'(v+z)
  +
  2\xi \frac{\e^{-\lambda}-1}{\Phi(0^-)}
  \int_{0}^\infty \dd z \:
  \Phi'(-z) \Phi'(v+z)
  \\
  + \frac{(\e^{\lambda}-1)(\e^{-\lambda}-1)}{\Phi(0^+)\Phi(0^-)}
  \int_0^\infty \dd z \int_0^\infty \dd z'
  \: \Phi''(-z-z') \Phi'(z+v) \Phi'(z')
  - \frac{(\e^{-\lambda}-1)^2}{\Phi(0^-)^2} \int_0^\infty \dd z \int_0^\infty \dd z'
  \: \Phi''(z+z'+v) \Phi'(-z) \Phi'(-z')
  \\
  + 2 \xi \frac{(\e^{\lambda}-1)(\e^{-\lambda}-1)}{\Phi(0^+)\Phi(0^-)}
  \int_0^\infty \dd z \int_0^\infty \dd z'
  \: \Phi'(-z-z') \Phi'(z+v) \Phi'(z')
  \\
  - 2 \xi \frac{(\e^{-\lambda}-1)^2}{\Phi(0^-)^2} \int_0^\infty \dd z \int_0^\infty \dd z'
  \: \Phi'(z+z'+v) \Phi'(-z) \Phi'(-z')
  + \O(\lambda^6)
  \:,
\end{multline}
\end{widetext}
properly reproduces all the known cumulant $\hat\kappa_n$ for $n \leq 6$~\cite{Imamura:2017}.
The structure of this equation leads us to introduce the two functions
\begin{equation}
  \label{eq:defNewFct}
    \Omega_\pm(v)
    = \mp \frac{\e^{\pm\lambda}-1}{\Phi(0^\pm)} \Phi'(v)
    = 2 \hat{\psi} \frac{\Phi'(v)}{\Phi'(0^\pm)}
     \quad \text{ for } v \gtrless 0 \:,
\end{equation}
where we have used the boundary condition~\eqref{eq:Bound_Phi} in the last equality. Equation~\eqref{eq:EqPhiOrder6p} then takes the more compact form
\begin{widetext}
\begin{multline}
  \label{eq:EqPhiOrder6pNew}
  \Omega_+'(v) + 2(v+\xi) \Omega_+(v)
  =
  \int_{0}^\infty \dd z \:
  \Omega_-'(-z) \Omega_+(v+z)
  +
  2\xi \int_{0}^\infty \dd z \:
  \Omega_-(-z) \Omega_+(v+z)
  \\
  - \int_0^\infty \dd z \int_0^\infty \dd z'
  \: \Omega_-'(-z-z') \Omega_+(z+v) \Omega_+(z')
  - 2 \xi
  \int_0^\infty \dd z \int_0^\infty \dd z'
  \: \Omega_-(-z-z') \Omega_+(z+v) \Omega_+(z')
  \\
  - \int_0^\infty \dd z \int_0^\infty \dd z'
  \: \Omega_+'(z+z'+v) \Omega_-(-z) \Omega_-(-z')
  - 2 \xi \int_0^\infty \dd z \int_0^\infty \dd z'
  \: \Omega_+(z+z'+v) \Omega_-(-z) \Omega_-(-z')
  + \O(\lambda^6)
\:.
\end{multline}
\end{widetext}
In order to extend this structure to arbitrary order in $\lambda$, it is convenient to introduce a matrix operator $\mathscr{L}$, defined as
\begin{equation}
  \label{eq:DefOpL}
  \mathscr{L} =
  \begin{pmatrix}
    \mathscr{L}_{++} & \mathscr{L}_{+-} \\
    \mathscr{L}_{-+} & \mathscr{L}_{--}
  \end{pmatrix}
  \:,
\end{equation}
where the four blocks are integral operators:
\begin{subequations}
  \label{eq:DefOpLOmega}
  \begin{align}
    (\mathscr{L}_{++}f)(v)
    &=
       \int_{0}^\infty \dd z \: \Omega_-(-z) f(v+z)
      \:,
    \\
    (\mathscr{L}_{+-}f)(v)
    &=
      -\int_{0}^\infty \dd z \: \Omega_+(v+z) f(-z) 
      \:,
    \\
    (\mathscr{L}_{-+}f)(v)
    &=
     -\int_{0}^\infty \dd z \: \Omega_-(v-z) f(z)
      \:,
    \\
    (\mathscr{L}_{--}f)(v)
    &=
      \int_{0}^\infty \dd z \: \Omega_+(z) f(v-z)
      \:.
  \end{align}
\end{subequations}
To show that this operator indeed produces the structure of Eq.~\eqref{eq:EqPhiOrder6pNew}, let us apply it to the column vector $( \Omega_+ \: 0)^{\mathrm{T}}$,
\begin{equation}
  \mathscr{L}
  \begin{pmatrix}
    \Omega_+ \\ 0
  \end{pmatrix}
  =
  \begin{pmatrix}
    \displaystyle
     \int_{0}^\infty \dd z \: \Omega_-(-z) \Omega_+(v+z)
     \\[0.3cm]
    \displaystyle
     -\int_{0}^\infty \dd z \: \Omega_-(v-z)\Omega_+(z)
  \end{pmatrix}
  \:.
\end{equation}
The first component is the second term in the r.h.s of~\eqref{eq:EqPhiOrder6pNew}.
Applying $\mathscr{L}$ a second time, we get for the first component,
\begin{multline}
  \int_0^\infty \dd z \int_0^\infty \dd z' \: \Omega_+(z+z'+v) \Omega_-(-z) \Omega_-(-z')
  \\
  +\int_0^\infty \dd z \int_0^\infty \dd z' \: \Omega_-(-z-z') \Omega_+(z+v) \Omega_+(z')
  \:,
\end{multline}
which corresponds to the last terms in the second and third line of Eq.~\eqref{eq:EqPhiOrder6pNew}. The other terms in~\eqref{eq:EqPhiOrder6pNew} can also be expressed similarly using integration by parts.
Finally, using that $\Omega_+(0) = \Omega_-(0)$ by definition (see Eq.~\eqref{eq:defNewFct}), we can rewrite~(\ref{eq:EqPhiOrder6pNew}) as
\begin{multline}
  \label{eq:ClosedEqOmegP0}
    2 v \Omega_+(v) +
    (\partial_v + 2\xi) \left[ (1+\mathscr{L})^{-1}
      \begin{pmatrix}
        \Omega_+ \\ 0
      \end{pmatrix} (v)
    \right]_1
    \\
    + \Omega_+(v)  \left[ (1+\mathscr{L})^{-1}
      \begin{pmatrix}
        \Omega_+ \\ 0
      \end{pmatrix} (0)
    \right]_1
    = \O(\lambda^6)
    \:.
\end{multline}
We assume that this structure holds true at all orders in $\lambda$, which leads us to propose the equation at all orders in $\lambda$,
\begin{multline}
  \label{eq:ClosedEqOmegP}
    2 v \Omega_+(v) +
    (\partial_v + 2\xi) \left[ (1+\mathscr{L})^{-1}
      \begin{pmatrix}
        \Omega_+ \\ 0
      \end{pmatrix} (v)
    \right]_1
    \\
    + \Omega_+(v)  \left[ (1+\mathscr{L})^{-1}
      \begin{pmatrix}
        \Omega_+ \\ 0
      \end{pmatrix} (0)
    \right]_1
    = 0
    \:.
\end{multline}
At this stage, the step from~\eqref{eq:ClosedEqOmegP0} to~\eqref{eq:ClosedEqOmegP} is a guess. We will argue below in Section~\ref{sec:ExpLambdaSEP} that~\eqref{eq:ClosedEqOmegP} is actually exact since, as a byproduct of this equation, we recover all the cumulants that were computed previously in~\cite{Imamura:2017,Imamura:2021}. We will further argue in Section~\ref{sec:Mallick}, that, since the discovery of this equation in~\cite{Grabsch:2022}, it has been proved recently in~\cite{Mallick:2022}, in an equivalent form written below. This confirms that our guess, in the step from~\eqref{eq:ClosedEqOmegP0} to~\eqref{eq:ClosedEqOmegP}, is not an approximation and is actually exact.
Following the same procedure for $v<0$, we obtain
\begin{multline}
  \label{eq:ClosedEqOmegM}
    2 v \Omega_-(v) +
    (\partial_v + 2\xi) \left[ (1+\mathscr{L})^{-1}
      \begin{pmatrix}
        0 \\ \Omega_-
      \end{pmatrix} (v)
    \right]_2
    \\
    +\Omega_-(v)  \left[ (1+\mathscr{L})^{-1}
       \begin{pmatrix}
        0 \\ \Omega_-
      \end{pmatrix} (0)
    \right]_2
    = 0
    \:.
\end{multline}
These equations are closed, and allow for a numerical determination of $\Omega_+$ and $\Omega_-$. Nevertheless, they are rather complicated, and we now look for a simpler version of these equations. To do so, it is instructive to look for a perturbative solution again. By definition~\eqref{eq:defNewFct}, $\Omega_\pm$ is small when $\hat{\psi}$ is small. Writing the first orders of the solution of~(\ref{eq:ClosedEqOmegP},\ref{eq:ClosedEqOmegM}) in terms of $\hat\psi$ instead of $\lambda$, we find that the solution can be conveniently expressed as
\begin{multline}
    \Omega_{\pm}(v)=
    \left( \omega \: \e^{\xi^2} \right) \Omega_\pm^{(1)}(v)
    +\left( \omega \: \e^{\xi^2} \right)^2 \Omega_\pm^{(2)}(v)
    \\
    +\left( \omega \: \e^{\xi^2} \right)^3 \Omega_\pm^{(3)}(v)
    + \O(\omega^4)
    \:,
\end{multline}
where we introduced the parameter $\omega$, which is defined from $\hat{\psi}$ by
\begin{equation}
\label{eq:expXpsi}
  \omega =  2 \sqrt{\pi} \hat{\psi} + \sqrt{2} \pi \: \hat{\psi}^2 +
  \frac{2 \pi^{3/2}}{9} (9 - 4 \sqrt{3}) \hat{\psi}^3 + \O(\hat{\psi}^4)
  \:.
\end{equation}
It takes a simpler form by inverting the series
\begin{equation}
\label{eq:expXpsiInv}
    \hat{\psi} = \frac{\omega}{2 \sqrt{\pi}}
  - \frac{\omega^2}{4 \sqrt{2 \pi}} 
  + \frac{\omega^3}{6 \sqrt{3 \pi}}  + \O(\omega^4)
  \:.
\end{equation}
In addition,
\begin{subequations}
\label{eq:FirstOrdersOmegaX}
  \begin{align}
    \label{eq:ExprOmPOrder1X}
    \Omega_\pm^{(1)}(v)
    &= \frac{1}{\sqrt{\pi}} \e^{-(v+\xi)^2}
      \:,
    \\
     \Omega_\pm^{(2)}(v)
    &= -\frac{1}{2 \sqrt{2 \pi}}
      \e^{-\frac{1}{2}(v+2\xi)^2} \erfc \left( \pm\frac{v}{\sqrt{2}} \right)
      \:,
    \\
    \nonumber
    \Omega_\pm^{(3)}(v)
    &=  \frac{1}{4 \sqrt{3 \pi}} \e^{- \frac{1}{3}(v+3\xi)^2}
      \left[
      \erfc\left( \pm\sqrt{\frac{2}{3}} v \right)
      \right.
      \\
      &\left.
      + \erfc\left( \pm\frac{v}{\sqrt{6}} \right)
      - 4 \mathrm{T} \left( \frac{v}{\sqrt{3}}, \sqrt{3} \right)
      \right]
      \:,
  \end{align}
\end{subequations}
where $\mathrm{T}$ is Owen's T-function defined by~\cite{Owen:1980}
\begin{equation}
  \label{eq:defOwenT}
  \mathrm{T}(h,a) = \frac{1}{2\pi}
  \int_0^a \frac{\e^{-\frac{h^2}{2}(1+x^2)}}{1+x^2} \dd x
  \:.
\end{equation}
From these expressions, we notice that we can express $\Omega_+^{(2)}$ from  $\Omega_\pm^{(1)}$ as
\begin{equation}
    \Omega_+^{(2)}(v) = - \int_0^\infty \Omega_+^{(1)}(v+z) \Omega_-^{(1)}(-z)
    \dd z
    \:.
\end{equation}
Similarly, $\Omega_+^{(3)}$ can be expressed in terms of the previous orders as
\begin{multline}
    \Omega_+^{(3)}(v) = 
    -\int_0^\infty \Omega_+^{(2)}(v+z) \Omega_-^{(1)}(-z) \dd z
    \\
    - \int_0^\infty \Omega_+^{(1)}(v+z) \Omega_-^{(2)}(-z) \dd z
    \:.
\end{multline}
We can thus write from these first orders a compact equation verified by $\Omega_+(v)$,
\begin{equation}
    \label{eq:BilinEqOmP}
    \Omega_+(v) + \int_{0}^\infty \Omega_+(v+z) \Omega_-(-z) \dd z
    = \omega \e^{\xi^2} \Omega_+^{(1)}(v)
    \:,
\end{equation}
with $\Omega_+^{(1)}(v)$ given by~\eqref{eq:ExprOmPOrder1X}. Proceeding similarly with $\Omega_-(v)$, we get
\begin{equation}
    \label{eq:BilinEqOmM}
    \Omega_-(v)+ \int_{0}^\infty \Omega_+(z) \Omega_-(v-z) \dd z
    = \frac{\omega}{\sqrt{\pi}} \: \e^{-(v+\xi)^2 + \xi^2} 
    \:.
\end{equation}
These two coupled equations can be mapped onto two independent linear equations, upon analytic continuation of $\Omega_+$ to $v<0$ and $\Omega_-$ to $v>0$~\cite{Arabadzhyan:1987}. This finally gives two Wiener-Hopf integral equations,
\begin{equation}
  \label{eq:IntegEqOmP}
    \Omega_+(v) + \int_{-\infty}^0 \dd z\: K(v-z) \: \Omega_+(z)  
    = K(v)
    \:,
\end{equation}
\begin{equation}
    \label{eq:IntegEqOmM}
    \Omega_-(v) +\int_0^{\infty} \dd z \: K(v-z) \: \Omega_-(z) 
    = K(v)
    \:,
\end{equation}
with the Gaussian kernel
\begin{equation}
    \label{eq:KernelK}
    K(v) \equiv \frac{\omega}{\sqrt{\pi}} \: \e^{-(v+\xi)^2 + \xi^2}
    \:.
\end{equation}
The parameter $\omega$ is determined by the boundary condition at the origin, which is a consequence of the definition of $\Omega_\pm$~\eqref{eq:defNewFct},
$\Omega_+(0) = \Omega_-(0) = 2 \hat{\psi}$.

\subsection{Summary of the equations}

We have found that the rescaled derivatives of the generalised density profiles,
\begin{equation}
    \label{eq:DefOmegaSummary}
    \Omega_\pm(v)
    = 2 \hat{\psi} \frac{\Phi'(v)}{\Phi'(0^\pm)}
     \quad \text{ for } v \gtrless 0 \:,
\end{equation}
obey the simple bilinear equation
\begin{equation}
    \label{eq:BilinEqOmegaSummary}
    \Omega_\pm(v) + \int_{0}^\infty \Omega_\pm(v \pm z) \Omega_\mp( \mp z) \dd z
    = K(v)
    \:,
\end{equation}
with the kernel
\begin{equation}
   \label{eq:KernelKSummary}
    K(v) \equiv \frac{\omega}{\sqrt{\pi}} \: \e^{-(v+\xi)^2 + \xi^2}
    \:,
\end{equation}
where the parameter $\omega$ is determined from the boundary condition
\begin{equation}
    \Omega_+(0) = \Omega_-(0) = 2 \hat{\psi}
    \:,
\end{equation}
which is a consequence of the definition~\eqref{eq:DefOmegaSummary}.
The bilinear equation~\eqref{eq:BilinEqOmegaSummary} is actually equivalent to the linear one~\cite{Arabadzhyan:1987}
\begin{equation}
    \label{eq:LinEqOmegaSummary}
    \Omega_\pm(v) + \int_{\mathbb{R}^\mp} \dd z \: K(v-z) \: \Omega_\pm(z)  
    = K(v)
    \:,
\end{equation}
upon analytical continuation of $\Omega_+$ to $v<0$ and $\Omega_-$ to $v>0$. This equation is completed by the boundary relations, derived above from microscopic considerations,
\begin{equation}
\label{eq:Bound_PhiSummary}
    \Phi'(0^\pm) \pm \frac{2\hat\psi}{\e^{\pm \lambda}-1} \Phi(0^\pm) = 0
    \:,
\end{equation}
\begin{equation}
    \label{eq:CondInf_PhiSummary}
    \lim_{v \to \pm \infty} \Phi(v) = \rho
    \:,
\end{equation}
\begin{equation}
   \label{eq:cancel_velocSummary}
    \frac{1-\Phi(0^-)}{1-\Phi(0^+)}
    = \e^{\lambda}
    \:.
\end{equation}
Together, these equations fully determine the generalised density profile $\Phi(v)$, as we discuss below. But first, let us make a few comments on these main results.

\subsection{Discussion}

First, although Eqs.~(\ref{eq:BilinEqOmegaSummary},\ref{eq:LinEqOmegaSummary}) have been shown to hold true up to order $3$ in $\lambda$ for the profiles $\Phi$ (and reproduce the cumulants up to order $6$), we argue below that these equations are valid at any order in $\lambda$ and are thus actually exact.
In fact, since the publication of~\cite{Grabsch:2022}, this has been shown by solving the MFT equations in~\cite{Mallick:2022} (see Section~\ref{sec:Mallick} below).

Second, we would like to stress that Wiener-Hopf equations arise in various contexts. A first example concerns the persistence (decay of the survival probability) of discrete time random walks~\cite{Bray:2013}. A second one can be found in the study of the large deviations of the KPZ equation~\cite{Krajenbrink:2021}. Actually, in this latter case, the Wiener-Hopf equation was interpreted in terms of the persistence of a random walk, giving the authors a simple way to prove the equivalence between two (linear and bilinear) integral equations. Here, a similar interpretation can be used to derive the linear equation~\eqref{eq:LinEqOmegaSummary} from the bilinear equation~\eqref{eq:BilinEqOmegaSummary}. More precisely, we can write
\begin{equation}
  \Omega_\pm(v) = - \sum_{n \geq 1} (-\omega \e^{\xi^2})^n p_n^\pm(v)
  \:,
\end{equation}
where
\begin{equation}
  p_n^+(v) = \mathbb{P}(X_0 = 0, X_1 <0, \ldots, X_{n-1}<0, X_n = v)
  \:,
\end{equation}
with $X_n$ the position after $n$ step of a random walker, whose Gaussian stationary increments are given by 
\begin{equation}
  X_{n+1} - X_n = \eta_n
  \:,
  \quad
  \eta_n \: \mathrm{i.i.d.},
  \quad
  \mathbb{P}(\eta) = \frac{\e^{-(\eta+\xi)^2}}{\sqrt{\pi}}
  \:.
\end{equation}
Importantly, we stress that this interpretation of $\Omega_\pm$ provides a clear meaning to the analytic continuation of $\Omega_\pm$ to $\mathbb{R}^\mp$ which appear in Eq.~\eqref{eq:LinEqOmegaSummary}.

\subsection{Solving the Wiener-Hopf equations for the profiles and the cumulants}
\label{sec:SolutionAndProcedure}

The solution of the Wiener-Hopf equations~\eqref{eq:LinEqOmegaSummary} can be expressed in terms of the half Fourier transforms $\hat{\Omega}_\nu^{(\pm)}$ of $\Omega_\pm$ (and their analytic continuations), defined as
\begin{equation}
    \label{eq:defFTOmega}
  \hat{\Omega}_\nu^{(\pm)}(k) \equiv
  \int_{\mathbb{R}^{\pm}} \dd v \: \Omega_\nu(v) \: \e^{\I k v}
  \:,
  \quad
  \nu = \pm
  \:.
\end{equation}
The solution of~\eqref{eq:LinEqOmegaSummary} is given in~\cite{Polyanin:2008} for the Fourier transforms $\hat{\Omega}_\mp^{(\pm)}$ of the analytic continuations of $\Omega_\pm$ to $\mathbb{R}^\mp$,
\begin{multline}
    \hat{\Omega}_\mp^{(\pm)}(k)
    = \exp \left[
    - \frac{1}{2} \ln (1+\hat{K}(k))
    \right.
    \\
    \left.
    \mp \frac{1}{2 \I \pi}
    \intVp_{-\infty}^\infty \frac{\ln(1+\hat{K}(u))}{u-k} \dd u 
    \right]
    -1
    \:,
\end{multline}
where $\intVp$ denotes a principal value integral, and
\begin{equation}
    \label{eq:FTkernel}
    \hat{K}(k) \equiv \int_{-\infty}^\infty K(v) \e^{\I k v} \dd v
    = \omega \: \e^{- \frac{1}{4}(k+2\I \xi)^2}
    \:.
\end{equation}
The Fourier transforms $\hat{\Omega}_\pm^{(\pm)}$ in the other domain are deduced by taking the Fourier transform of~\eqref{eq:LinEqOmegaSummary},
\begin{equation}
    \hat\Omega_\pm^{(\pm)}(k)  = \hat{K}(k) + \hat\Omega_\pm^{(\mp)}(k) (1+\hat{K}(k))
    \:.
\end{equation}
By using that $1+\hat{K} = \exp \ln (1+\hat{K})$, we obtain
\begin{multline}
    \hat\Omega_\pm^{(\pm)}(k)  = \exp \left[
    \frac{1}{2} \ln (1+\hat{K}(k))
    \right.
    \\
    \left.
    \mp \frac{1}{2 \I \pi}
    \intVp_{-\infty}^\infty \frac{\ln(1+\hat{K}(u))}{u-k} \dd u 
    \right]
    -1
    \:.
\end{multline}
We can rewrite this solution in a more compact form as
\begin{equation}
    \label{eq:IntegSolOmega}
    \hat{\Omega}_\mp^{(\pm)}(k)
    = \exp \left[
    \mp \frac{1}{2 \I \pi}
    \int_{-\infty}^\infty \frac{\ln(1+\hat{K}(u))}{u-k \pm \I 0^+} \dd u 
    \right] -1
    \:,
\end{equation}
or alternatively
\begin{equation}
    \label{eq:IntegSolOmega2}
    \hat{\Omega}_\pm^{(\pm)}(k)
    = \exp \left[
    \int_{\mathbb{R}^\pm} \frac{\dd x}{2\pi} \e^{\I k x}
    \int_{-\infty}^\infty \dd u \:  \e^{-\I u x} \ln(1+\hat{K}(u))
    \right]-1
    \:.
\end{equation}
With this last form, inserting the expression of $\hat{K}$~\eqref{eq:FTkernel} and expanding in powers of $\omega$, we get
\begin{multline}
    \int_{\mathbb{R}^\pm} \frac{\dd x}{2\pi} \e^{\I k x}
    \int_{-\infty}^\infty \dd u \:  \e^{\I u x} \ln(1+\hat{K}(u))
    \\
    = -Z_\pm \left(\omega, \xi - \frac{\I k}{2} \right)
      \:,
\end{multline}
with
\begin{equation}
    \label{eq:DefZpm}
    Z_\pm \left(\omega, \xi \right)
    = \frac{1}{2} \sum_{n \geq 1}
      \frac{(-\omega \: \e^{\xi^2})^n}{n}
      \erfc \left( \pm \sqrt{n} \: \xi \right)
      \:.
\end{equation}
With the definitions~\eqref{eq:defFTOmega} of the Fourier transforms, this explicitly gives
\begin{equation}
  \label{eq:SolOmPFourier}
    \int_0^\infty  \Omega_+(v) \e^{\I k v} \dd v
    = \exp \left[- Z_+ \left( \omega, \xi - \frac{\I k}{2} \right) \right]-1
    \:,
\end{equation}
\begin{equation}
  \label{eq:SolOmMFourier}
    \int_{-\infty}^0  \Omega_-(v) \e^{\I k v} \dd v
    = \exp \left[- Z_- \left( \omega, \xi - \frac{\I k}{2} \right) \right]-1
    \:.
\end{equation}
These expressions are convenient to get perturbative expansions in powers of $\lambda$, as we will see below. For arbitrary values of $\lambda$, the expressions~(\ref{eq:IntegSolOmega},\ref{eq:IntegSolOmega2}) are more practical. For instance, setting $k = \pm \I s$ in~\eqref{eq:IntegSolOmega} and letting $s \to \infty$, we get
\begin{equation}
    \hat\Omega_\pm^{(\pm)}( \pm \I s) 
    \underset{s \to \infty}{\simeq} 
    \frac{1}{2\pi s} 
    \int_{-\infty}^\infty \ln(1+\hat{K}(u)) \dd u
    \:,
\end{equation}
while on the other hand, from the definition~\eqref{eq:defFTOmega} we have
\begin{equation}
    \hat\Omega_\pm^{(\pm)}( \pm \I s) 
    \underset{s \to \infty}{\simeq}
    \frac{1}{s} \: \Omega_\pm(0)
    = \frac{1}{s} \: 2 \hat\psi
    \:,
\end{equation}
where we have used that $\Omega_\pm(0) = 2 \hat\psi$ from the definition of $\Omega_\pm$~\eqref{eq:DefOmegaSummary}. Combining these two asymptotic results, we get
\begin{equation}
    \hat\psi = \frac{1}{4 \pi} \int_{-\infty}^\infty \ln(1+\hat{K}(u)) \dd u
    \:.
\end{equation}
Using the expression of $\hat{K}$~\eqref{eq:FTkernel} and expanding in powers of $\omega$, this becomes
\begin{equation}
    \label{eq:RelPsiOmega}
    \hat\psi = 
    -\frac{1}{2 \sqrt{\pi}} \mathrm{Li}_{\frac{3}{2}}(- \omega)
    \:,
\end{equation}
where $\mathrm{Li}_\nu(z) = \sum_{n \geq 1} z^n/n^\nu$ is the polylogarithm function.
This expression is consistent with the first orders obtained previously~\eqref{eq:expXpsi}.
It relates the parameter $\omega$ and the cumulant generating function $\hat\psi$. Since $\xi = \frac{\dd \hat\psi}{\dd \lambda}$, one can thus think of $\Omega_\pm$ as parametrised by $\hat\psi$. At this stage, the function $\hat\psi(\lambda)$ is not known. It can be determined in the following way:
\begin{enumerate}
    \item The integration of $\Omega_\pm$ on $\mathbb{R^\pm}$ with the boundary conditions at infinity~\eqref{eq:CondInf_PhiSummary} gives a relation between $\Phi(0^+)$ and $\Phi'(0^+)$, and between $\Phi(0^-)$ and $\Phi'(0^-)$. It can be obtained by setting $k=0$ in~(\ref{eq:SolOmPFourier},\ref{eq:SolOmMFourier}):
    \begin{equation}
    \int_0^\infty \Omega_+ 
    = 2 \hat\psi \frac{\rho - \Phi(0^+)}{\Phi'(0^+)}
    = \exp\left[ - Z_+(\omega,\xi)\right]-1
    \:,
    \end{equation}
    \begin{equation}
    \int_{-\infty}^0 \Omega_-
    = 2 \hat\psi \frac{\Phi(0^-)-\rho}{\Phi'(0^-)}
    = \exp\left[ - Z_-(\omega,\xi)\right]-1
    \:.
    \end{equation}
    \item Combining these relations with the boundary conditions~\eqref{eq:Bound_PhiSummary} yield $\Phi(0^+)$ and $\Phi(0^-)$, parametrised by $\lambda$ and $\hat{\psi}$ (via $\omega$ and $\xi$):
    \begin{equation}
    \label{eq:Phi0P}
    \Phi(0^+) = \rho \frac{\e^\lambda-1}{\e^\lambda - \e^{-Z_+(\omega,\xi)}}
    \:,
    \end{equation}
    \begin{equation}
    \label{eq:Phi0M}
        \Phi(0^-) = \rho \frac{\e^{-\lambda}-1}{\e^{-\lambda} - \e^{-Z_-(\omega,\xi)}}
        \:.
    \end{equation}
    \item Using finally the relation~\eqref{eq:cancel_velocSummary} which relates $\Phi(0^+)$ and $\Phi(0^-)$, we obtain the cumulant generating function $\hat\psi(\lambda)$.
\end{enumerate}
We now illustrate how this procedure can be applied to obtain the cumulants and the profiles $\Phi_n$ at lowest orders in $\lambda$.

\subsection{Expansions in powers of \texorpdfstring{$\lambda$}{lambda}}
\label{sec:ExpLambdaSEP}

\subsubsection{For the cumulants}

Inserting the expansion of $\hat{\psi}$ in powers of $\lambda$~\eqref{eq:scaling_psi} into Eq.~\eqref{eq:RelPsiOmega}, we obtain the expansion of $\omega$ in powers of $\lambda$,
\begin{equation}
    \omega = \sqrt{\pi} \hat\kappa_2 \lambda^2
    + \frac{\sqrt{\pi}}{12} (\hat\kappa_4 + 3 \sqrt{2\pi} \hat\kappa_2^2) \lambda^4
    + \cdots
    \:.
\end{equation}
Note that the odd order cumulants vanish, as expected. We also have by definition
\begin{equation}
    \xi = \frac{\dd \hat{\psi}}{\dd \lambda} 
    = \hat\kappa_2 \lambda + \hat\kappa_4 \frac{\lambda^3}{6} + \cdots
    \:.
\end{equation}
Plugging these expansions into the expressions~(\ref{eq:Phi0P},\ref{eq:Phi0M}) give $\Phi(0^\pm)$ in terms of the cumulants. Inserting these results into the last relation~\eqref{eq:cancel_veloc}, we obtain the cumulants
\begin{equation}
  \hat\kappa_2 = \frac{1-\rho}{\rho \sqrt{\pi}}
  \:,
\end{equation}
\begin{widetext}
\begin{equation}
  \hat\kappa_4 = \frac{(1-\rho)}{\pi^{3/2} \rho^3}
  \left(
    12 (1-\rho)^2 - \pi (3 - 3 (4 - \sqrt{2})\rho + (8-3\sqrt{2}) \rho^2)
  \right)
  \:.
\end{equation}
\begin{multline}
  \hat\kappa_6 = 
  \frac{(1-\rho )}{\pi^{5/2} \rho^5}
  \left(
    30 \pi  \left(2 \left(9 \sqrt{2}-20\right) \rho^2+\left(60-18 \sqrt{2}\right) \rho -15\right) (1-\rho )^2
  \right.
  \\
  -\pi^2 \left(
    8 \left(-17+15 \sqrt{2}-5 \sqrt{3}\right) \rho^4
    +\left(480-300 \sqrt{2}+80 \sqrt{3}\right) \rho ^3
  \right.
  \\
  \left.
    \left.
      +5 \left(-114+45 \sqrt{2}-8 \sqrt{3}\right) \rho^2
      -45 \left(\sqrt{2}-6\right) \rho -45\right)+1020 (1-\rho )^4
  \right)
  \:.
\end{multline}
\end{widetext}
These expressions coincide with the cumulants computed by Bethe ansatz in Ref.~\cite{Imamura:2017}. This is expected, since we have constructed our starting equation~\eqref{eq:EqPhiOrder6p} to reproduce these cumulants. We have computed the next cumulants, up to order $10$ (this number is arbitrary, one can go further at the cost of longer computational time), by implementing the procedure described above with Mathematica. These cumulants also coincide with those obtained from Ref.~\cite{Imamura:2017}. This provides a strong nontrivial validation of our integral equations~(\ref{eq:BilinEqOmegaSummary},\ref{eq:LinEqOmegaSummary}). We have thus found an alternative parametrization for the cumulant generating function to the one obtained in~\cite{Imamura:2017}. We will show below that we can actually recover the exact same parametrization of Ref.~\cite{Imamura:2017}, and thus providing automatically the exact same cumulants as in~\cite{Imamura:2017}, at arbitrary order.

\subsubsection{For the generalised profiles}
\label{sec:ProfSEP}

Having obtained the cumulant generating function $\hat\psi(\lambda)$, we can go further than~\cite{Imamura:2017} and obtain the profiles $\Phi_n$ which encode the bath-tracer correlations. Indeed, we have the expression of $\xi = \frac{\dd \hat\psi}{\dd \lambda}$ and $\omega$ in terms of $\lambda$, via Eq.~\eqref{eq:RelPsiOmega}. We thus have $\Omega_\pm(v)$ in terms of $\lambda$ from the solution of~\eqref{eq:LinEqOmegaSummary}. We then obtain $\Phi(v)$ by integration of $\Omega_\pm$, with the definition~\eqref{eq:DefOmegaSummary}:
\begin{equation}
    \label{eq:ExprPhiP}
    \Phi(v>0) = \rho + \frac{\Phi(0^+)}{\e^\lambda-1} \int_{v}^\infty \Omega_+(z) \dd z
    \:,
\end{equation}
\begin{equation}
    \label{eq:ExprPhiM}
    \Phi(v<0) = \rho + \frac{\Phi(0^-)}{\e^{-\lambda}-1} \int_{-\infty}^v \Omega_-(z) \dd z
    \:,
\end{equation}
where we have used the boundary conditions at infinity~\eqref{eq:CondInf_PhiSummary}. Note that we have already obtained the expressions of $\Phi(0^\pm)$, given by~(\ref{eq:Phi0P},\ref{eq:Phi0M}), in the derivation of the cumulants. Therefore, expanding~\eqref{eq:ExprPhiP}, we get the profiles
\begin{widetext}
\begin{align}
    \label{eq:Phi0}
  \Phi_0(v>0)
  &= \rho \:,
  \\
    \label{eq:Phi1}
  \Phi_1(v>0)
  &= \frac{1-\rho}{2} \erfc(v) \:,
  \\
    \label{eq:Phi2}
  \Phi_2(v>0)
  &= \frac{(1-\rho)(1-2\rho)}{2\rho} \erfc(v)
    - \frac{2}{\pi} \frac{(1-\rho)^2}{\rho} \: \e^{-v^2}
    \:,
\end{align}
\begin{multline}
  \label{eq:Phi3}
  \Phi_3(v>0)
  =
  (1-\rho )\frac{2 (3+\pi ) \rho ^2-(12+\pi ) \rho +6 + \pi \rho(1-\rho)}
  {2 \pi  \rho ^2}
  \erfc(v)
  \\
  +3 (1-\rho)^2  \frac{2 (1-\rho) v - \sqrt{\pi } (1-2 \rho)}
  {\pi^{3/2} \rho ^2} \e^{-v^2}
  -\frac{3(1-\rho)^2}{4\rho} \erfc \left( \frac{v}{\sqrt{2}} \right)^2
  \:,
\end{multline}
\begin{figure*}
    \centering
    \includegraphics[width=\textwidth]{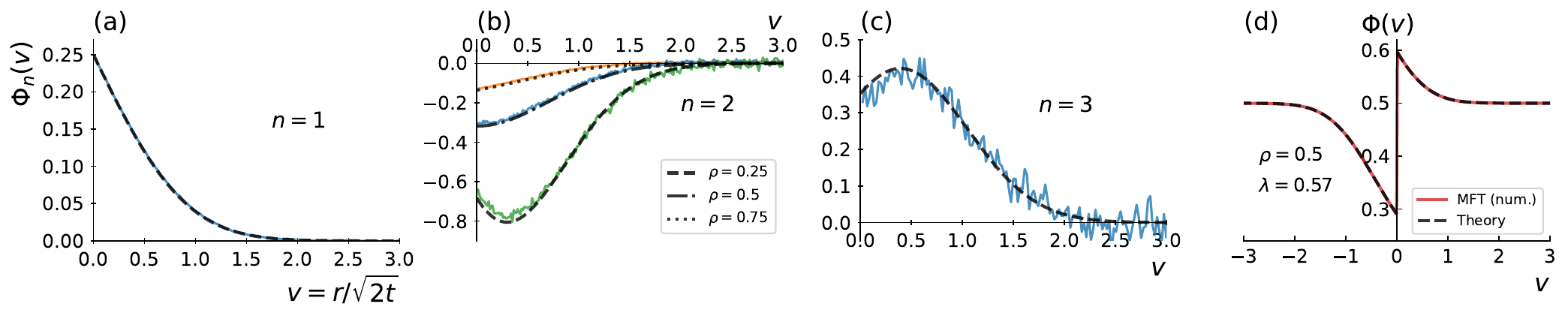}
    \caption{\textbf{SEP.} Generalised density profiles (GDP) of order \textbf{(a)} $n=1$ at density $\rho=0.5$, \textbf{(b)} $n=2$ at densities $\rho=0.25$, $0.5$ and $0.75$, and \textbf{(c)} $n=3$ at density $\rho=0.5$. The solid lines correspond to simulations of the SEP (see Appendix~\ref{sec:AppNumSim}), performed at time $t=3000$ on a lattice of $5000$ sites. The averaging is performed over $10^8$ realisations. The dashed lines are the theoretical predictions~(\ref{eq:Phi1},\ref{eq:Phi2},\ref{eq:Phi3}). \textbf{(d)} GDP-generating function at $\rho=0.5$ and $\lambda=0.57$ (actually, in the algorithm we specified $\xi=0.4$ as input to go beyond the perturbative regime near $\xi=0$ but keep a reasonable convergence time), obtained from solving numerically the Wiener-Hopf equation~\eqref{eq:LinEqOmegaSummary} (dashed line, see Section~\ref{sec:ConsRel}), compared to the numerical solution (red solid line) of the MFT equations (see Section~\ref{sec:CompInvScat} below) obtained from the algorithm described in~\cite{Krapivsky:2012}.}
    \label{fig:Sim_SEP}
\end{figure*}
\begin{multline}
  \label{eq:Phi4}
  \Phi_4(v>0) =(1-\rho)(1-2\rho) \frac{24 (1-\rho)^2 - \pi (3 (1-2\rho)^2 + 4
    (1-\rho)(1-2\rho)+18\rho (1-\rho)-4)}{2\pi \rho^3} \erfc(v)
  \\
  + 4(1-\rho)^2 \frac{3 \sqrt{\pi} (1-\rho)(1-2\rho)v
    -4(1-\rho)^2 (4+v^2) + \pi(5 (1-\rho)(1-2\rho) + 3 (1+\sqrt{2}) \rho(1-\rho)-2)
  }{\pi^2 \rho^3}\e^{-v^2}
  \\
  + 12 \sqrt{2} \frac{(1-\rho)^3}{\pi \rho^2} \e^{-\frac{v^2}{2}}
  \erfc \left( \frac{v}{\sqrt{2}} \right)
  - 3 \frac{(1-\rho)^2(1-2\rho)}{2 \rho^2}  \erfc \left( \frac{v}{\sqrt{2}} \right)^2
  \:.
\end{multline}
\begin{multline}
    \Phi_5(v>0) = -\frac{(1-\rho) \left(\pi ^2 \rho ^4
    +440 (1-\rho)^4+30 \pi  \left(2 \rho  \left(\left(2 \sqrt{2}-3\right) \rho-2\sqrt{2}+6\right)-3\right) 
    (1-\rho)^2\right)}{240 \pi^2 \rho^4}
   \erfc(v)
   \\
   + \frac{(1-\rho)^2}{24 \pi ^{5/2} \rho ^4} \left[
      4 \sqrt{\pi } (1-\rho)^2 \left(19 \rho +(4 \rho -2) v^2-11\right)
      +12 \pi  \left(\rho \left(\left(2 \sqrt{2}-5\right) \rho -2 \sqrt{2}+8\right)-2\right) (1-\rho) v
   \right.
   \\
   \left.
   -3 \pi ^{3/2} (1-2 \rho) \left(2 \rho \left(\left(\sqrt{2}-3\right) \rho -\sqrt{2}+5\right)-3\right)
   +4 (1-\rho)^3 v \left(2 v^2+27\right)
   \right]
   \e^{-v^2}
   \\
   + \frac{(1-\rho)^3\left(\sqrt{\pi } (1-2 \rho)-2 (1-\rho) v\right)}{2 \sqrt{2} \pi ^{3/2} \rho ^3}
   \e^{-\frac{v^2}{2}}
   \erfc \left(\frac{v}{\sqrt{2}}\right) 
   + \frac{(1-\rho)^2 \left(\pi  \left(2 \rho ^2-4 \rho +1\right)-12 (1-\rho)^2\right)}{32 \pi \rho ^3}
   \erfc\left(\frac{v}{\sqrt{2}}\right)^2
   \\
   - \frac{(1-\rho)^3}{4 \sqrt{3 \pi} \: \rho^2} 
   \int_v^{+\infty} \dd z \: \e^{-\frac{z^2}{3}} 
    \left[
      \erfc\left( \sqrt{\frac{2}{3}} z \right)
      + \erfc\left( \frac{z}{\sqrt{6}} \right)
      - 4 \mathrm{T} \left( \frac{z}{\sqrt{3}}, \sqrt{3} \right)
      \right]
   \:,
\end{multline}
\end{widetext}
with the Owen-T function defined in~\eqref{eq:defOwenT}. The expressions for $v<0$ can be deduced by the left/right symmetry $\Phi(-v,-\lambda) = \Phi(v,\lambda)$ which implies $\Phi_n(-v) = (-1)^n \Phi_n(v)$. The first few profiles $\Phi_n$ are represented in Fig.~\ref{fig:Sim_SEP}, compared to numerical simulations of the SEP. The small discrepancy in Fig.~\ref{fig:Sim_SEP}(b) for $\rho=0.25$ is due to finite size effects in the numerical simulations. This effect becomes smaller when increasing the system size and the time of the simulation, which both strongly impact the computational time.

The first order profile $\Phi_1(v)$ gives the long-time asymptotics of the covariance $\mathrm{Cov}(X_t, \eta_{X_t}+r)$, with $v = r/\sqrt{2t}$, see Eq.~\eqref{eq:Phi1Cov}. For $v>0$, this covariance is positive, indicating that an increase of $X_t$ (displacement towards the right) is correlated with an increase of the occupation of the sites in front of the tracer. The profile $\Phi_1$ thus provides a quantitative measurement of the "jam" that forms in front of the tracer when it moves in a given direction (see Fig.~\ref{fig:Sim_SEP}(a)).

Similarly, the second order profile $\Phi_2(v)$ is the long time limit of $\mathrm{Cov}(X_t^2, \eta_{X_t+r})$. It measures the correlations between the amplitude of the fluctuations of the tracer, and the density around it. This function is negative, meaning that these two quantities are anti-correlated. This can be interpreted in the following way: a decrease of the occupation of the sites around the tracer gives more space for the tracer to fluctuate and thus increases its fluctuations. Surprisingly, when the mean density of particles becomes less than $1/2$, $\Phi_2$ becomes non-monotonic, indicating that this anti-correlation effect is stronger at a given distance (rescaled by $\sqrt{t}$) from the tracer.

\subsubsection{Conservation relation and numerical resolution}
\label{sec:ConsRel}

Using the expressions above for the profiles, one can check that the conservation relation
\begin{equation}
\label{eq:ConsRel}
    \int_0^\infty (\Phi(v) - \rho) \dd v 
    - \int_{-\infty}^0 (\Phi(v) - \rho) \dd v
    = \rho \xi
\end{equation}
holds up to $\O(\lambda^6)$. We have further checked numerically that this relation holds non-perturbatively in $\lambda$.

This relation is particularly useful to implement a numerical computation of the profile $\Phi(v)$ from the closed Wiener-Hopf equation~\eqref{eq:LinEqOmegaSummary}. Indeed, we have an explicit analytical solution~(\ref{eq:SolOmPFourier},\ref{eq:SolOmMFourier}) only in Fourier space, and inverting it to real space can be difficult. Numerically, it is much more stable and faster to solve directly in real space by using the following procedure:
\begin{enumerate}
    \item Select an initial "guess" for the values of $\omega$ and $\xi$;
    \item Discretise the Wiener-Hopf equations~\eqref{eq:LinEqOmegaSummary} and solve them for $\Omega_\pm(v)$; 
    \item We then need to determine the parameters $\lambda$ and $\rho$. They can be deduced from $\Omega_\pm(v)$ from~(\ref{eq:ExprPhiP},\ref{eq:ExprPhiM}), but that introduces two new parameters, $\Phi(0^+)$ and $\Phi(0^-)$. One relation between them is given by the cancellation of the velocity~\eqref{eq:cancel_velocSummary}. In principle, the last relation needed is $\xi = \frac{\dd \hat{\psi}}{ \dd\lambda}$, since $\hat\psi = \Omega_+(0)/2$ is known. However, this relation is not practical to use because we cannot easy compute the derivative with respect to $\lambda$. This last equation is however conveniently replaced by the conservation relation~\eqref{eq:ConsRel}, which can be used straightforwardly;
    \item Finally, having determined all the parameters, the profile $\Phi(v)$ is obtained from~(\ref{eq:ExprPhiP},\ref{eq:ExprPhiM}).
\end{enumerate}
This procedure can be implemented easily thanks to the conservation relation~\eqref{eq:ConsRel}. We have used it to plot the profile $\Phi(v)$ for an arbitrary value of $\lambda$, as shown in Fig.~\ref{fig:Sim_SEP}(d).

\section{Extensions to other situations and observables in the SEP}
\label{sec:ExtOtherSit}

Although we have focused on the example of tracer diffusion in the SEP with a mean density $\rho$, our closed equations~\eqref{eq:LinEqOmegaSummary} can be applied to various other situations involving single-file diffusion, as we announced in Ref.~\cite{Grabsch:2022}. Here, we describe explicitly the extension to other situations and observables within the SEP, give exact expression for the profiles $\Phi$ and discuss some of the physical consequences of these results. We present extensions to other models than the SEP in the next Section.

\subsection{SEP with an initial step density profile}

We consider a SEP with a mean initial step density $\rho_+$ for $v>0$ and $\rho_-$ for $v<0$, which constitutes a paradigmatic example of a system that remains out-of-equilibrium at all times. The tracer is initially placed at the origin. The microscopic equations of Section~\ref{sec:MicroEq} are unchanged, and so is the hydrodynamic limit of Section~\ref{sec:HydroLim}. The boundary relations at the origin~(\ref{eq:Bound_PhiSummary},\ref{eq:cancel_velocSummary}) are unchanged. Only the boundary condition at infinity~\eqref{eq:CondInf_PhiSummary} is changed into
\begin{equation}
    \lim_{v \to \pm \infty} = \rho_\pm
\end{equation}
to take into account the imbalance of density.

In this case, we found that the closed equations~\eqref{eq:LinEqOmegaSummary} still apply. Indeed, by following the procedure described at the end of Section~\ref{sec:SolutionAndProcedure}, we obtain the cumulants $\hat\kappa_n$ of the position of the tracer, which coincide with those computed using Bethe ansatz in Ref.~\cite{Imamura:2017}. In this case, the left/right symmetry is broken by the difference of density on the two domains $x>0$ and $x<0$, so the odd cumulants are now nonzero. For instance, the first cumulant is obtained from the solution of the equation
\begin{equation}
    \frac{\rho_-}{1 + \sqrt{\pi} \: \hat\kappa_1 \e^{\hat\kappa_1^2} 
    \erfc(-\hat\kappa_1)}
    =
    \frac{\rho_+}{1 - \sqrt{\pi} \: \hat\kappa_1 \e^{\hat\kappa_1^2} 
    \erfc(\hat\kappa_1)}
    \:,
\end{equation}
and the higher order cumulants have explicit expressions in terms of $\hat\kappa_1$. For instance,
\begin{multline}
    \hat\kappa_2 = 
    \hat\kappa _1^2 \Bigg(
     2 \pi  \e^{2 \hat\kappa_1^2} 
     \hat\kappa_1 \erfc\left(\sqrt{2} \hat\kappa_1\right) -\sqrt{2 \pi }
     \\
    +\frac{4 \pi  \e^{2 \hat\kappa_1^2} \hat\kappa_1 \rho_+ 
    \left(\rho_+^2-3 \rho_- \rho_+ +2 \rho_-\right)} {\left(\rho_- -\rho_+\right)^3}
    \Bigg)
    \:,
\end{multline}
in agreement with~\cite{Imamura:2017}.

\begin{figure}
    \centering
    \includegraphics[width=\columnwidth]{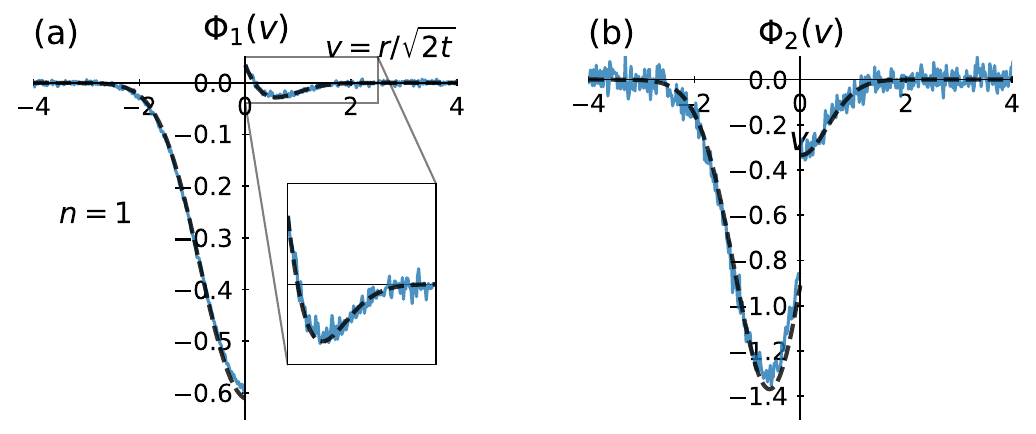}
    \caption{\textbf{SEP with an initial step density.} Generalised density profiles (GDP) of order \textbf{(a)} $n=1$ and \textbf{(b)} $n=2$ for a step of density $\rho_-=0.7$ and $\rho_+=0.2$. Solid lines: result of the simulations, computed at $t=1500$ on a lattice with $2000$ sites, with the averaging performed over $10^8$ realisations. Dashed lines: analytical predictions obtained from the resolution of the Wiener-Hopf equation~\eqref{eq:LinEqOmegaSummary}. For instance $\Phi_1(v)$ is given by~\eqref{eq:Phi1Step}.}
    \label{fig:Sim_SEP_step}
\end{figure}

Our procedure additionally yields the generalised density profiles, for instance
\begin{widetext}
\begin{equation}
    \Phi_0(v) = \frac{\rho_+}{2} \erfc(-v-\hat\kappa_1) 
    + \frac{\rho_-}{2} \erfc(v+\hat\kappa_1)
    \:,
\end{equation}
\begin{multline}
    \label{eq:Phi1Step}
    \Phi_1(v) = \frac{2 \sqrt{\pi } \e^{\hat\kappa _1^2} \hat\kappa _1 \rho _-
   \left(1-\rho_+\right)-\left(\rho _-\rho _+\right)^2}{2 \left(\rho _--\rho_+\right)}
   \erfc \left( \hat\kappa _1+v\right)
   \\
   -\frac{\e^{-\left(\hat\kappa _1+v\right)^2} \left(4 \sqrt{\pi } \e^{2\hat\kappa _1^2} \hat\kappa _1^3 \left(\left(\rho _- -\rho _+\right)^3
   \erfc\left(\sqrt{2} \hat\kappa _1\right)+2 \rho _+ \left(\rho
   _+^2+\rho _- \left(2-3 \rho _+\right)\right)\right)-2 \sqrt{2}
   \hat\kappa _1^2 \left(\rho _--\rho _+\right)^3\right)}{2 \left(\rho _--\rho _+\right)^2}
   \\
   + \frac{1}{2} \sqrt{\pi } \e^{\hat\kappa _1^2} \hat\kappa _1 \left(\rho _+-\rho_-\right) 
   \left(4 \mathrm{T} \left(\sqrt{2} \hat\kappa _1,\frac{\hat\kappa_1+v}{\hat\kappa _1}\right)
   -4 \mathrm{T}\left(2 \hat\kappa_1+v,\frac{v}{2 \hat\kappa_1+v}\right)
   + \erfc\left(\frac{2 \hat\kappa_1+v}{\sqrt{2}}\right) 
   - \erfc\left(\hat\kappa_1\right)\right)
   \:,
\end{multline}
\end{widetext}
for $v>0$ and $\hat\kappa_1>0$. Similar expressions can be written for $\hat\kappa_1 < 0$. The values of $\Phi_n(v < 0)$ can be obtained by the symmetry $\Phi(-v,\lambda,\rho_+,\rho_-) = \Phi(v, - \lambda, \rho_-, \rho_+)$. The first two profiles $\Phi_n$ for $n=1$ and $2$ are represented in Fig.~\ref{fig:Sim_SEP_step}.

Unlike the case of the flat initial density, the profiles $\Phi_n$ are no longer symmetric or anti-symmetric, but their physical meaning remains the same. $\Phi_1(v)$ again measures the covariance between $X_t$ and $\eta_{X_t+r}$ at large times. For $v>0$ and close to $0$, it is still positive, indicating that a displacement of the tracer towards the right leads to an increase of the density of particles in front of the tracer. Surprisingly, unlike the case of the flat density, $\Phi_1(v>0)$ changes sign at a given distance (rescaled by $\sqrt{t}$) from the tracer. This indicates that for $v$ larger than this critical value, the effect is inverted: a displacement of the tracer towards the right is correlated with a \textit{decrease} of the density. This unexpected behaviour is fully quantified by our computation of the generalised profile $\Phi_1$, and is stressed by the inset in Fig.~\ref{fig:Sim_SEP_step}(a).

\subsection{Another observable: the integrated current through the origin}
\label{sec:Current}

We consider another observable which has been the focus of many studies in single-file diffusion~\cite{Derrida:2009,Derrida:2009a,Krapivsky:2012}: the integrated current $Q_t$ thought the origin. It corresponds to the total number of particles that jumped across the origin from left to right, minus the number from right to left, up to time $t$. Microscopically, it is defined as
\begin{equation}
  Q_t = \sum_{r \geq 1}
  \left(
    \eta_r(t) - \eta_r(0)
  \right)
  \:.
\end{equation}
We are interested in the cumulant generating function of this observable,
\begin{equation}
  \label{eq:PsiQ}
  \psi_Q(\lambda) = \sqrt{2t} \: \hat{\psi}_Q = \ln \moy{\e^{\lambda Q_t}}
  \:,
\end{equation}
and the generalised density profiles for the current, which we define as
\begin{equation}
  w_{Q;r}(t) = \frac{\moy{\eta_r \: \e^{\lambda Q_t}}}{\moy{\e^{\lambda Q_t}}}
  \:.
\end{equation}
These profiles measure the correlation between the integrated current $Q_t$ and the density at a distance $r$ from the origin. At large time, they behave as
\begin{equation}
  \label{eq:PhiQ}
  w_{Q;r}(t) \underset{t \to \infty}{\simeq} \Phi_Q(v)
    = \sum_{n\geq 0} \Phi_{Q;n}(v) \frac{\lambda^n}{n!}
  \:,
  \quad
  v = \frac{r-\frac{1}{2}}{\sqrt{2t}}
  \:,
\end{equation}
where the $-\frac{1}{2}$ has been introduced such that $w_{Q;1}(t) \simeq \Phi(0^+)$ while $w_{Q;0}(t) \simeq \Phi(0^-)$ for $t \to \infty$.

We again define the functions $\Omega_\pm(v)$ by
\begin{equation}
\label{eq:defOmegaQ}
  \Omega_\pm(v) = 2 \hat{\psi}_Q \frac{\Phi_Q'(v)}{\Phi_Q'(0^{\pm})}
  \:.
\end{equation}
We have checked that these two functions still verify the integral equations~\eqref{eq:LinEqOmegaSummary}, but with the kernel~\eqref{eq:KernelKSummary} with $\xi = 0$:
\begin{equation}
    \label{eq:KernelQ}
    K(v) = \frac{\omega_Q}{\sqrt{\pi}} \: \e^{-v^2}
    \:.
\end{equation}
This still implies
\begin{equation}
    \label{eq:PsiQomega}
    \hat{\psi}_Q = - \frac{1}{2 \sqrt{\pi}}
    \mathrm{Li}_{\frac{3}{2}}(- \omega_Q)
    \:.
\end{equation}
The functions $\Omega_\pm$ are thus given by the solution~(\ref{eq:SolOmPFourier},\ref{eq:SolOmMFourier}), with $\xi = 0$.

In order to use this solution to deduce the profiles $\Phi_Q(v)$, we need the boundary conditions satisfied by $\Phi_Q$. In order to derive them, we follow the same approach we used for the tracer in Sections~\ref{sec:MicroEq} and~\ref{sec:HydroLim}. First, we write the time evolution of the cumulant generating function from the master equation,
\begin{multline}
    \label{eq:EvolPsiQ}
      \partial_t \ln \moy{\e^{\lambda Q_t}}
  = \frac{1}{2} \left[
    (\e^\lambda -1 )
    \frac{\moy{\eta_0(1-\eta_1) \e^{\lambda Q_t}}}{\moy{\e^{\lambda Q_t}}}
    \right.
    \\
    \left.
    +(\e^{-\lambda} -1 )
    \frac{\moy{\eta_1(1-\eta_0) \e^{\lambda Q_t}}}{\moy{\e^{\lambda Q_t}}}
  \right]
  \:.
\end{multline}
Similarly, we obtain the microscopic equations for the generalised profiles, which we write as
\begin{equation}
  \partial_t w_{Q;0} = (\e^{-\lambda} - (\e^{-\lambda}-1)w_{Q;0})
  \frac{\partial_t \ln \moy{\e^{\lambda Q_t}}}{\e^{-\lambda} -1}
  + \frac{w_{Q;-1} - w_{Q,0}}{2}
  \:,
\end{equation}
\begin{equation}
  \partial_t w_{Q;1} = (\e^{\lambda} - (\e^{\lambda}-1)w_{Q;1})
  \frac{\partial_t \ln \moy{\e^{\lambda Q_t}}}{\e^{\lambda} -1}
  + \frac{w_{Q;2} - w_{Q;1}}{2}
  \:.
\end{equation}
Taking the long time limit of these equations, with the scalings~(\ref{eq:PsiQ},\ref{eq:PhiQ}), we obtain at leading order
\begin{equation}
    \label{eq:CancelVQ}
    \frac{\Phi_Q(0^+) (1-\Phi_Q(0^-))}{\Phi_Q(0^-) (1-\Phi_Q(0^+))} = \e^{\lambda}
    \:,
\end{equation}
\begin{align}
    \label{eq:BoundZeroPQ}
    \Phi_Q'(0^-) 
    &= 2 \hat{\psi}_Q \left( \frac{1}{1-\e^{\lambda}} -  \Phi_Q(0^-) \right)
    \:,
    \\
    \label{eq:BoundZeroMQ}
    \Phi_Q'(0^+) 
    &= -2 \hat{\psi}_Q \left( \frac{1}{1-\e^{-\lambda}} - \Phi_Q(0^+) \right)
    \:.
\end{align}

\begin{figure}
    \centering
    \includegraphics[width=\columnwidth]{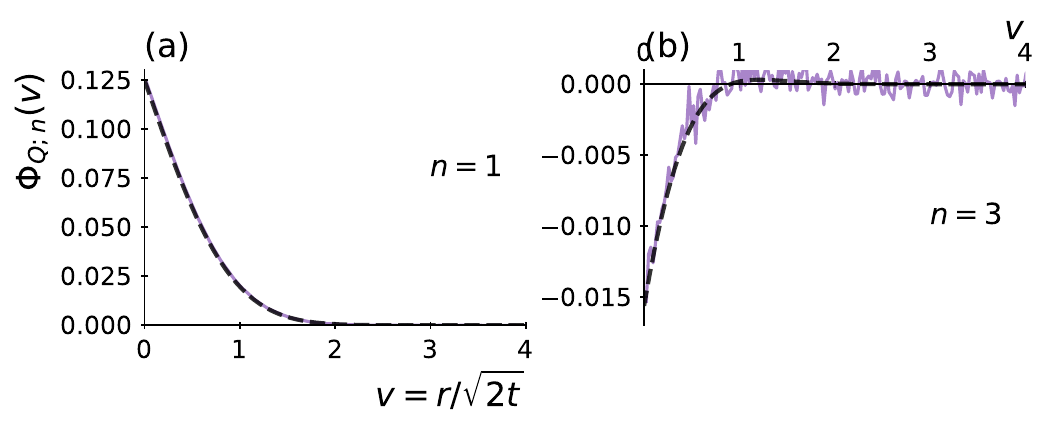}
    \caption{\textbf{Current in the SEP.} Generalised density profiles (GDP) $\Phi_{Q;n}(v)$ describing the correlations between the density and the current, at orders \textbf{(a)} $n=1$ and \textbf{(b)} $n=3$, at density $\rho = 0.5$. We do not show the profile $n=2$ because it is zero for this density. Solid lines: GDP computed from the simulations of the SEP, at time $t=900$. The averaging is performed over $10^8$ realisations. Dashed lines: theoretical predictions~(\ref{eq:Phi1Q},\ref{eq:Phi3Q}).}
    \label{fig:Sim_SEP_current}
\end{figure}

In order to get the cumulants of the current, we follow the procedure described at the end of Section~\ref{sec:SolutionAndProcedure}. From the solution~(\ref{eq:SolOmPFourier},\ref{eq:SolOmMFourier}) of the integral equations~(\ref{eq:IntegEqOmP},\ref{eq:IntegEqOmM}) with $\xi=0$, we get
\begin{equation}
    \int_0^\infty \Omega_+ = 2 \hat{\psi}_Q
    \frac{\rho_+ - \Phi_Q(0^+)}{\Phi_Q'(0^+)} =  \sqrt{1 + \omega_Q}-1
    \:,
\end{equation}
\begin{equation}
    \int_0^\infty \Omega_- = 2 \hat{\psi}_Q
    \frac{\Phi_Q(0^-) - \rho_-}{\Phi_Q'(0^-)} =  \sqrt{1 + \omega_Q}-1
    \:.
\end{equation}
Combined with the boundary conditions~(\ref{eq:BoundZeroPQ},\ref{eq:BoundZeroMQ}), we obtain
\begin{align}
    \Phi_Q(0^+)
    &= \frac{1}{\sqrt{1+\omega_Q}}
    \left(
        \rho_+ + \frac{1 - \sqrt{1+\omega_Q}}{\e^{-\lambda}-1}
    \right)
    \:,
    \\
    \Phi_Q(0^-)
    &= \frac{1}{\sqrt{1+\omega_Q}}
    \left(
        \rho_- + \frac{1-\sqrt{1+\omega_Q}}{\e^{\lambda}-1}
    \right)
    \:.
\end{align}
Using these expressions into the last boundary condition~\eqref{eq:CancelVQ}, we obtain a simple expression for $\omega_Q$,
\begin{equation}
\label{eq:OmegaQ}
  \omega_Q = \rho_- (1-\rho_+) (\e^{\lambda} - 1) + \rho_+ (1-\rho_-) (\e^{-\lambda} - 1)
  \:.
\end{equation}
Together with~\eqref{eq:PsiQomega}, this recovers the result of Derrida and Gerschenfeld~\cite{Derrida:2009,Derrida:2009a} for the cumulant generating function $\hat{\psi}_Q$ of the current through the origin.
Furthermore, we additionally obtain the profiles $\moy{\eta_r(t) Q_t^n}_c \underset{t \to \infty}{\simeq}\Phi_{Q;n}(v)$ which measure the correlations between the current and the density in the hydrodynamic limit. For instance,
\begin{subequations}
  \label{eq:PhiFirstOrdersQ}
  \begin{align}
    \label{eq:Phi1Q}
  \Phi_{Q;1}(v)
  =& \frac{\rho (1-\rho)}{2} \erfc(v) \:,
  \\
    \label{eq:Phi2Q}
  \Phi_{Q;2}(v)
  =& \frac{\rho(1-\rho)(1-2\rho)}{2} \erfc(v) \:,
  \\
  \nonumber
  \Phi_{Q;3}(v)
  =& \frac{\rho(1-\rho)(1-3\rho + 3 \rho^2)}{2} \erfc(v)
  \\
  \label{eq:Phi3Q}
    &- 3\frac{\rho^2(1-\rho)^2}{4} \erfc \left( \frac{v}{\sqrt{2}} \right)^2
    \:.
\end{align}
\end{subequations}
These profiles are shown in Fig.~\ref{fig:Sim_SEP_current}, and present similar features to the ones obtained above for the tracer.

The closed integral equations~\eqref{eq:LinEqOmegaSummary}, which we have obtained in the context of the study of a tracer, therefore still apply for another observable, which is the integrated current through the origin. It is only required to change the boundary conditions satisfied by $\Phi$ at the origin.

\subsection{Another observable: a generalized current}
\label{sec:GenCurr}

We now consider another observable, which is the generalised current studied in~\cite{Imamura:2017,Imamura:2021}, defined as
\begin{equation}
\label{eq:defGenFlux}
  J_t(x) = \sum_{r \geq 1}
  \left(
    \eta_{r+x}(t) - \eta_r(0)
  \right)
  \:.
\end{equation}
It measures the difference between the number of particles to the right of the position $x$ at time $t$ and the number of particles initially on the positive axis at $t=0$. This quantity can be used to locate a tracer~\cite{Imamura:2017,Imamura:2021}, since $J_t(X_t) = 0$, which means that the number of particles to the right of the tracer is conserved. 

In order to apply our formalism to this observable, we consider the time evolution of its cumulant generating function, $\partial_t \ln \moy{ \e^{\lambda J_t(x)} }$.
However, when doing so we encounter one difficulty: for $x \neq 0$, $J_0(x) \neq 0$ and is random. This differs from the case of the integrated current through the origin $Q_t$, for which $Q_0=0$, and the case of the displacement $X_t$ of a tracer, taken by definition to be $X_0=0$. In order to circumvent this problem, we consider a slightly different observable,
\begin{equation}
    J_t(x_t) \:,
    \quad \text{with} \quad
    x_t = \lfloor \xi \sqrt{2 t} \rfloor
\end{equation}
the integer part of $\xi \sqrt{2t}$, with a constant $\xi$ which corresponds to the large time limit of $x_t/\sqrt{2t}$. Since at $t=0$, $x_0 = 0$, this new observable vanishes at initial time, $J_0(x_0) = 0$. We thus consider the cumulant generating function
\begin{equation}
    \hat\psi_J(\lambda) = \lim_{t \to \infty}
    \frac{1}{\sqrt{2t}} \ln \moy{\e^{\lambda J_t(x_t)}}
    \:,
\end{equation}
and the generalised profiles
\begin{equation}
    w_{J;r}(t) = \frac{\moy{\eta_{x_t+r}(t) \e^{\lambda J_t(x_t)}} }{\moy{\e^{\lambda J_t(x_t)} }}
    \:.
\end{equation}
Let us first consider the time evolution of the cumulant generating function. It is the sum of two contributions:
\begin{enumerate}[label=(\roman*)]
    \item At times $t_n = (n/\xi)^2/2$, the increment of $x_{t_n^+} = x_{t_n^-} + 1$ causes a change of $J_t(x_t)$, depending on the occupation $\eta_{x_{t_n^-} + 1}$ of the site that is "leaving" the sum~\eqref{eq:defGenFlux} at $t_n$. Therefore,
    \begin{equation}
        \moy{\e^{\lambda J_t(x_{t^+})}}
        = \moy{\e^{\lambda J_t(x_{t^-}) - \lambda \eta_{x_{t}+1}}}
        \:,
        \text{ for } t = t_n
        \:.
    \end{equation}
    Taking the logarithm, and using that $\eta_{x} = 0$ or $1$, we get
    \begin{multline}
        \left. \ln \moy{\e^{\lambda J_t(x_t)}} \right|_{t=t_n^+}
        = \left. \ln \moy{\e^{\lambda J_t(x_t)}} \right|_{t=t_n^-}
        \\
        + \ln ( 1 + (\e^{-\lambda}-1) w_{J;1})
        \:.
    \end{multline}
    \item Between two increments of $x_t$, the stochastic dynamics of the SEP, given by
    \begin{multline}
        \partial_t \ln \moy{\e^{\lambda J_t(x_t)}}
    = \frac{1}{2} \left[
    (\e^\lambda -1 )
    \frac{\moy{\eta_x(1-\eta_{x+1}) \e^{\lambda J_t(x_t)}}}{\moy{\e^{\lambda J_t(x_t)}}}
    \right.
    \\
    \left.
    +(\e^{-\lambda} -1 )
    \frac{\moy{\eta_{x+1}(1-\eta_x) \e^{\lambda J_t(x_t)}}}{\moy{\e^{\lambda J_t(x_t)}}}
    \right]
    \:,
    \end{multline}
    for $t \neq t_n$.
\end{enumerate}
We can combine these two contributions, by writing,
\begin{multline}
   \partial_t \ln \moy{\e^{\lambda J_t(x_t)}}
  = \frac{1}{2} \left[
    (\e^\lambda -1 )
    \frac{\moy{\eta_{x_t}(1-\eta_{x_t+1}) \e^{\lambda J_t(x_t)}}}{\moy{\e^{\lambda J_t(x_t)}}}
    \right.
    \\
    \left.
    +(\e^{-\lambda} -1 )
    \frac{\moy{\eta_{x_t+1}(1-\eta_{x_t}) \e^{\lambda J_t(x_t)}}}{\moy{\e^{\lambda J_t(x_t)}}}
  \right]
  \\
  + \sum_n \delta(t-t_n) \ln [1 + (\e^{-\lambda}-1) w_{J;1}(t)]
  \:,
\end{multline}
At large times, we can replace the sum over the delta functions by the continuous density of jumps, which is $\xi/\sqrt{2t}$:
\begin{multline}
  \label{eq:EvolCGFfluxKirone}
   \partial_t \ln \moy{\e^{\lambda J_t(x_t)}}
  = \frac{1}{2} \left[
    (\e^\lambda -1 )
    \frac{\moy{\eta_x(1-\eta_{x+1}) \e^{\lambda J_t(x_t)}}}{\moy{\e^{\lambda J_t(x_t)}}}
    \right.
    \\
    \left.
    +(\e^{-\lambda} -1 )
    \frac{\moy{\eta_{x+1}(1-\eta_x) \e^{\lambda J_t(x_t)}}}{\moy{\e^{\lambda J_t(x_t)}}}
  \right]
  \\
  + \frac{\xi}{\sqrt{2t}} \ln [1 + (\e^{-\lambda}-1) w_{J;1}(t)]
  \:.
\end{multline}
Similarly, we can write the equation satisfied by the time evolution of the profiles. For instance,
\begin{multline}
\label{eq:EvoW0GenFlux}
    \partial_t w_{J;0} = \frac{\e^{-\lambda} - (\e^{-\lambda}-1)w_{J;0}}{\e^{-\lambda} -1}
    \Bigg( \partial_t \ln \moy{\e^{\lambda J_t(x_t)}}
    \\
    - \frac{\xi}{\sqrt{2t}} \ln [1 + (\e^{-\lambda}-1) w_{J;1}(t)]
    \Bigg)
    + \frac{w_{J;-1} - w_{J;0}}{2}
    \\
    + \frac{\xi}{\sqrt{2t}} \left(
    \frac{\e^{-\lambda} w_{J;1}}{1+(\e^{-\lambda}-1)w_{J;1}} - w_{J;0}
    \right)
    \:.
\end{multline}
We can now take the hydrodynamic limit, with the scalings
\begin{equation}
    \ln \moy{\e^{\lambda J_t(x_t)}} \simeq \sqrt{2t} \: \hat{\psi}_J(\lambda,\xi)
    \:,
\end{equation}
\begin{equation}
    w_{J;r}(t) \underset{t \to \infty}{\simeq}
    \Phi_J \left(v = \frac{r-\frac{1}{2}}{\sqrt{2t}}, \xi, \lambda \right)
    \:,
\end{equation}
which we will denote by $\Phi_J(v)$ for simplicity.
Note that we have again shifted the positions by $\frac{1}{2}$ so that $w_{J;0}$ corresponds to $\Phi_J(0^-)$ and $w_{J;1}$ to $\Phi_J(0^+)$.
With these scalings, Eqs.~(\ref{eq:EvolCGFfluxKirone},\ref{eq:EvoW0GenFlux}) yield
\begin{equation}
    \label{eq:CancelVJ}
    \frac{\Phi_J(0^+) (1-\Phi_J(0^-))}{\Phi_J(0^-) (1-\Phi_J(0^+))} = \e^{\lambda}
    \:,
\end{equation}
\begin{equation}
  \label{eq:DerZeroMfluxKirone}
  \Phi_J'(0^\pm) = \mp 2 \Psi \left( \frac{1}{1-\e^{\mp \lambda}} - \Phi_J(0^\pm) \right)
  \:,
\end{equation}
where we have denoted
\begin{equation}
    \Psi = \hat{\psi}_J - \xi \ln[ 1 + (\e^{-\lambda}-1) \Phi_J(0^+)]
    \:.
\end{equation}
We have noticed that, if we use this new $\Psi$ to define
\begin{equation}
\label{eq:DefOmGenFlux}
  \Omega_\pm(v) = 2 \Psi \frac{\Phi_J'(v)}{\Phi_J'(0^\pm)}
  \:,
\end{equation}
these functions again verify the closed integral equations~\eqref{eq:LinEqOmegaSummary}, with the same kernel~\eqref{eq:KernelKSummary}. Indeed, the solution~(\ref{eq:SolOmPFourier},\ref{eq:SolOmMFourier}), combined with the boundary conditions for $\Phi_J$ yield
\begin{equation}
    \omega_J \e^{\xi^2} =
    \rho_- (1-\rho_+) (\e^{\lambda} - 1) + \rho_+ (1-\rho_-) (\e^{-\lambda} - 1)
    \:,
\end{equation}
and thus
\begin{equation}
  \hat{\psi}_J = 
  - \frac{1}{2 \sqrt{\pi}} \mathrm{Li}_{\frac{3}{2}}(- \omega_J) 
  + \xi \ln [1 + (\e^{-\lambda}-1) \Phi_J(0^+)]
  \:.
\end{equation}
Combining this expression with the value
\begin{equation}
    (\e^{-\lambda}-1)
    \Phi_J(0^+) =
    (1+\rho_+ (\e^{-\lambda}-1) ) \e^{Z_+(\omega_J,\xi)} - 1
\end{equation}
obtained from~\eqref{eq:SolOmPFourier} at $k=0$, we finally get
\begin{multline}
    \label{eq:CGFgenflux}
    \hat{\psi}_J(\lambda,\xi) = \xi \ln [ 1 + (\e^{-\lambda}-1) \rho_+]
    \\
    - \sum_{n \geq 1} \frac{(-\omega_J \e^{\xi^2})^n}{2n} \left(
    \frac{\e^{-n \xi^2}}{\sqrt{n \pi}} - \xi \erfc(\sqrt{n} \xi)
    \right)
    \:.
\end{multline}
This expression coincides exactly with the result of Ref.~\cite{Imamura:2017,Imamura:2021}. That supports the exactness of our main equations~\eqref{eq:LinEqOmegaSummary}, nonperturbatively in $\lambda$.

\section{Extension to other single-file systems}
\label{sec:ExtOtherModels}

We now argue that the closed integral equations~\eqref{eq:LinEqOmegaSummary}, which we have derived for the SEP, can be applied to other models of single-file systems.
The applicability of this equation to the KMP model and the random average process (see below) was already mentioned in Ref.~\cite{Grabsch:2022}. Here, we give a detailed analysis of these two models, including explicit analytical expressions, as well as a new application to another model of single-file diffusion: the double exclusion process.

\begin{figure}
    \centering
    \includegraphics[width=0.8\columnwidth]{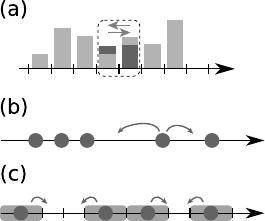}
    \caption{The different models of single-file diffusion considered in this paper. \textbf{(a) The Kipnis-Marchioro-Presutti (KMP) model.} Each site hosts a continuous variable, which represents an energy or a mass. At random times, a site can exchange energy with its neighbour such that the total energy of the two sites is randomly redistributed between the two. \textbf{(b) The random average process (RAP).} Particles on a continuous line can hop, at random times, to a random fraction of the distance to the next particle (either to the left or to the right). \textbf{(c) The double exclusion process (DEP).} Particles on a lattice can hop, at random times, in either direction, only if the two nearest neighbouring sites in that direction are free. It corresponds to an exclusion model in which the particles have a volume that occupies two sites (illustrated by the light gray area represented around the particles).}
    \label{fig:Models}
\end{figure}

\subsection{Hydrodynamic description of single-file systems in terms of two transport coefficients}
\label{sec:GenSingFile}

At large time and large distances, a single-file system can be described within the framework of fluctuating hydrodynamics~\cite{Spohn:1983}. The basic idea of this approach is to describe the system by a density $\rho(x,t)$ and a current $j(x,t)$, which are both random due to the underlying stochastic dynamics of the model (e.g. the random hops of the particles in the SEP). Then, the current $j$ is approximated by a stochastic perturbation of the average current,
\begin{equation}
    j = - D(\rho) \partial_x \rho - \sqrt{\sigma(\rho)} \: \eta
    \:,
\end{equation}
where $\eta(x,t)$ is a Gaussian white noise, with unit variance, uncorrelated in space and time. The density then evolves according to the continuity equation
\begin{equation}
    \label{eq:FlucHydroCont}
    \partial_t \rho = - \partial_x j 
    = \partial_x \left[
    D(\rho) \partial_x \rho + \sqrt{\sigma(\rho)} \: \eta
    \right]
    \:.
\end{equation}
All the microscopic details of the models are encoded into two transport coefficients: the diffusion coefficient $D(\rho)$ and the mobility $\sigma(\rho)$. These coefficients can be conveniently defined for a system of finite size $L$, placed between two reservoirs at density $\rho_1$ and $\rho_2$~\cite{Derrida:2007}. We denote $Q_t$ the number of particles transferred from left to right up to time $t$. The transport coefficients are defined from the first two moments of $Q_t$ as
\begin{align}
    \lim_{t\to\infty} \frac{\left\langle Q_t\right\rangle}{t}
    &= \frac{D(\rho)}{L}(\rho_1 - \rho_2) 
    \text{ for } \rho_1 - \rho_2 \text{ small,} 
    \\
    \lim_{t\to\infty} \frac{\left\langle Q_t^2\right\rangle}{t}
    &= \frac{\sigma(\rho)}{L} 
    \text{ for } \rho_1=\rho_2 = \rho.
\end{align}
These coefficients have been computed for various models of single-file systems. We list in Table~\ref{tab:tr_coefs} their expressions for the different models we consider in this paper, and represented in Fig.~\ref{fig:Models}.

\begin{table}[]
\[
\renewcommand{\arraystretch}{2}
\begin{array}{l*2{|>{\displaystyle}c}}
\text{Model} & D(\rho)  & \sigma(\rho)\\ \hline
\text{Symmetric exclusion process~\cite{Krapivsky:2015}} & D_0 & 2 D_0 \rho(1-\rho)  \\
\text{Hard Brownian particles~\cite{Krapivsky:2015}} & D_0 & 2D_0 \rho \\
\text{Kipnis-Marchioro-Presutti~\cite{Zarfaty:2016}} & D_0 & \sigma_0 \rho^2\\
\text{Random average process~\cite{Krug:2000,Kundu:2016}} & \frac{\mu_1}{2 \rho^2} & \frac{1}{\rho} \frac{ \mu_1 \mu_2}{\mu_1 - \mu_2}\\[0.2cm]
\text{Double exclusion process~\cite{Hager:2001,Baek:2017}} &                           \frac{D_0}{(1-\rho)^2} & \frac{2 D_0 \rho(1-2\rho)}{1-\rho}
\end{array}
\]
    \caption{The transport coefficients $D(\rho)$ and $\sigma(\rho)$ for the different models presented in this paper. $D_0$ is the diffusion coefficient of an individual particle, $\sigma_0 = 2a D_0$ with $a$ the lattice constant of the KMP model~\cite{Zarfaty:2016}, and $\mu_k$ are the moments of the probability law of the jumps in the RAP~\cite{Kundu:2016}. }
    \label{tab:tr_coefs}
\end{table}

\bigskip

We consider a tracer, at position $X_t$, which can be determined from the density $\rho(x,t)$ as~\cite{Krapivsky:2015a}
\begin{equation}
  \label{eq:defXtfromRho}
  \int_0^{X_t} \rho(x,t) \dd x
  = \int_0^{\infty} \left( \rho(x,t) - \rho(x,0) \right) \dd x
  \:.
\end{equation}
This equation expresses that the number of particles to the right of the tracer is conserved. We define the cumulant generating function and the generalised profiles
\begin{equation}
\label{eq:defwrGenSF}
    \psi(\lambda) = \ln \moy{ \e^{\lambda X_t}}
    \:,
    \quad
    w_r(\lambda,t) = 
    \frac{\moy{\rho(X_t + r,t) \e^{\lambda X_t}}}{ \moy{\e^{\lambda X_t}} }
    \:.
\end{equation}

We also consider another observable, the integrated current through the origin
\begin{equation}
  \label{eq:defQfromRho}
  Q_t
  = \int_0^{\infty} \left( \rho(x,t) - \rho(x,0) \right) \dd x
  \:.
\end{equation}
Its cumulant generating function and the corresponding profiles are given by
\begin{equation}
\label{eq:defwrGenSFQ}
    \psi_Q(\lambda) = \ln \moy{ \e^{\lambda Q_t}}
    \:,
    \quad
    w_{Q;r}(\lambda,t) = 
    \frac{\moy{\rho(r,t) \e^{\lambda Q_t}}}{ \moy{\e^{\lambda Q_t}} }
    \:.
\end{equation}

\subsection{Modified equations in the case of constant diffusion and quadratic mobility}
\label{sec:ExtEqQuadSig}

We have obtained a closed equation for the SEP, corresponding to $D(\rho) = \frac{1}{2}$ and $\sigma(\rho) = \rho(1-\rho)$. These results can be extended to any single-file system with $D(\rho) = D_0$ constant and $\sigma''(\rho)$ constant with $\sigma(0) = 0$. A procedure to deduce this model from the SEP is given in Ref.~\cite{Rizkallah:2022}.
Here, we only give the resulting equations.

\subsubsection{For the position of the tracer}

At large times, the cumulant generating function of the position of the tracer behaves as
\begin{equation}
    \psi(\lambda, t) \equi{t\to\infty} \hat{\psi}(\lambda)\sqrt{4 D_0 t}
    \:.
\end{equation}
The generalised profiles~\eqref{eq:defwrGenSF} also have a diffusive scaling
\begin{equation} \label{eq:sm_scale_w_ext}
    w_r(\lambda, t) \equi{t\to\infty}  \Phi\left(v = \frac{r}{\sqrt{4 D_0 t}}, \lambda\right)
    \:.
\end{equation}
We again define the functions
\begin{equation}
    \Omega_\pm(v) = 2 \hat{\psi} \frac{\Phi'(v)}{\Phi'(0^\pm)}
     \:,
\end{equation}
which now verify the bilinear integral equations
\begin{equation}
\label{eq:BilinIntegEqOmGen}
    \Omega_\pm(v) = K(v) +
    \frac{\sigma''(0)}{4D_0} \int_{\mathbb{R}^{\mp}} \Omega_\pm(v\pm z) \Omega_\mp(\mp z) \dd z
    \:,
\end{equation}
equivalent to the linear ones
\begin{equation}
\label{eq:IntegEqOmGen}
    \Omega_\pm(v) = K(v) +
    \frac{\sigma''(0)}{4D_0} \int_{\mathbb{R}^{\mp}} \Omega_\pm(z) \: K(v-z) \dd z
    \:,
\end{equation}
with the kernel still given by Eq.~\eqref{eq:KernelKSummary}. The profile $\Phi$ can then be deduced by integration of $\Omega_\pm$, with the boundary conditions
\begin{equation}
    \Phi'(0^\pm) \mp \hat{\psi}\frac{\sigma''(0)}{2 D_0} 
    \frac{\Phi(0^{\pm})}{\e^{\mp \frac{\sigma''(0) \lambda}{4 D_0}}-1}
    = 0
    \:,
\end{equation}
\begin{equation}
    \frac{2 \sigma'(0) + \sigma''(0) \Phi(0^+)}{2 \sigma'(0) + \sigma''(0) \Phi(0^-)} 
    = \e^{\frac{\sigma''(0) \lambda}{4 D_0}}
    \:,
    \quad
    \Phi(\pm \infty) = \rho_\pm
    \:.
\end{equation}

The solution of the integral equations~\eqref{eq:IntegEqOmGen} can be straightforwardly deduced from~(\ref{eq:SolOmPFourier},\ref{eq:SolOmMFourier}), and reads
\begin{multline}
    \int_0^\infty  \Omega_+(v) \e^{\I k v} \dd v
    = \frac{4 D_0}{\sigma''(0)} \Bigg(
        1-
        \\
        \exp \left[- Z_+ \left( - \frac{\sigma''(0)}{4D_0} \omega, \xi - \frac{\I k}{2} \right) \right]
    \Bigg)
    \:,
\end{multline}
\begin{multline}
    \int_{-\infty}^0  \Omega_-(v) \e^{\I k v} \dd v
    = \frac{4 D_0}{\sigma''(0)} \Bigg(
        1-
        \\
        \exp \left[- Z_- \left( - \frac{\sigma''(0)}{4D_0} \omega, \xi - \frac{\I k}{2} \right) \right]
    \Bigg)
    \:,
\end{multline}
with $Z_\pm$ given by Eq.~\eqref{eq:DefZpm}. In particular, setting $k= \pm \I s$ and letting $s \to +\infty$, we obtain the expression of $\Omega_\pm(0) = 2\hat{\psi}$, thus,
\begin{equation}
    \label{eq:RelPsiOmegaGen}
    \hat{\psi} = \frac{2 D_0}{\sigma''(0)\sqrt{\pi}}
    \mathrm{Li}_{\frac{3}{2}} \left( \frac{\sigma''(0)}{4 D_0} \omega \right)
    \:.
\end{equation}

\subsubsection{For the current through the origin}

Similarly as we did in Section~\ref{sec:Current}, we can also obtain similar equations for the study of the integrated current through the origin~\eqref{eq:defQfromRho}.
The cumulant generating function scales as
\begin{equation}
    \psi_Q(\lambda, t) 
    = \ln \moy{ \e^{\lambda Q_t}}
    \equi{t\to\infty} \hat{\psi}_Q(\lambda)\sqrt{4 D_0 t}
    \:,
\end{equation}
and the profiles~\eqref{eq:defwrGenSFQ} as
\begin{equation}
    w_{Q;r}(\lambda, t) \equi{t\to\infty}  
    \Phi_Q\left(v = \frac{r}{\sqrt{4 D_0 t}}, \lambda\right)
    \:.
\end{equation}
Defining $\Omega_\pm$ as in~\eqref{eq:defOmegaQ}, these functions again satisfy the integral equation~\eqref{eq:IntegEqOmGen}, with a kernel $K(v)$ deduced from~\eqref{eq:KernelKSummary} by setting $\xi=0$ and replacing $\omega$ by $\omega_Q$:
\begin{equation}
    K(v) = \frac{\omega_Q}{\sqrt{\pi}} \: \e^{-v^2}
    \:.
\end{equation}
This parameter $\omega_Q$ can be related to $\hat\psi_Q$ by
\begin{equation}
    \label{eq:RelPsiOmegaQGen}
    \hat{\psi}_Q = \frac{2 D_0}{\sigma''(0)\sqrt{\pi}}
    \mathrm{Li}_{\frac{3}{2}} \left( \frac{\sigma''(0)}{4 D_0} \omega_Q \right)
    \:.
\end{equation}
The profiles can again be deduced by integration of the solution $\Omega_\pm$ of~\eqref{eq:IntegEqOmGen}, with the boundary conditions
\begin{equation}
    \label{eq:BoundPhiGenQ}
    \Phi_Q'(0^\pm) = \mp 2 \hat{\psi}_Q
    \left(
        \frac{\sigma'(0)}{2 D_0} \frac{1}{1 - \e^{\mp \frac{\sigma'(0)}{2 D_0} \lambda}}
        + \frac{\sigma''(0)}{4 D_0} \Phi_Q(0^\pm)
    \right)
    \:,
\end{equation}
\begin{equation}
    \label{eq:CancelVGenQ}
    \frac{\Phi_Q(0^+)(2 \sigma'(0) + \sigma''(0) \Phi_Q(0^-))}
    {\Phi_Q(0^-)(2 \sigma'(0) + \sigma''(0) \Phi_Q(0^+))} 
    = \e^{\frac{\sigma'(0)}{2D_0} \lambda}
    \:.
\end{equation}

\subsection{The Kipnis Marchioro Presutti model}
\label{sec:KMP}

As a first application of the generalised equation~\eqref{eq:IntegEqOmGen}, we consider the Kipnis Marchioro Presutti (KMP) model~\cite{Kipnis:1982,Hurtado:2009}. It is a mass transfer model, which describes a one dimensional lattice where each site contains a continuous variable which represents a mass. At random times, the total mass of two neighbouring sites is randomly redistributed (uniformly) between the two sites. See Fig.~\ref{fig:Models}(a). This system is described by the transport coefficients
\begin{equation}
    \label{eq:TrCoefsKMP}
    D(\rho) = D_0 
    \:,
    \quad
    \sigma(\rho) = \sigma_0 \rho^2
    \:,
\end{equation}
with $\sigma_0 = 2a D_0$ where $a$ is the lattice spacing~\cite{Zarfaty:2016}. It is a model with constant diffusion and quadratic mobility, so it falls into the category studied in Section~\ref{sec:ExtEqQuadSig} and we can directly apply the results obtained there.

\subsubsection{Position of a tracer}

The KMP model is not a particle model, but one can still define the position of a tracer using Eq.~\eqref{eq:defXtfromRho}. It represents a fictitious wall that separates the system into two regions in which the mass is conserved.

\begin{figure*}
    \centering
    \includegraphics[width=\textwidth]{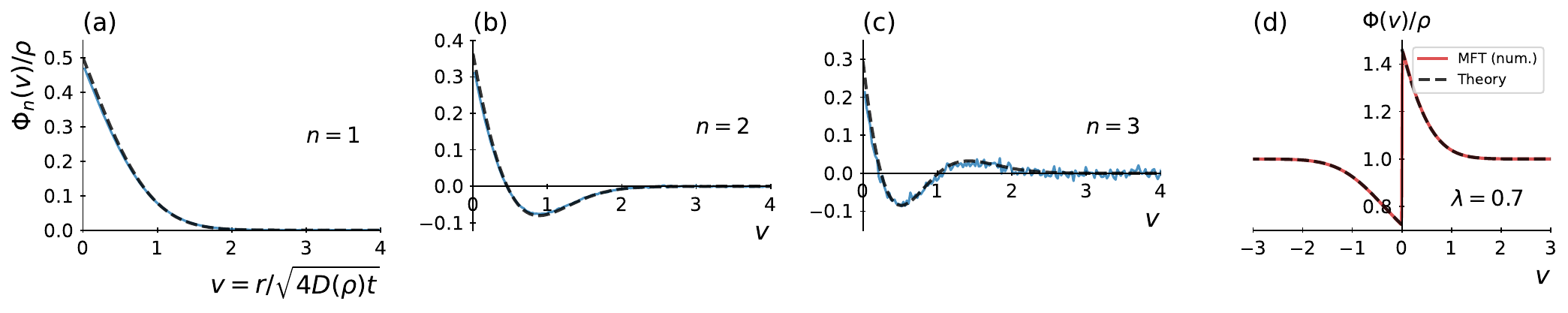}
    \caption{\textbf{KMP.} Generalised density profiles $\Phi_n^{\mathrm{(KMP)}}(v)/\rho$ at orders \textbf{(a)} $n=1$, \textbf{(b)} $n=2$ and \textbf{(c)} $n=3$ (the density $\rho$ plays no role in this model). Solid lines: result of the simulations of the KMP model (see Appendix~\ref{sec:AppNumSim}) at time $t=900$ on $500$ sites. Dashed lines: theoretical predictions~(\ref{eq:Phi1KMP},\ref{eq:Phi2KMP},\ref{eq:Phi3KMP}). \textbf{(d)} GDP-generating function at $\rho=1$ and $\lambda=0.7$, obtained from solving numerically the Wiener-Hopf equation~\eqref{eq:IntegEqOmGen} (dashed line), compared to the numerical solution (red solid line) of the MFT equations (see Section~\ref{sec:CompInvScat} below).}
    \label{fig:KMP_sim}
\end{figure*}

Following the procedure described in Section~\ref{sec:SolutionAndProcedure}, we obtain the cumulants of the position of this tracer. For instance, in the case $\rho_+ = \rho_- = \rho$,
\begin{equation}
    \hat\kappa_2^{(\mathrm{KMP})} = \frac{\sigma_0}{2 D_0\sqrt{\pi}}
    \:,
\end{equation}
\begin{equation}
    \hat\kappa_4^{(\mathrm{KMP})} = \frac{\left(12+\left(3 \sqrt{2}-8\right) \pi \right) \sigma_0^3}
    {8 D_0^3\pi ^{3/2}}
    \:.
\end{equation}
We additionally get the generalised density profiles,
\begin{align}
    \label{eq:Phi1KMP}
      \Phi_1^{(\mathrm{KMP})}(v)
      &= \frac{\rho \sigma_0}{4 D_0} \erfc(v)
      \:,
      \\
    \label{eq:Phi2KMP}
      \Phi_2^{(\mathrm{KMP})}(v)
      &=
      \frac{\rho \sigma_0^2}{4 D_0^2} \left( \erfc(v) - \frac{2}{\pi} \e^{-v^2} \right)
      \:,
\end{align}
\begin{multline}
    \label{eq:Phi3KMP}
    \Phi_3^{(\mathrm{KMP})}(v)
      =
      \frac{\rho \sigma_0^3}{32 D_0^3} 
      \Bigg( 
      2 \left( 1 + \frac{6}{\pi} \right)\erfc(v) 
      - 24\frac{\sqrt{\pi}-v}{\pi^{3/2}} \e^{-v^2} 
      \\
      + 3 \erfc\left( \frac{v}{\sqrt{2}} \right)^2
      \Bigg)
      \:.
\end{multline}
These profiles are represented in Fig.~\ref{fig:KMP_sim}.

\subsubsection{Integrated current through the origin}

Following an approach similar to Section~\ref{sec:Current}, we find that
\begin{equation}
    \omega_Q = \frac{\sigma_0 \rho^2 \lambda^2}{2 D_0}
    \:,
\end{equation}
which, combined with~\eqref{eq:RelPsiOmegaQGen} yields
\begin{equation}
\label{eq:CGFfluxKMP}
    \hat{\psi}_Q^{(\mathrm{KMP})}(\lambda) = \frac{D_0}{\sqrt{\pi} \sigma_0}
    \mathrm{Li}_{\frac{3}{2}} \left( \left(\frac{\sigma_0 \rho \lambda}{2 D_0}\right)^2 \right)
    \:.
\end{equation}
This expression coincides with the one given in~\cite{Derrida:2009a}. We additionally obtain the profiles
\begin{subequations}
\label{eq:ProfKMPflux}
  \begin{align}
      \Phi_{Q;1}^{(\mathrm{KMP})}(v)
      &= \frac{\rho^2 \sigma_0}{4 D_0} \erfc(v)
      \:,
      \\
      \Phi_{Q;2}^{(\mathrm{KMP})}(v)
      &=
      \frac{\rho^3 \sigma_0^2}{4 D_0^2} \erfc(v)
      \:,
      \\
      \Phi_{Q;3}^{(\mathrm{KMP})}(v)
      &=
      \frac{3 \rho^4 \sigma_0^4}{32 D_0^3} 
      \left( 
      2 \erfc(v) 
      + \erfc\left( \frac{v}{\sqrt{2}} \right)^2
      \right)
      \:.
  \end{align}
\end{subequations}
Note that, for this model, $\sigma'(0)=0$, so the boundary condition~\eqref{eq:BoundPhiGenQ} is ill-defined. We have used the boundary condition deduced from~\eqref{eq:BoundPhiGenQ} by taking the limit $\sigma'(0) \to 0$,
\begin{equation}
    \Phi_Q'(0^\pm) = -2 \hat{\psi}_Q
    \left(\frac{1}{\lambda}
        \pm \frac{\sigma''(0)}{4 D_0} \Phi_Q(0^\pm)
    \right)
    \:.
\end{equation}
With this relation, the definitions of $\Omega_\pm$ take the form
\begin{equation}
\label{eq:OmegaFluxKMP}
    \Omega_\pm(v) = 2 \hat{\psi}_Q \frac{\Phi_Q'(v)}{\Phi_Q'(0^\pm)}
    = -\frac{\Phi_Q'(v)}{\frac{1}{\lambda} \pm \frac{\sigma''(0)}{4 D_0} \Phi_Q(0^\pm)}
    \:.
\end{equation}

\subsection{The random average process}

The random average process (RAP) is a model of particles on an infinite line, at positions $x_k(t)$, with an initial mean density $\rho$~\cite{Ferrari:1998,Krug:2000,Rajesh:2001}. Each particle can randomly jump to a fraction of the distance to the next particle, either to the left or to the right, with rate $\frac{1}{2}$. See Fig.~\ref{fig:Models}(b). In the hydrodynamic limit, only the first two moments $\mu_1$ and $\mu_2$ of the distribution of this random fraction are relevant, so that the transport coefficients only depend on these moments~\cite{Krug:2000,Kundu:2016}:
\begin{equation}
    D(\rho) = \frac{\mu_1}{2 \rho^2}
    \:,
    \quad
    \sigma(\rho) = \frac{1}{\rho} \frac{\mu_1 \mu_2}{\mu_1 - \mu_2}
    \:.
\end{equation}
These coefficients do not fall into the category of Section~\ref{sec:ExtEqQuadSig}, so we cannot use the integral equation~\eqref{eq:IntegEqOmGen} in this case. However, this model is known to be equivalent to a mass transfer model~\cite{Ferrari:1998,Krug:2000,Cividini:2016,Cividini:2016a}, which, in the hydrodynamic limit, becomes equivalent to the KMP model~\cite{Kundu:2016}, associated with the transport coefficients~\eqref{eq:TrCoefsKMP}, with
\begin{equation}
    D_0 = \frac{\mu_1}{2}
    \:,
    \quad
    \sigma_0 = \frac{\mu_1 \mu_2}{\mu_1-\mu_2}
    \:.
\end{equation}
This duality relation that allows to map a model of single-file system onto another has recently been extended to any transport coefficients $D$ and $\sigma$~\cite{Rizkallah:2022}. Here, it allows us to obtain results on the RAP by studying the KMP model, which obeys the closed equation~\eqref{eq:IntegEqOmGen}.

Under this duality, the density field $\rho^{\mathrm{(RAP)}}(x,t)$ of the RAP can be expressed in terms of the one of the KMP $\rho^{\mathrm{(KMP)}}(k,t)$ as~\cite{Rizkallah:2022}
\begin{equation}
    \label{eq:MappingRAPtoKMP}
    \rho^{\mathrm{(RAP)}}(x_k(t),t) = \frac{1}{\rho^{\mathrm{(KMP)}}(k,t)}
    \:,
\end{equation}
with $x_k(t)$ the position of the particle with label $k$ at time $t$, which can be expressed from the density as
\begin{equation}
    x_k(t) = x_0(t) + \int_0^k \rho^{\mathrm{(KMP)}}(k',t) \dd k'
    \:,
\end{equation}
with $x_0$ the position of the tracer, defined from~\eqref{eq:defXtfromRho}. This transformation can be inverted, to express the position $x_0(t)$ of the tracer in terms of the density $\rho^{\mathrm{(KMP)}}$~\cite{Kundu:2016,Rizkallah:2022}:
\begin{equation}
    x_0(t) = \int_{-\infty}^0 
    (\rho^{\mathrm{(KMP)}}(k',t) - \rho^{\mathrm{(KMP)}}(k',0))\dd k'
    \:.
\end{equation}
Thanks to the conservation of the total number of particles, it can be expressed in terms of the integrated current $Q_t^{\mathrm{(KMP)}}$ through the origin in the KMP model,
\begin{equation}
    x_0(t) = - Q_t^{\mathrm{(KMP)}}
    \:.
\end{equation}
Therefore, one can easily relate the cumulant generating functions since
\begin{align}
    \nonumber
    \hat\psi^{\mathrm{(RAP)}}(\lambda)
    &= \lim_{t \to \infty} \frac{1}{\sqrt{4 D(\rho) t}}
    \ln \moy{ \e^{\lambda x_0(t)}}
    \\
    \nonumber
    &= \sqrt{\frac{D_0}{D(\rho)}} \lim_{t \to \infty} \frac{1}{\sqrt{4 D_0 t}}
    \ln \moy{ \e^{-\lambda Q_t^{\mathrm{(KMP)}}}}
    \\
    &= \rho \: \hat\psi_Q^{\mathrm{(KMP)}}(-\lambda)
    \:.
\end{align}
Note that, due to the relation~\eqref{eq:MappingRAPtoKMP}, and the RAP having mean density $\rho$, the KMP model has mean density $1/\rho$. The cumulant generating function $\hat\psi_Q^{\mathrm{(KMP)}}$ must thus be evaluated at this density. From~\eqref{eq:CGFfluxKMP}, this gives
\begin{equation}
    \hat{\psi}^{(\mathrm{RAP})}(\lambda)
    = \frac{\rho (\mu_1-\mu_2)}{2 \mu_2\sqrt{\pi} }
    \mathrm{Li}_{\frac{3}{2}} \left( \left(\frac{\mu_2 \lambda}{\rho(\mu_1-\mu_2)}\right)^2 \right)
    \:.
\end{equation}

The generalised profiles can be also obtained from the relation~\eqref{eq:MappingRAPtoKMP}. Indeed, in the large time limit, the averages in Eqs.~(\ref{eq:defwrGenSF},\ref{eq:defwrGenSFQ}) are dominated by a single realisation of the field $\rho(x,t)$, the typical realisation (see Appendix~\ref{sec:AppendixMFT}). Since all realisations verify~\eqref{eq:MappingRAPtoKMP}, so does the typical one, which we denote $\rho_\star$. Hence,
\begin{align}
    \nonumber
    \Phi^{\mathrm{(RAP)}}
    &\left( v = \frac{r(k)}{\sqrt{4 D(\rho) t}} \right)
    \underset{t \to \infty}{\simeq} \rho_\star^{\mathrm{(RAP)}}(x_0(t) + r(k), t)
    \\
    \nonumber
    &= \frac{1}{\rho_\star^{\mathrm{(KMP)}}(k,t)}
    \\
    &\underset{t \to \infty}{\simeq} 
    \frac{1}{\Phi_Q^{\mathrm{(KMP)}}(u = k/\sqrt{4D_0 t})}
    \label{eq:MappingPhiRAPtoKMP}
    \:,
\end{align}
where we have defined
\begin{equation}
    r(k) \equiv x_k(t) - x_0(t) = \int_0^k \rho_\star^{\mathrm{(KMP)}}
    \:,
\end{equation}
which becomes
\begin{equation}
    v(u) = \frac{r(k)}{\sqrt{4 D(\rho) t}}
    = \sqrt{\frac{D_0}{D(\rho)}} \int_0^u \Phi_Q^{\mathrm{(KMP)}}(u')\dd u'
    \:.
\end{equation}
Together with~\eqref{eq:MappingPhiRAPtoKMP}, expanding this relation in powers of $\lambda$, we can compute the profiles of the RAP from the ones obtained above on the KMP model~\eqref{eq:ProfKMPflux}, but evaluated at the density $1/\rho$. This gives for instance
\begin{figure*}
    \centering
    \includegraphics[width=\textwidth]{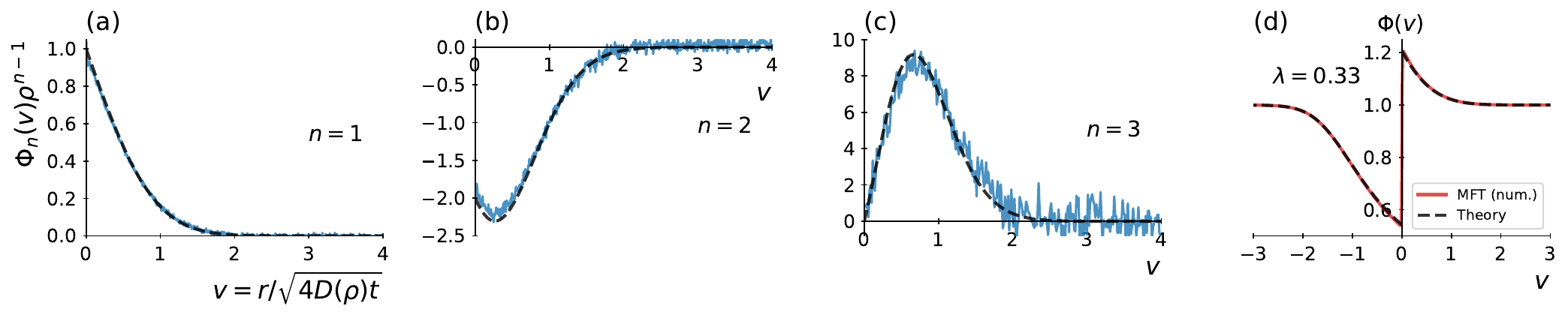}
    \caption{\textbf{RAP.} Generalised density profiles $\Phi_n^{\mathrm{(RAP)}}(v) \rho^{n-1}$ at orders \textbf{(a)} $n=1$, \textbf{(b)} $n=2$ and \textbf{(c)} $n=3$ (the density $\rho$ plays no role in this model). Solid lines: result of the simulations of the RAP (see Appendix~\ref{sec:AppNumSim}) at time $t=4000$ with $5000$ particles. Dashed lines: theoretical predictions~(\ref{eq:Phi1RAP},\ref{eq:Phi2RAP},\ref{eq:Phi3RAP}). \textbf{(d)} GDP-generating function at $\rho=1$ and $\lambda=0.33$, obtained from solving numerically the Wiener-Hopf equation~\eqref{eq:IntegEqOmGen} and using the mapping~\eqref{eq:MappingPhiRAPtoKMP} (dashed line), compared to the numerical solution (red solid line) of the MFT equations (see Section~\ref{sec:CompInvScat} below).}
    \label{fig:RAP_sim}
\end{figure*}
\begin{equation}
    \label{eq:Phi1RAP}
    \Phi_1^{(\mathrm{RAP})}(v)
    = \frac{\mu_2}{2(\mu_1-\mu_2)} \erfc(v)
    \:,
\end{equation}
\begin{multline}
    \label{eq:Phi2RAP}
    \Phi_2^{(\mathrm{RAP})}(v)
    =
      \frac{\mu_2^2}{2\pi \rho(\mu_1-\mu_2)^2} \Bigg(
       \pi \erfc(v)^2 - 2 \e^{-v^2}
       \\
       - 2 \pi \left( 1 + v\frac{\e^{-v^2}}{\sqrt{\pi}} \right) \erfc(v)
        + 2 \e^{-2v^2}
      \Bigg)
      \:,
\end{multline}
\begin{multline}
    \label{eq:Phi3RAP}
    \Phi_3^{(\mathrm{RAP})}(v)
      = \frac{3}{4 \pi^2} \frac{\mu_1^3}{\rho^2 (\mu_1-\mu_2)^3}
      \Bigg(
        \pi^2 \erfc(v)^3 + 2 \sqrt{\pi} v \e^{-3v^2} 
        \\
        - \pi^2 \left(4 +  \frac{2v(3-v^2)}{\sqrt{\pi}} \e^{-v^2} \right) \erfc(v)^2
        + 2(4 \pi + \sqrt{\pi} v) \e^{-v^2}
        \\
        + (2 \pi^2 + 2 \pi (3-2v^2) \e^{-2v^2} + 2\pi (2 v^2 + 4 \sqrt{\pi} v -3) \e^{-v^2}) \erfc(v)
       \\
       - 4 (2\pi + \sqrt{\pi} v) \e^{-2v^2}
      \Bigg)
      \:.
\end{multline}
These profiles are represented in Fig.~\ref{fig:RAP_sim}.

\subsection{The double exclusion process}
\label{sec:DEP}

The double exclusion process (DEP) is an exclusion process in which the particle on site $k$ can jump to site $k+1$ only if sites $k+1$ and $k+2$ are empty or to site $k-1$ only if sites $k-1$ and $k-2$ are empty~\footnote{This model can also be defined by taking a limit of the more general Katz–Lebowitz–Spohn (KLS) model~\cite{Hager:2001,Baek:2017}}. It thus corresponds to an exclusion process with particles that occupy the volume of two sites. See Fig.~\ref{fig:Models}(c).
By imposing that the $n$ neighbouring sites must be empty, we can also more generally define a $n$-exclusion process. These kind of models have for instance been used to model biological systems~\cite{Shaw2003}. Here we focus on the double exclusion case, but the following can be adapted to $n$-exclusion. We consider annealed initial conditions for the DEP of mean density $\rho$, namely particles are placed at random among configurations where there is at least one empty site in front of each particle (note that necessarily $\rho \leq 1/2$). The transport coefficients of the DEP are~\cite{Hager:2001,Baek:2017}
\begin{equation}
    D(\rho) = \frac{D_0}{(1-\rho)^2}
    \:,
    \quad
    \sigma(\rho) = 2 D_0 \frac{\rho(1-2\rho)}{1-\rho}
    \:.
\end{equation}

We can map the SEP to the DEP at the microscopic level by adding a site in front of every particle. This mapping can also be performed at the macroscopic level \cite{Rizkallah:2022}. Here we summarize the main ideas of the mapping, which we perform in the reference frame of the tracer particle.

At the macroscopic level, let us denote, with $x = r/\sqrt{t}$,
\begin{align}
    \label{eq:DefPhiSEPtoDEP}
    \Phi^{\mathrm{(SEP)}}(x) 
    & \underset{t \to \infty}{\simeq} \frac{\moy{\eta_{X_t + r}^{\mathrm{(SEP)}}\e^{\lambda X_t}}}{\moy{\e^{\lambda X_t}}}
    \:,
    \\
    \label{eq:DefPhiDEP}
    \Phi^{\mathrm{(DEP)}}(x) 
    & \underset{t \to \infty}{\simeq} 
    \frac{\moy{\eta_{X_t + r}^{\mathrm{(DEP)}}\e^{\lambda X_t}}}{\moy{\e^{\lambda X_t}}}
    \:. 
\end{align}
A relation between these two profiles can be obtained in the following way. If we look at a position $x$ in the SEP (in the reference frame of the tracer particle), it will correspond to a position $x + n(x)$ in the DEP, with $n(x)$ the number of particles in the SEP between the tracer (at position $X_t$) and the position $X_t+x$, which is thus the number of sites added to go to the DEP. It can be written as
\begin{equation}
    n(x) = \int_0^x \Phi^{\mathrm{(SEP)}}(z) \dd z
    \:.
\end{equation}
Next, we need to relate the density of particles in the SEP and in the DEP at that position. If the microscopic SEP is at a local density $\rho_{\mathrm{SEP}} = \frac{N}{L}$ ($N$ particles on $L$ sites), then when we map to the DEP by adding a site in front of each particle, the density becomes $\rho_{\mathrm{DEP}} = \frac{N}{L + N} = \frac{\rho_{\mathrm{SEP}}}{1+\rho_{\mathrm{SEP}}}$. Writing this relation for the generalised profiles yields
\begin{equation}
    \Phi^{\mathrm{(DEP)}}(x + n(x)) = \frac{\Phi^{\mathrm{(SEP)}}(x)}{1 + \Phi^{\mathrm{(SEP)}}(x)}
    \:.
    \label{eq:relDEPSEP}
\end{equation}
This relation holds because, in the long time limit, the averages in~(\ref{eq:DefPhiSEPtoDEP},\ref{eq:DefPhiDEP}) are dominated by a single realisation of the density for each model, and these densities verify~\eqref{eq:relDEPSEP}.

We can expand~\eqref{eq:relDEPSEP} in orders of $\lambda$ to get the generalized density profiles for the DEP from those of the SEP given in Section~\ref{sec:ProfSEP}. At lowest orders, we get
\begin{figure*}
    \centering
    \includegraphics[width=\textwidth]{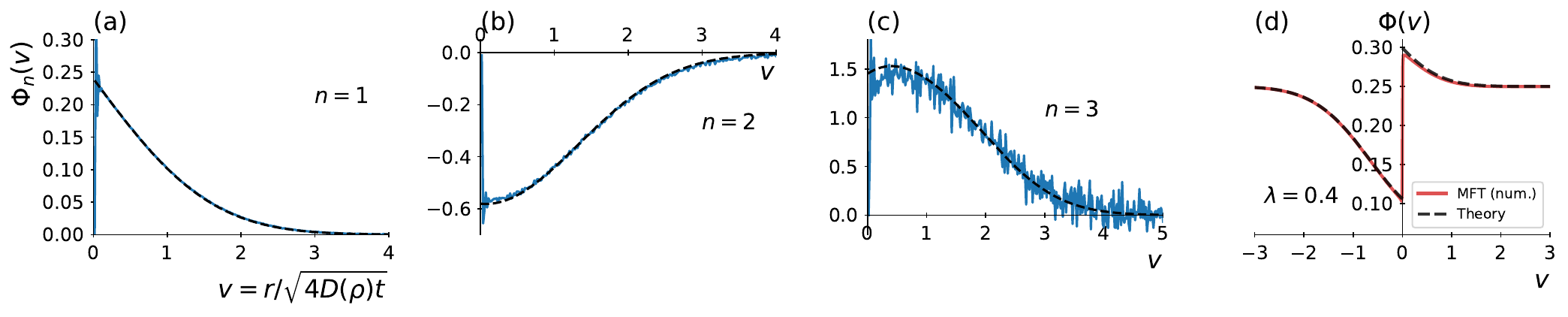}
    \caption{Generalised density profiles $\Phi_n^{\mathrm{(DEP)}}(v)$ at orders \textbf{(a)} $n=1$, \textbf{(b)} $n=2$ and \textbf{(c)} $n=3$ at density $\rho=0.25$. Solid lines: result of the simulations of the DEP (see Appendix~\ref{sec:AppNumSim}) at time $t=3000$ on $20000$ sites. Dashed lines: theoretical predictions~(\ref{eq:Phi1DEP},\ref{eq:Phi2DEP}). \textbf{(d)} GDP-generating function at $\rho=0.25$ and $\lambda=0.4$, obtained from solving numerically the Wiener-Hopf equation~\eqref{eq:IntegEqOmGen} and using the mapping~\eqref{eq:relDEPSEP} (dashed line), compared to the numerical solution (red solid line) of the MFT equations (see Section~\ref{sec:CompInvScat} below).}
    \label{fig:DEP_sim}
\end{figure*}
\begin{align}
    \Phi^{\mathrm{(DEP)}}_0(x)  = & \rho
    \:,\\
    \label{eq:Phi1DEP}
    \Phi^{\mathrm{(DEP)}}_1(x)  = & \frac{1}{2} (1-\rho) (1 - 2 \rho) \text{erfc}(v) 
    \:,\\
    \nonumber
    \Phi^{\mathrm{(DEP)}}_2(x)  = & \frac{(1-\rho) (1 - 2 \rho)}{4 \pi  \rho }
    \left(
    2 \sqrt{\pi } \rho (1-2 \rho) v \: \e^{-v^2} \erfc(v)
    \right.
    \\
    \nonumber
    &-2 \rho  (1-2 \rho) \e^{-2 v^2}
    - 2 (1-2 \rho) (2-\rho) \e^{-v^2}
    \\
    &\left.
    +\pi  \erfc(v)
    ((2 \rho -1) \rho  \erfc(v)-3 \rho +1)
    \right)
    \label{eq:Phi2DEP}
    \:,
\end{align}
where $v = x/\sqrt{4 D(\rho)} = x (1-\rho)/\sqrt{4 D_0}$. The profile $\Phi_3^{(\mathrm{DEP})}$ can be written similarly, but the expression is rather cumbersome so we do not reproduce it here.
These profiles are represented in Fig.~\ref{fig:DEP_sim}.

\section{Comparison with the results obtained by inverse scattering}
\label{sec:CompInvScat}

Since the publication of \cite{Grabsch:2022}, several works have obtained exact results for the integrated currents in different single-file systems~\cite{Bettelheim:2022,Bettelheim:2022a,Mallick:2022} with a quadratic mobility $\sigma(\rho)$. As we show below, these solutions can also be used to obtain information about the generalised density profiles, which characterise the correlations between the current and the density of particles.
Importantly, all these results on single-file diffusion can be recast into the equation~\eqref{eq:BilinIntegEqOmGen}. In this Section we demonstrate that the details of the problem under consideration are only encoded in the kernel $K$.

Note that this equation in the Fourier domain also appeared in Ref.~\cite{Krajenbrink:2021} in a different context, for the study of the Kardar-Parisi-Zhang (KPZ) equation (see Section~\ref{sec:Kraj}).

The works~\cite{Bettelheim:2022,Bettelheim:2022a,Mallick:2022,Krajenbrink:2022} rely on the formalism of Macroscopic fluctuation theory (MFT), which is a deterministic rewriting of the fluctuating hydrodynamics presented in Section~\ref{sec:GenSingFile}. In this formalism, all amounts to the resolution of the MFT equations
\begin{align}
  \label{eq:MFT_q}
  \partial_t q &= \partial_x[D(q) \partial_x q] - \partial_x[\sigma(q)\partial_x p]
  \:,
  \\
  \label{eq:MFT_p}
  \partial_t p &= - D(q) \partial_x^2 p - \frac{1}{2}  \sigma'(q) (\partial_x p)^2 
  \:,
\end{align}
where $p$ is a conjugate field introduced to enforce the continuity relation~\eqref{eq:FlucHydroCont}. Actually, the MFT solution at final time $q(x,T)$ coincides with the generalised density profile~\eqref{eq:defwrGenSFQ}, as we show in Appendix~\ref{sec:AppendixMFT},
\begin{equation}
    \Phi(v) = q(x = \sqrt{2} v, t=T)
    \:.
\end{equation}
The knowledge of the MFT solution $q(x,t)$ therefore allows the determination of the profile $\Phi(v)$.
Until recently, the MFT equations had only been solved for the model of hard Brownian particles~\cite{Krapivsky:2015a} (corresponding to a linear $\sigma$). The recent works~\cite{Bettelheim:2022,Bettelheim:2022a,Mallick:2022} constitute major achievements in the context of MFT: by applying the inverse scattering method~\cite{Ablowitz:1981}, the authors have solved these equations for a quadratic $\sigma$. For a related system of equations, a breakthrough had been performed previously in the study of the weak noise theory of the KPZ equation~\cite{Krajenbrink:2021}.
Actually, we show in Section~\ref{sec:Kraj} that the solution obtained in the recent work~\cite{Krajenbrink:2022}, which encompass the previous results both on MFT~\cite{Bettelheim:2022,Bettelheim:2022a,Mallick:2022} and on the KPZ equation~\cite{Krajenbrink:2021}, can also be rewritten in terms of the equation~\eqref{eq:BilinIntegEqOmGen}. Related equations also arise in mean field games, see for instance~\cite{Bonnemain:2021} in which the conserved quantities of similar equations are characterised.

\subsection{Comparison with Bettelheim, Smith and Meerson}
\label{sec:Bettel}

In Ref.~\cite{Bettelheim:2022}, Bettelheim, Smith and Meerson have solved the MFT equations~(\ref{eq:MFT_q},\ref{eq:MFT_p}) for the KMP model (see Section~\ref{sec:KMP}), corresponding to
\begin{equation}
    D(\rho) = 1
    \:,
    \quad
    \sigma(\rho) = 2 \rho^2
    \:,
\end{equation}
for a specific initial condition
\begin{equation}
    \label{eq:InitCondBettel}
    q(x,0) = W \: \delta(x)
    \:.
\end{equation}
Unlike our study on the SEP, which was performed on a fluctuating initial condition picked from the equilibrium distribution (\textit{annealed}), this is a fixed initial condition (\textit{quenched}), which corresponds to having a mass $W$ placed on site $0$ at $t=0$.

They have obtained that the Fourier transforms of the MFT profile at final time $T$ (taken to be $T=1$ without loss of generality),
\begin{equation}
    \label{eq:DefQBettel}
    Q_\pm(k) = - \lambda \int_{\mathbb{R}^\pm} q(-z,1) \e^{\I k z} \dd z
\end{equation}
satisfy the following closed equation~\cite{Bettelheim:2022}
\begin{equation}
    \label{eq:EqFourierBettel}
    \I k (Q_+(k) + Q_-(k)) - (\I k)^2 Q_+(k) Q_-(k)
    = - \lambda \I k \e^{-k^2}
    \:.
\end{equation}
In order to make the link with Eq.~\eqref{eq:BilinIntegEqOmGen}, we perform an integration by parts in the definition~\eqref{eq:DefQBettel},
\begin{equation}
    \label{eq:DefOmegaFromQ}
    \I k Q_\pm(k) = \pm \lambda q(0^\mp,1)
    + (1 \mp \lambda q(0^\mp, 1) ) \hat{\Omega}_{\mp} \left(-2k\right)
    \:,
\end{equation}
with
\begin{equation}
     \hat{\Omega}_{\pm}(k) = \int_{\mathbb{R}^\pm} \e^{\I k v} \Omega_\pm(v)
     \dd v
     \:,
\end{equation}
where we have used the definition given previously for
\begin{equation}
     \Omega_\pm(v) = -2
     \frac{\partial_v q(x = 2v,1)}{\frac{1}{\lambda} \pm q(0^\pm,1) }
     \:,
\end{equation}
which is identical to our previous definition of $\Omega_\pm$ for the KMP model~\eqref{eq:OmegaFluxKMP}. With the relation~\eqref{eq:DefOmegaFromQ}, the equation~\eqref{eq:EqFourierBettel} becomes
\begin{equation}
    \label{eq:OmegaBettel}
    \hat{\Omega}_+(k) + \hat{\Omega}_-(k) - \hat{\Omega}_+(k) \hat{\Omega}_-(k)
    = \frac{1}{2} \lambda \I k \e^{-k^2/4}
    \:,
\end{equation}
where we have used that
\begin{equation}
    (1+\lambda q(0^+,1))(1-\lambda q(0^-,1)) = 1
    \:,
\end{equation}
which is given in the Supplementary Material of Ref.~\cite{Bettelheim:2022} and is equivalent to~\eqref{eq:CancelVGenQ}. Taking the inverse Fourier transform of~\eqref{eq:OmegaBettel}, we obtain the bilinear equations~\eqref{eq:BilinIntegEqOmGen}, with the kernel
\begin{equation}
    K(v) = \lambda \frac{v \:  \e^{-v^2}}{\sqrt{\pi}}
    \:.
\end{equation}
Finally, the work of Bettelheim and collaborators is a proof that the Eq.~\eqref{eq:BilinIntegEqOmGen} also describes the generalised density profiles for the flux in the KMP model, with the \textit{quenched} initial condition~\eqref{eq:InitCondBettel}.

Recently Bettelheim et al have extended their inverse scattering resolution of the MFT equations to the case of the generalised current defined in Section~\ref{sec:GenCurr}. Following the same derivation as above, we can show that the Eq.~\eqref{eq:BilinIntegEqOmGen} still holds for this observable, with the kernel
\begin{equation}
    K(v) = \lambda (v-\xi)\frac{\:  \e^{-(v-\xi)^2}}{\sqrt{\pi}}
    \:,
\end{equation}
with $\xi$ defined in Section~\ref{sec:GenCurr}. This further extends the range of applications of Eq.~\eqref{eq:BilinIntegEqOmGen}.

\subsection{Comparison with Mallick, Moriya and Sasamoto}
\label{sec:Mallick}

Soon after the work of Bettelheim et al, Mallick, Moriya and Sasamoto have solved the MFT equations for the SEP~\cite{Mallick:2022}, corresponding to
\begin{equation}
    D(\rho) = 1
    \:,
    \quad 
    \sigma(\rho) = 2 \rho(1-\rho)
    \:,
\end{equation}
using the inverse scattering method. As we proceed to show, they provide a proof that our Eq.~\eqref{eq:BilinEqOmegaSummary} for the integrated current in the SEP is exact, as claimed in Section~\ref{sec:Current}.

In~\cite{Mallick:2022}, an equation is obtained for the derivative of the MFT profile at final time
\begin{equation}
    \hat{u}_\pm(k) = \int_{\mathbb{R}^\mp} u(x,T) \e^{-2\I k x}
    \:,
    \quad
    u(x,T) \propto  \partial_x q(x,T)
    \:,
\end{equation}
which reads
\begin{equation}
    \label{eq:MallickU}
    \hat{u}_+(k) + \hat{u}_-(k) 
    + \hat{u}_+(k) \hat{u}_-(k) 
    = \omega_Q \e^{-4 k^2 T}
    \:,
\end{equation}
with $\omega_Q$ given by~\eqref{eq:OmegaQ}. In fact, it can be shown that
\begin{equation}
    u(x,T) = \pm \frac{\partial_x q(x,T)}
    {\frac{1}{1-\e^{\mp \lambda}}- q(0^\pm,T)}
    \:,
    \text{ for } x \gtrless 0
    \:,
\end{equation}
which corresponds to
\begin{equation}
    u(x,T) = \frac{1}{2 \sqrt{T}} \Omega_\pm \left( v = \frac{x}{\sqrt{4T}} \right)
    \:,
\end{equation}
with the definition of $\Omega_\pm$~\eqref{eq:defOmegaQ} and the boundary conditions~(\ref{eq:BoundZeroPQ},\ref{eq:BoundZeroMQ}).
In the Fourier domain, this yields
\begin{equation}
    \hat\Omega_\pm(k) =
    \int_{\mathbb{R}^\pm} \e^{\I k v} \Omega_\pm(v) \dd v
    = \hat{u}_\mp \left( - \frac{k}{4 \sqrt{T}} \right)
    \:.
\end{equation}
With these relations, Eq.~\eqref{eq:MallickU} becomes
\begin{equation}
    \hat{\Omega}_+(k) + \hat{\Omega}_-(k) + \hat{\Omega}_+(k) \hat{\Omega}_-(k)
    = \omega_Q \e^{-k^2/4}
    \:.
\end{equation}
Taking the inverse Fourier transform, we get that $\Omega_\pm$ satisfy the equations~(\ref{eq:BilinEqOmegaSummary}), equivalent to~\eqref{eq:LinEqOmegaSummary}, with the kernel $K$ given by~\eqref{eq:KernelQ}.

The work of Mallick and collaborators~\cite{Mallick:2022} therefore proves that the closed integral equations~\eqref{eq:LinEqOmegaSummary} which we have obtained from a guess (see Section~\ref{sec:CloseEqSEP} and Ref.~\cite{Grabsch:2022}), are indeed exact.

Interestingly, Mallick et al have also obtained an exact expression for a different observable: the MFT profile at initial time $t=0$. Indeed, it is not an initial condition because we consider an annealed situation: the SEP is at equilibrium at $t=0$, so the occupations of the sites $\eta_i(0)$ are random. These occupations are also correlated with the current at time $T$. For instance, a fluctuation of the initial condition that has more particles on the left of $0$ than on the right will relax with time, leading to a higher current than on average. This is what measures the MFT profile at $t=0$. More explicitly, it can be shown (see Appendix~\ref{sec:AppendixMFT}) that
\begin{equation}
    q(x,t=0) = \frac{\moy{\eta_r(0) \e^{\lambda Q_T}}}{\moy{\e^{\lambda Q_T}}}
    \:,
    \quad
    x = \frac{r}{\sqrt{T}}
    \:.
\end{equation}
Defining similarly
\begin{equation}
    \bar{\Omega}(v) = \mp \frac{\omega_Q}{A_\pm} \partial_x q(x,0)
    \text{ for } v \gtrless 0
    \:,
    \quad
    v = \frac{x}{\sqrt{4T}}
    \:,
\end{equation}
with $\omega_Q$ given by~\eqref{eq:OmegaQ} and
\begin{equation}
    A_\pm = \sigma(\rho_\pm)\frac{\e^{\mp \lambda}-1}{2}
    \sqrt{\frac{1 + (\e^{\pm \lambda}-1)\rho_\mp}{1+(\e^{\mp \lambda}-1)\rho_\pm}}
    \:.
\end{equation}
It can be shown that this function obeys the exact same equation~\eqref{eq:BilinEqOmegaSummary}, with the same kernel $K$~\eqref{eq:KernelQ}.
This provides one more observable for which this Wiener-Hopf equation holds.

\subsection{Comparison with Krajenbrink and Le Doussal}
\label{sec:Kraj}

Before the inverse scattering technique was applied to the MFT equations, Krajenbrink and Le Doussal have successfully used it to solve the weak noise theory of the Kardar-Parisi-Zhang (KPZ) equation~\cite{Krajenbrink:2021,Krajenbrink:2022a}. More recently, they have constructed a system of equations which interpolates between the one of the weak noise KPZ equation and the MFT equations of the KMP model (or more generally, models with a quadratic $\sigma(\rho)$)~\cite{Krajenbrink:2022},
\begin{align}
    \label{eq:KLq}
    \partial_t Q
    &= \partial_x^2 Q + 2 \beta \partial_x (Q^2 R) + 2 g Q^2 R
    \:,
    \\
    \label{eq:KLr}
    \partial_t R
    &= - \partial_x^2 R - 2 \beta \partial_x (Q R^2) + 2g Q R^2
    \:,
\end{align}
with initial and terminal boundary conditions
\begin{equation}
    Q(x,0) = \delta(x)
    \:,
    R(x,1) = \Lambda \: \delta(x)
    \:.
\end{equation}
In the MFT context, these conditions arise for any model with quadratic $\sigma(\rho)$ for annealed initial conditions around a step density profile~\cite{Krajenbrink:2022}.
For $\beta=0$ these equations appear in the weak noise theory of the KPZ equation, solved in~\cite{Krajenbrink:2021}, while for $g=0$ and $\beta=-1$, these correspond to the MFT equations of the KMP model solved in~\cite{Bettelheim:2022}.

We now check that the solution obtained in~\cite{Krajenbrink:2022} can be rephrased into the Wiener-Hopf equation~\eqref{eq:BilinIntegEqOmGen}, with a kernel written below.
In Ref.~\cite{Krajenbrink:2022}, it is shown that the half Fourier transforms,
\begin{equation}
    \hat{Q}_\pm(k) = \int_{\mathbb{R}^\pm} Q(x,1) \e^{-\I k x} \dd x
    \:,
\end{equation}
satisfy the equation
\begin{equation}
    \label{eq:KrajFourier}
    (1 - \alpha \hat{Q}_-(k))(1 - \alpha \hat{Q}_+(k))
    = 1 - \alpha \: \e^{-k^2}
    \:,
\end{equation}
with $\alpha = (g + \I \beta k) \Lambda$.
Since the system of equations~(\ref{eq:KLq},\ref{eq:KLr}) no longer describes only MFT equations, we have to extend the definition of $\Omega_\pm$ involved in the desired Wiener-Hopf equation, as
\begin{equation}
    \Omega_\pm(v) = -
    \Lambda \frac{g + \beta \partial_x}{1 \mp \Lambda \beta Q(0^\pm,1)}
    Q(x = 2v,1)
    \:.
\end{equation}
Taking the Fourier transform of this definition, it gives the relation
\begin{multline}
    \Lambda (g + \I \beta k) \hat{Q}_\pm(k)
    = \pm \Lambda \beta Q(0^\pm,1)
    \\
    - (1 \mp \Lambda \beta Q(0^\pm,1))
    \hat{\Omega}_\pm(-2k)
    \:.
\end{multline}
Which, combined with Eq.~\eqref{eq:KrajFourier}, yields
\begin{equation}
    \hat{\Omega}_+(k) + \hat{\Omega}_-(k) + \hat{\Omega}_+(k) \hat{\Omega}_-(k)
    = -\Lambda \left(g - \I \beta \frac{k}{2} \right) \e^{-\frac{k^2}{4}}
    \:,
\end{equation}
where we have used that~\cite{Krajenbrink:2022}
\begin{equation}
    (1-\beta \Lambda Q(0^+,1))(1+\beta \Lambda Q(0^-,1))
    = 1
    \:.
\end{equation}
Taking the inverse Fourier transform, we obtain that $\Omega_\pm$ are solution of the bilinear equations~\eqref{eq:BilinIntegEqOmGen}, with the kernel
\begin{equation}
    K(v) = -\frac{\Lambda}{\sqrt{\pi}}(g-\beta v) \e^{-v^2}
    \:.
\end{equation}

In particular, for $\beta = 0$, describing the weak noise theory of the KPZ equation, the Wiener-Hopf equations~\eqref{eq:BilinIntegEqOmGen} are also involved, as previously obtained in~\cite{Krajenbrink:2021}.

\section{Conclusion}

In this paper, we have provided details on the derivation of the Wiener-Hopf equations for the correlation profiles in the SEP, either in their bilinear form~\eqref{eq:BilinEqOmegaSummary} or their linear form~\eqref{eq:LinEqOmegaSummary}, obtained in~\cite{Grabsch:2022}. We have showed that this exact same equation applies to other situations and models of single-file diffusion, such as the KMP model which is not a model of particles but a mass transfer model, and the random average process which is a model of particles hopping on the continuous line. We have also presented a new application to the double exclusion process, which is a lattice model in which the particles have a volume which occupies two sites. The fact that this equation applies to such a variety of models points towards its universality.

We have compared our approach with the recent results obtained by using the inverse scattering method~\cite{Krajenbrink:2021,Bettelheim:2022,Bettelheim:2022a,Mallick:2022,Krajenbrink:2022}, and shown that these results can be rephrased into the same Wiener-Hopf equation~\eqref{eq:LinEqOmegaSummary}, upon modification of the kernel $K$. This further emphasises the wide range of applications of this equation, and makes it a new promising tool to investigate further questions in single-file diffusion and beyond.

\appendix

\section{Computations at lowest orders from Macroscopic Fluctuation Theory}
\label{sec:AppendixMFT}

Macroscopic Fluctuation Theory~\cite{Bertini:2001,Bertini:2002,Bertini:2005,Bertini:2009,Bertini:2015} is a coarsed-grained description of a single-file system at large time and large distances. The system is then described by a continuous density field $\rho(x,t)$, which evolves stochastically in time. All the microscopic details of the model are encoded in the two transport coefficients $D(\rho)$ and $\sigma(\rho)$ defined in Section~\ref{sec:GenSingFile}. For the SEP, $D(\rho) = \frac{1}{2}$ and $\sigma(\rho) = \rho(1-\rho)$. The probability to evolve from an initial profile $\rho(x,0)$ to a final profile $\rho(x,T)$ at time $T$ is given by the functional integral over the all the possible time evolutions, and over a conjugate field $H$ which enforces the continuity relation between the current and the density~\cite{Derrida:2009a}:
\begin{equation}
  \mathbb{P}(\rho(x,0) \rightarrow \rho(x,T))
  = \int \Df [\rho(x,t)] \Df[H(x,t)] \: \e^{-S[\rho,H]}
  \:,
\end{equation}
where the action $S$ reads
\begin{multline}
  \label{eq:ActS}
  S[\rho,H] =
  \int_{-\infty}^\infty \dd x \int_0^T \dd t
  \Bigg(
    H \partial_t \rho 
    + D(\rho) \partial_x \rho \partial_x H
    \\
    - \frac{\sigma(\rho)}{2} (\partial_x H)^2
  \Bigg)
  \:.
\end{multline}
The initial condition is also random, and distributed as
\begin{equation}
  \mathbb{P}[\rho(x,0)] \simeq \e^{- F[\rho(x,0)]}
  \:,
\end{equation}
where
\begin{equation}
  \label{eq:ActF}
  F[\rho(x,0)] = \int_{-\infty}^{\infty} \dd x \int_{\rho_0(x)}^{\rho(x,0)} \dd z \: 
  \frac{2 D(z)}{\sigma(z)}( \rho(x,0)-z)
  \:,
\end{equation}
and $\rho_0(x)$ is the mean density of the system at position $x$ at $t=0$.

This formalism has been used to compute the first cumulants of a tracer in Ref.~\cite{Krapivsky:2015a}. The cumulant generating function can indeed be expressed as
\begin{multline}
\label{eq:DefCumulGenFctMFT}
  \moy{ \e^{\lambda X_T} } \simeq \int \Df [\rho(x,0)]
  \int \Df [\rho(x,t)] \Df[H(x,t)] \\
  \exp \left[-(S[\rho,H] + F[\rho(x,0)] - \lambda X_T[\rho]) \right]
  \:,
\end{multline}
where $X_T[\rho]$ is the position of the tracer at time $T$, obtained from the density field $\rho(x,t)$ by imposing conservation of the density to the right of the tracer,
\begin{equation}
  \label{eq:FoncXt}
  \int_0^{X_T[\rho]} \rho(x,T) \dd x
  = \int_0^{\infty} \left( \rho(x,T) - \rho(x,0) \right) \dd x
  \:.
\end{equation}

By rescaling the time $t$ by $T$ and the position $x$ by $\sqrt{T}$, one can see that the terms in the exponential in Eq.~\eqref{eq:DefCumulGenFctMFT} are proportional to $\sqrt{T}$. Therefore, for large $T$, the functional integrals are dominated by the optimal realisation $(q,p)$ of the fields $(\rho,H)$ which minimize the action $S+F-\lambda X_T$. This yields the equation~\cite{Krapivsky:2015a}
\begin{align}
  \label{eq:MFT_qApp}
  \partial_t q &= \partial_x[D(q) \partial_x q] - \partial_x[\sigma(q)\partial_x p]
  \:,
  \\
  \label{eq:MFT_pApp}
  \partial_t p &= - D(q) \partial_x^2 p - \frac{1}{2}  \sigma'(q) (\partial_x p)^2 
  \:,
\end{align}
with the condition at $t=T$ for $p$,
\begin{equation}
  \label{eq:MFT_limitP}
  p(x,t=T) = B \: \Theta(x-X_T[q])
  \:,
  \quad B = \frac{\lambda}{q(Y,T)}
  \:,
\end{equation}
and the initial condition for $q$, expressed in terms of $p(x,t=0)$:
\begin{equation}
  \label{eq:MFT_limitQ}
  p(x,0) = B \: \Theta(x) + \int_{\rho_0(x)}^{q(x,0)} \dd r \frac{2 D(r)}{\sigma(r)}
  \:,
\end{equation}
with $\rho_0(x)$ the mean initial profile.

As shown in Ref.~\cite{Poncet:2021}, the generalised density profiles~\eqref{eq:def_wr} can be also be computed using MFT, since
\begin{widetext}
\begin{equation}
  \label{eq:FctIntegW}
  w_r(\lambda,T)
  \simeq \frac{
    \displaystyle
    \int \Df [\rho(x,0)]
    \int \Df [\rho(x,t)] \Df[H(x,t)] \: \rho(X_T[\rho] + r,T) \: 
    \e^{- (S[\rho,H] + F[\rho(x,0)] - \lambda X_T[\rho])}
  }
  {
    \displaystyle
     \int \Df [\rho(x,0)]
    \int \Df [\rho(x,t)] \Df[H(x,t)] \:
    \e^{-(S[\rho,H] + F[\rho(x,0)] - \lambda X_T[\rho])}
  }
  \:.
\end{equation}
\end{widetext}
Evaluating again the integrals with a saddle point estimate for $T \to \infty$, we find the same optimal fields $(q,p)$, and thus
\begin{equation}
  \label{eq:EquivMFTprof}
  w_r(\lambda,T)
  \simeq
  q(X_T[q]+r,T)
  =
  \Phi \left(v = \frac{r}{\sqrt{2T}} , \lambda \right)
    \:.
\end{equation}
The goal is to compute the profile $q(x,T)$ of MFT at lowest orders in $\lambda$ and deduce the corresponding profiles $\Phi$. We only consider the case of the SEP, which corresponds to the coefficients $D(\rho) = 1/2$ and $\sigma(\rho) = \rho(1-\rho)$. As in Ref.~\cite{Krapivsky:2015a}, we solve perturbatively the MFT equations by expanding them in powers of the parameter $B$, defined in~\eqref{eq:MFT_limitP}, as
\begin{equation}
\label{eq:ExpMFTB}
    p(x,t) = \sum_{n \geq 1}B^n p_n(x,t)
    \:,
    \quad
    q(x,t) = \sum_{n \geq 0}B^n q_n(x,t)
    \:.
\end{equation}
We also define
\begin{equation}
\label{eq:defYfromXT}
    Y = X_T[q] = \sum_{n \neq 1} B^n Y_n
    \:.
\end{equation}
We now distinguish two cases, depending on the initial condition, and we set $T=1$ without loss of generality.

\subsection{Flat initial density}

We consider the case of a flat initial mean density $\rho_0(x) = \rho$. The solution for the lowest orders has been computed in~\cite{Krapivsky:2015a}, and reads
\begin{align}
    q_0(x,t) = \rho \:,
    \\
    p_1(x,t) 
    &= \frac{1}{2} \erfc \left( \frac{-x}{\sqrt{2(1-t)}} \right)
    \:,
    \\
    q_1(x,t) 
    &= \frac{\rho(1-\rho)}{2}
  \left[
    \erfc \left( \frac{-x}{\sqrt{2(1-t)}} \right)
    - \erfc \left( \frac{-x}{\sqrt{2t}} \right)
  \right]
  \:,
  \\
  Y_1 &= \sqrt{\frac{2}{\pi}}(1-\rho)
  \:,
\end{align}
\begin{widetext}
\begin{equation}
  p_2(x,t) = - Y_1 K(x|1-t)
  + \frac{1-2\rho}{8} \erfc \left( \frac{x}{\sqrt{2(1-t)}} \right)
  \erfc \left( \frac{-x}{\sqrt{2(1-t)}} \right)
  \:,
  \quad
  Y_2 = 0
  \:,
\end{equation}
\begin{equation}
    q_2(x,t) = \frac{\rho(1-\rho)(1-2\rho)}{4}
  \left[
    \erfc \left( \frac{x}{\sqrt{2t}} \right)
    + 
    \erfc \left( \frac{x}{\sqrt{2t}} \right)
    \erfc \left( \frac{x}{\sqrt{2(1-t)}} \right)
    - \frac{4 Y_1}{1-2\rho} K(x|1-t)
  \right]
  \:,
\end{equation}
\begin{multline}
  p_3(x,t) =
  \frac{(1-2\rho)^2}{24} \left[
    -\erf \left( \frac{x}{\sqrt{2(1-t)}} \right)
    \erfc \left( \frac{-x}{\sqrt{2(1-t)}} \right)
    -\frac{12 Y_1}{1-2\rho}K(x|1-t)
  \right]
  \erfc \left( \frac{x}{\sqrt{2(1-t)}} \right)
  \\
  -\frac{1}{2} \left(
    \frac{Y_1^2 x}{1-t} + (2\rho - 1) Y_1
  \right)
  K(x|1-t)
  + u(x,t)
  \:,
\end{multline}
\begin{multline}
    \label{eq:SolQ3MFT}
  q_3(x,t) =
  \frac{\rho(1-\rho)(1-2\rho)^2}{12}
  \left[
    \erfc \left( \frac{x}{\sqrt{2t}} \right)
    -\erfc \left( \frac{x}{\sqrt{2(1-t)}} \right)
  \right]
  \\
  + \frac{\rho(1-\rho)}{2} Y_1 K(x|1-t)
  \left(
    -\frac{x}{1-t} Y_1
    + (1-2\rho)
    - (1-2\rho) \erfc \left( \frac{x}{\sqrt{2t}} \right)
  \right)
  + h(x,t)
  \:,
\end{multline}
\end{widetext}
where
\begin{equation}
    K(x|t) =  \frac{\e^{-\frac{x^2}{2t}}}{\sqrt{2\pi t}}
\end{equation}
is the heat kernel, $u(x,t)$ and $h(x,t)$ satisfy the following inhomogeneous heat equations
\begin{equation}
\label{eq:MFTu}
  \partial_t u = -\frac{1}{2} \partial_x^2 u
  + q_1 (\partial_x p_1)^2
  \:,
\end{equation}
\begin{equation}
\label{eq:MFTh}
  \partial_t h = \frac{1}{2} \partial_x^2 h
  - \rho(1-\rho) \partial_x^2 u
  + \partial_x(q_1^2 \partial_x p_1)
  \:,
\end{equation}
with the boundary conditions
\begin{equation}
\label{eq:FinCondU}
  u(x,1) = 0
  \:,
\end{equation}
\begin{equation}
\label{eq:InitCondH}
    h(x,0) =
  - \frac{1}{3\rho(1-\rho)}q_1(x,0)^3 + \rho(1-\rho) u(x,0)
  \:.
\end{equation}
In Ref.~\cite{Krapivsky:2015a}, these equations were not computed analytically, but instead studied numerically in order to compute the fourth cumulant $\hat\kappa_4$. These equations can actually be studied to obtain the solution $u(x,t=1)$ at final time, which is the only time we need to get $\Phi$, as shown in Eq.~\eqref{eq:EquivMFTprof}. We make the transformation
\begin{multline}
    \label{eq:DefFctHq3}
    h(x,t) = - \frac{1}{3\rho(1-\rho)}q_1(x,t)^3 + \rho(1-\rho) u(x,t) 
    \\
    + \rho^2(1-\rho)^2\tilde{h}(x,t)
    \:,
\end{multline}
inspired by the form of the initial condition~\eqref{eq:InitCondH}. By combining the equations~(\ref{eq:MFTu},\ref{eq:MFTh}), we obtain an equation on $\tilde{h}$ only:
\begin{equation}
    \partial_t \tilde{h} = \frac{1}{2} \partial_x^2 \tilde{h}
    - \frac{1}{\rho^3(1-\rho)^3} q_1 \left( \partial_x q_1 - \rho(1-\rho) \partial_x p_1 \right)^2
    \:,
\end{equation}
which explicitly gives
\begin{multline}
    \partial_t \tilde{h} = \frac{1}{2} \partial_x^2 \tilde{h}
    \\
    -\frac{\e^{-\frac{x^2}{t}}}{4 \pi t}
    \left[ \erfc \left( \frac{x}{\sqrt{2(1-t)}} \right) -
    \erfc \left( \frac{x}{\sqrt{2t}} \right)\right]
    \:,
\end{multline}
with $\tilde{h}(x,0) = 0$ by construction. The solution can be written as the sum of two terms $\tilde{h}(x,t) = \tilde{h}_1(x,t) + \tilde{h}_2(x,t)$, where
\begin{multline}
    \tilde{h}_1(x,t) =  - \int_{-\infty}^{\infty} \dd y \int_0^t \dd t'
    K(x-y|t-t')
    \\
    \frac{\e^{-\frac{y^2}{t'}}}{4 \pi t'} 
    \erfc \left( \frac{y}{\sqrt{2(1-t')}} \right)
    \:,
\end{multline}
\begin{multline}
    \label{eq:DefHt2}
    \tilde{h}_2(x,t) =  \int_{-\infty}^{\infty} \dd y \int_0^t \dd t'
    K(x-y|t-t')
    \\
    \frac{\e^{-\frac{y^2}{t'}}}{4 \pi t'} 
    \erfc \left( \frac{y}{\sqrt{2t'}} \right)
    \:.
\end{multline}
Let us first consider the function $\tilde{h}_1$. In order to compute it, we introduce another function,
\begin{multline}
    \tilde{H}_1(x,t;Y) =  - \int_{-\infty}^{\infty} \dd y \int_0^t \dd t'
    K(x-y|t-t')
    \\
    \frac{\e^{-\frac{y^2}{t'}}}{4 \pi t'} 
    \erfc \left( \frac{y-Y}{\sqrt{2(1-t')}} \right)
    \:,
\end{multline}
so that $\tilde{H}_1(x,t;0) = \tilde{h}_1(x,t)$. Since we only need to compute the profile at final time $t=1$ thanks to~\eqref{eq:EquivMFTprof}, we will focus on $\tilde{H}_1(x,t;Y)$. Taking a derivative with respect to $Y$ and performing the Gaussian integrals over $y$, we obtain
\begin{equation}
    \partial_Y \tilde{H}_1(x,1;Y) = 
    -\int_0^1 \dd t' \frac{\e^{-\frac{(2-t')x^2 + 2t' x Y + (2-t') Y^2}{4(1-t')}}}{4 \pi^{3/2} \sqrt{t'(1-t')}}
    \:.
\end{equation}
This integral can be evaluated by \texttt{Mathematica} (upon performing the change of variables $t' \to 1-t'$), and reads
\begin{equation}
    \partial_Y \tilde{H}_1(x,1;Y) = 
    -\frac{\e^{- \frac{1}{4}(x+Y)^2}}{4 \sqrt{\pi}}
    \erfc \left(
        \frac{\abs{x-Y}}{2}
    \right)
    \:.
\end{equation}
We can then integrate over $Y$ to obtain
\begin{equation}
    \tilde{h}_1(x,1) = -\int_{-\infty}^0 \frac{\e^{- \frac{1}{4}(x+Y)^2}}{4 \sqrt{\pi}}
    \erfc \left(
        \frac{\abs{x-Y}}{2}
    \right)
    \dd Y
    \:.
\end{equation}
This last integral can be performed using the using the tables given in~\cite{Owen:1980}, and we obtain, for $x>0$,
\begin{equation}
    \tilde{h}_1(x,1) = \frac{1}{4} \erfc \left( \frac{x}{\sqrt{2}} \right)
    - \frac{1}{8} \erfc \left( \frac{x}{2} \right)^2
    \:.
\end{equation}
For $\tilde{h}_2$~\eqref{eq:DefHt2}, we can perform the integral over $y$ using again~\cite{Owen:1980},
\begin{multline}
    \label{eq:IntegRepresHt2}
    \tilde{h}_2(x,1) = \int_0^1 \dd t'
    \frac{\e^{-\frac{x^2}{2-t'}}}{4 \pi \sqrt{t'(2-t')}}
    \\
    \erfc \left(
        x \sqrt{\frac{t'}{(2-t')(6-4t')}}
    \right)
    \:.
\end{multline}
We have not been able to evaluate this integral analytically. However, this representation allows for a very precise numerical computation, for various values of $x$. We have thus used a semi-numerical procedure to evaluate this integral, which we now discuss. We have computed a list of values of $\tilde{h}_2(x,1)$ for $100$ points between $0.1$ and $2$. Fitting this list of points with a few functions we expect to find, such that powers of $\erfc(\frac{x}{\sqrt{2}})$, we obtain
\begin{multline}
    \tilde{h}_2(x,1) \simeq 0.16666666666666677
    \erfc \left( \frac{x}{\sqrt{2}} \right)
    \\
    + 8.21189 \times 10^{-16} \erfc \left( \frac{x}{\sqrt{2}} \right)^2
    \\
    -0.041666666666667136 \erfc \left( \frac{x}{\sqrt{2}} \right)^3
    \:.
\end{multline}
This is a numerical result, which we can consider to yield an analytical result, since
\begin{equation}
    0.16666666666666677 - \frac{1}{6} \simeq 1.11 \times 10^{-16}
    \:,
\end{equation}
\begin{equation}
    0.041666666666667136 - \frac{1}{24} \simeq 4.72 \times 10^{-16}
    \:,
\end{equation}
which are of the same order as the precision of the numerical evaluation of the integral~\eqref{eq:IntegRepresHt2}. Therefore, we consider that
\begin{equation}
    \tilde{h}_2(x,1) = \frac{1}{6} \erfc \left( \frac{x}{\sqrt{2}} \right)
    - \frac{1}{24} \erfc \left( \frac{x}{\sqrt{2}} \right)^3
    \:,
\end{equation}
for $x>0$ (since the fit is performed on positive values of $x$). Combining these expressions with the definition~\eqref{eq:DefFctHq3}, we obtain for $x>0$,
\begin{equation}
    h(x,1) = \frac{1}{12} \erfc \left( \frac{x}{\sqrt{2}} \right)
    - \frac{1}{8} \erfc \left( \frac{x}{2} \right)^2
    \:.
\end{equation}
Using a similar approach for $x<0$, we finally find that
\begin{equation}
    h(x,1) = \sign(x) \left(
    \frac{1}{12} \erfc \left( \frac{\abs{x}}{\sqrt{2}} \right)
    - \frac{1}{8} \erfc \left( \frac{\abs{x}}{2} \right)^2
    \right)
    \:.
\end{equation}
Combined with the expression~\eqref{eq:SolQ3MFT}, this completes the resolution of the MFT equations (at $t=1$) at order $3$ in $\lambda$.

Note that, in this procedure, we have used a fit to determine an integral, so the result is not fully analytic. This expression is used to obtain~\eqref{eq:PhiFlatOrder3} in the main text, from which is deduced~\eqref{eq:FirstGuessRHSflat}. Nevertheless, this last equation can be obtained fully analytically by considering the case of a step initial density (see below), from which~\eqref{eq:RHSOrder1Step} is deduced.

The last step is to relate $B$ and $\lambda$. This cannot be done from the relation~\eqref{eq:MFT_limitP} because $q(x,T)$ is discontinuous at $x=Y$. In Ref.~\cite{Krapivsky:2015a}, the relation between these parameters was obtained by optimising the action with respect to $B$. This procedure is rather complex, so here we use a shortcut. $Y = X_T[q]$ is the position of the tracer in the optimal configuration $q$. From the fonctional integral, it is given by
\begin{widetext}
\begin{equation}
\label{eq:RelXTdpsidlambda}
  X_T[q]
  \simeq \frac{
    \displaystyle
    \int \Df [\rho(x,0)]
    \int \Df [\rho(x,t)] \Df[H(x,t)] \: X_T[\rho] \: 
    \e^{- (S[\rho,H] + F[\rho(x,0)] - \lambda X_T[\rho])}
  }
  {
    \displaystyle
     \int \Df [\rho(x,0)]
    \int \Df [\rho(x,t)] \Df[H(x,t)] \:
    \e^{-(S[\rho,H] + F[\rho(x,0)] - \lambda X_T[\rho])}
  }
  \equiv \frac{\moy{X_T \: \e^{\lambda X_T}}}{\moy{\e^{\lambda X_T}}}
  = \frac{\dd }{\dd \lambda} \ln \moy{\e^{\lambda X_T}}
  \:,
\end{equation}
\end{widetext}
which is the derivative of the cumulant generating function of $X_T$, computed in~\cite{Imamura:2017,Imamura:2021}. This result directly gives the expansion of $Y = X_T[q]$ in powers of $\lambda$, which combined with~\eqref{eq:defYfromXT} yields
\begin{equation}
    B = \frac{\lambda}{\rho} + \left(
    \frac{(1-\rho)^2}{\pi \rho^3}
    - \frac{(1-\rho)(1-2\rho)}{6 \rho^3}
    \right) \lambda^3
    + \O(\lambda^4)
    \:.
\end{equation}
Finally, using the relation~\eqref{eq:EquivMFTprof} between $q$ and $\Phi$, we obtain for the first orders
\begin{multline}
  \label{eq:PhiMFTflat}
  \Phi(v) = \rho + \lambda \frac{1-\rho}{2} \erfc(v)
  \\
  + \lambda^2 \left(
  \frac{(1-\rho)(1-2\rho)}{4 \rho} \erfc(v)
  - \frac{1-\rho}{\pi \rho} \e^{-v^2}
  \right)
  \\
  + \lambda^3 \Bigg(
  (1-\rho )\frac{2 (3+\pi ) \rho ^2-(12+\pi ) \rho +6 + \pi \rho(1-\rho)}
  {12 \pi  \rho ^2}
  \erfc(v)
  \\
  +(1-\rho)^2  \frac{2 (1-\rho) v - \sqrt{\pi } (1-2 \rho)}
  {2\pi^{3/2} \rho ^2} \e^{-v^2}
  \\
  -\frac{(1-\rho)^2}{8\rho} \erfc \left( \frac{v}{\sqrt{2}} \right)^2
  \Bigg) + \O(\lambda^4)
  \:,
\end{multline}
for $v>0$. The expression for $v<0$ can be deduced by replacing $v \to -v$ and $\lambda \to -\lambda$.

\subsection{Step initial density}

We now consider the case of an initial step density
\begin{equation}
    \rho_0(x) = \left\lbrace
    \begin{array}{ll}
        \rho_+ & \text{for } x > 0 \:,  \\
        \rho_- & \text{for } x < 0 \:.
    \end{array}
    \right.
\end{equation}
At lowest order, the MFT equations give
\begin{equation}
    q_0(x,t) = \frac{\rho_+}{2} \erfc \left( - \frac{x}{\sqrt{2t}} \right)
    + \frac{\rho_-}{2} \erfc \left(\frac{x}{\sqrt{2t}} \right)
    \:,
\end{equation}
from which we deduce $Y_0$ solution of
\begin{equation}
    \int_0^{Y_0} q_0(x,1) \dd x
  = \int_0^{\infty} \left( q_0(x,1) - q_0(x,0) \right) \dd x
  \:.
\end{equation}
At first order, we get
\begin{equation}
    p_1(x,t) = \frac{1}{2} \erfc \left( \frac{Y_0-x}{\sqrt{2(1-t)}} \right)
    \:,
\end{equation}
while $q_1$ is solution of
\begin{equation}
    \partial_t q_1 = \frac{1}{2} \partial_x^2 q_1
    - \partial_x (q_0(1-q_0) \partial_x p_1)
    \:,
\end{equation}
with
\begin{equation}
    q_1(x,0) = q_0(x,0) (1-q_0(x,0)) (p_1(x,0) - \Theta(x))
    \:.
\end{equation}
We define
\begin{equation}
    q_1(x,t) = q_0(x,t) (1-q_0(x,t)) p_1(x,t) + \tilde{q}_1(x,t)
    \:,
\end{equation}
which is solution of
\begin{equation}
    \partial_t \tilde{q}_1 = \frac{1}{2} \partial_x^2 \tilde{q}_1
    - (\partial_x q_0)^2 p_1
    \:,
\end{equation}
with
\begin{equation}
    \tilde{q}_1(x,0) = - \rho_+ (1-\rho_+) \Theta(x)
    \:.
\end{equation}
The equation can be solved by treating $Y_0$, which appears in $p_1$, as a parameter, and differentiating with respect to $Y_0$. This gives, for $t=1$:
\begin{equation}
    \partial_{Y_0} \tilde{q}_1(x,1) = 
    \frac{(\rho_+-\rho_-)^2}{4 \sqrt{\pi}}
    \e^{- \frac{1}{4} (x+ Y_0)^2} \erfc \left( \frac{\abs{x-Y_0}}{2} \right)
    \:.
\end{equation}
Integrating over $Y_0$, and adding the contribution of the initial condition, we finally get
\begin{multline}
    q_1(x,t=1) = q_0(x,1) (1-q_0(x,1)) \Theta(x-Y_0)
    \\
    - (\rho_+-\rho_-)^2
    \int_{Y_0}^\infty \frac{\dd z}{4 \sqrt{\pi}}
    \e^{- \frac{1}{4}(x+z)^2} \erfc\left( \frac{\abs{x-z}}{2} \right)
    \\
    - \frac{\rho_+(1-\rho_+)}{2} \erfc \left( - \frac{x}{\sqrt{2t}} \right)
    \:.
\end{multline}
Since $q_0$ is continuous, we can actually use~\eqref{eq:MFT_limitP} to deduce
\begin{equation}
    B = \frac{\lambda}{q_0(Y_0,1)} + \O(\lambda^2)
    \:,
\end{equation}
which combined with~\eqref{eq:RelXTdpsidlambda} yields $Y_1 = \sqrt{2} \hat\kappa_2 q_0(Y_0,1)$. Finally, from~\eqref{eq:EquivMFTprof} we get
\begin{equation}
    \Phi_0(v) = \frac{\rho_+}{2} \erfc(-v-\hat\kappa_1)
    + \frac{\rho_-}{2}\erfc(v+\hat\kappa_1)
    \:,
\end{equation}
with $\hat\kappa_1 = Y_0/\sqrt{2}$, and
\begin{multline}
\label{eq:Phi1StepMFT}
    \Phi_1(v) =
    \hat\kappa_2 \Phi_0'(v)
    + \frac{1}{\Phi_0(0)} \Phi_0(v) (1-\Phi_0(v))
    \\
    - \frac{\rho_+(1-\rho_+)}{2 \Phi_0(0)} \erfc(-v-\hat\kappa_1)
    \\
    - \frac{(\rho_+-\rho_-)^2}{\Phi_0(0)} \int_0^\infty \frac{\dd z}{2 \sqrt{2\pi}} 
    \e^{- \frac{1}{2}(v+2\hat\kappa_1 + z)^2} \erfc \left( \frac{\abs{v-z}}{\sqrt{2}} \right)
    \:,
\end{multline}
where the second cumulant $\hat\kappa_2$ is given in~\cite{Imamura:2017,Imamura:2021}, but its exact value is not required here, since the term involving $\hat\kappa_2$ will cancel in~\eqref{eq:RHSOrder1Step}.

\section{Numerical simulations}
\label{sec:AppNumSim}

Here we provide some details on the microscopic simulations performed for the different systems presented in the paper. Most of these explanations were already presented in Refs.~\cite{Poncet:2021,Grabsch:2022}.

\subsection{Symmetric exclusion process} 

The SEP simulations are performed on a periodic ring of size $L=5000$ sites, with $N=\rho L$ particles at average density $\rho$. The particles are initially placed uniformly at random. The particle jumps are implemented as follows: a particle is chosen uniformly at random, with one direction (left and right with equal probabilities). If the chosen particle has no neighbour in this direction, the jump is performed, otherwise it is rejected. In both cases, the simulation time is incremented by a random number chosen from an exponential distribution of rate $N$. 

We iterate this process until the wanted final time is reached. We keep track of a particle (the tracer) and calculate the moments of its displacement and the generalized density profiles.

\subsection{Kipnis-Marchioro-Presutti model}

On each site of a periodic lattice of length $L = 500$, we place a continuous energy variable $\varepsilon_i > 0$. The variables are initialized according to the equilibrium measure, which is a product of exponential laws of parameter $\rho=1$ (the variables represent energies, so this corresponds to a Boltzmann distribution). Similarly as in the SEP, we choose a site $n$ according to the uniform distribution on integers between 1 and $L$, then we compute the sum of energy variables at sites $n$ and $n+1$ and we randomly redistribute it between these sites. The first site receives a random fraction of the energy sampled from the uniform distribution on the real interval $[0,1]$ and the other site receives the remainder. The simulation time is increment by a random number following an exponential distribution of rate $L$.

Initially we define the position of the tracer to be $0$. We fix a site $b = \lfloor L/2 \rfloor$ and we denote $e$ the sum of the variables on sites between $0$ and $b$ at initial time. At further times, the position of the tracer is the site $x$ such that the sum of the variables on sites between $x$ and $b$ plus the total current (since initial time) between sites $b$ and $b+1$ is equal to $e$. If $x$ is greater than $b+1$ we count the variables negatively. This corresponds to the definition~\eqref{eq:FoncXt} in the hydrodynamic description in the limit of large $L$. We averaged the observables of interest over $10^9$ simulations.

\subsection{Random-average process}

To simulate the RAP, we place $N=10000$ particles at positions $x_i$ on a periodic ring of length $L=10000$. When a particle jumps in a given direction, it travels a random fraction following the uniform law on $[0,1]$ of the distance to the next particle. At equilibrium, the particles are distributed according to the following law~\cite{Cividini:2016a}, where we define the gaps $g_i = x_{i+1}- x_i$.
\begin{equation}
  P_{N,L}(\{ g_n \} ) \propto \prod_{n=1}^N \frac{1}{\sqrt{g_n}}
  \:
  \delta \left(
    \sum_{n=1}^N g_n - L
  \right)
  \:.
\end{equation}
This can be rephrased in terms of variables $G_n = \sqrt{g_n}$. The vector $(G_1,\ldots,G_N)$ is uniformly distributed on the $N$-dimensional sphere of radius $\sqrt{L}$. To sample this distribution, we simply draw $N$ independent gaussian variables $X_i$ with zero mean and unit variance and we consider
\begin{equation}
    g_i = G_i^2 = L \frac{X_i^2}{\displaystyle \sum_{n=1}^N X_n^2}
    \:.
\end{equation}
The data we use in our plots have been averaged from $8 \cdot 10^6$ simulations.

\subsection{The double exclusion process}
We simulate a periodic system of $L=2000$ sites and $N$ particles. Initially, the equilibrium measure is sampled the following way: we place at random $N$ particles on a ring of $L-N$ sites uniformly among all possible configurations, then we add a site in front of each particle. The dynamics is then simulated the same way as the SEP. We chose a particle according to the uniform law among all particles, and a direction (left or right with probability $1/2$). The difference compared to the SEP is that here the jump is performed only if the \textit{two} neighbouring sites in the chosen direction are empty, otherwise it is rejected. In both cases the simulation time is incremented by a random number sampled from an exponential distribution of rate $N$.

Initially, one particle is chosen to be the tracer (defining the position 0 on the ring) and its displacements are followed during all the dynamics. We perform $10^8$ simulations.


\begin{thebibliography}{53}%
\makeatletter
\providecommand \@ifxundefined [1]{%
 \@ifx{#1\undefined}
}%
\providecommand \@ifnum [1]{%
 \ifnum #1\expandafter \@firstoftwo
 \else \expandafter \@secondoftwo
 \fi
}%
\providecommand \@ifx [1]{%
 \ifx #1\expandafter \@firstoftwo
 \else \expandafter \@secondoftwo
 \fi
}%
\providecommand \natexlab [1]{#1}%
\providecommand \enquote  [1]{``#1''}%
\providecommand \bibnamefont  [1]{#1}%
\providecommand \bibfnamefont [1]{#1}%
\providecommand \citenamefont [1]{#1}%
\providecommand \href@noop [0]{\@secondoftwo}%
\providecommand \href [0]{\begingroup \@sanitize@url \@href}%
\providecommand \@href[1]{\@@startlink{#1}\@@href}%
\providecommand \@@href[1]{\endgroup#1\@@endlink}%
\providecommand \@sanitize@url [0]{\catcode `\\12\catcode `\$12\catcode
  `\&12\catcode `\#12\catcode `\^12\catcode `\_12\catcode `\%12\relax}%
\providecommand \@@startlink[1]{}%
\providecommand \@@endlink[0]{}%
\providecommand \url  [0]{\begingroup\@sanitize@url \@url }%
\providecommand \@url [1]{\endgroup\@href {#1}{\urlprefix }}%
\providecommand \urlprefix  [0]{URL }%
\providecommand \Eprint [0]{\href }%
\providecommand \doibase [0]{http://dx.doi.org/}%
\providecommand \selectlanguage [0]{\@gobble}%
\providecommand \bibinfo  [0]{\@secondoftwo}%
\providecommand \bibfield  [0]{\@secondoftwo}%
\providecommand \translation [1]{[#1]}%
\providecommand \BibitemOpen [0]{}%
\providecommand \bibitemStop [0]{}%
\providecommand \bibitemNoStop [0]{.\EOS\space}%
\providecommand \EOS [0]{\spacefactor3000\relax}%
\providecommand \BibitemShut  [1]{\csname bibitem#1\endcsname}%
\let\auto@bib@innerbib\@empty
\bibitem [{\citenamefont {Levitt}(1973)}]{Levitt:1973}%
  \BibitemOpen
  \bibfield  {author} {\bibinfo {author} {\bibfnamefont {D.~G.}\ \bibnamefont
  {Levitt}},\ }\href {http://link.aps.org/doi/10.1103/PhysRevA.8.3050}
  {\bibfield  {journal} {\bibinfo  {journal} {Phys. Rev. A}\ }\textbf {\bibinfo
  {volume} {8}},\ \bibinfo {pages} {3050} (\bibinfo {year} {1973})}\BibitemShut
  {NoStop}%
\bibitem [{\citenamefont {Arratia}(1983)}]{Arratia:1983}%
  \BibitemOpen
  \bibfield  {author} {\bibinfo {author} {\bibfnamefont {R.}~\bibnamefont
  {Arratia}},\ }\href {http://www.jstor.org/stable/2243693} {\bibfield
  {journal} {\bibinfo  {journal} {Ann. Probab.}\ }\textbf {\bibinfo {volume}
  {11}},\ \bibinfo {pages} {362} (\bibinfo {year} {1983})}\BibitemShut
  {NoStop}%
\bibitem [{\citenamefont {Harris}(1965)}]{Harris:1965}%
  \BibitemOpen
  \bibfield  {author} {\bibinfo {author} {\bibfnamefont {T.~E.}\ \bibnamefont
  {Harris}},\ }\href {http://www.jstor.org/stable/3212197} {\bibfield
  {journal} {\bibinfo  {journal} {J. Appl. Probab.}\ }\textbf {\bibinfo
  {volume} {2}},\ \bibinfo {pages} {323} (\bibinfo {year} {1965})}\BibitemShut
  {NoStop}%
\bibitem [{\citenamefont {Hahn}\ \emph {et~al.}(1996)\citenamefont {Hahn},
  \citenamefont {K{\"a}rger},\ and\ \citenamefont {Kukla}}]{Hahn:1996}%
  \BibitemOpen
  \bibfield  {author} {\bibinfo {author} {\bibfnamefont {K.}~\bibnamefont
  {Hahn}}, \bibinfo {author} {\bibfnamefont {J.}~\bibnamefont {K{\"a}rger}}, \
  and\ \bibinfo {author} {\bibfnamefont {V.}~\bibnamefont {Kukla}},\ }\href
  {http://link.aps.org/doi/10.1103/PhysRevLett.76.2762} {\bibfield  {journal}
  {\bibinfo  {journal} {Phys. Rev. Lett.}\ }\textbf {\bibinfo {volume} {76}},\
  \bibinfo {pages} {2762} (\bibinfo {year} {1996})}\BibitemShut {NoStop}%
\bibitem [{\citenamefont {Wei}\ \emph {et~al.}(2000)\citenamefont {Wei},
  \citenamefont {Bechinger},\ and\ \citenamefont {Leiderer}}]{Wei:2000}%
  \BibitemOpen
  \bibfield  {author} {\bibinfo {author} {\bibfnamefont {Q.-H.}\ \bibnamefont
  {Wei}}, \bibinfo {author} {\bibfnamefont {C.}~\bibnamefont {Bechinger}}, \
  and\ \bibinfo {author} {\bibfnamefont {P.}~\bibnamefont {Leiderer}},\ }\href
  {\doibase 10.1126/science.287.5453.625} {\bibfield  {journal} {\bibinfo
  {journal} {Science}\ }\textbf {\bibinfo {volume} {287}},\ \bibinfo {pages}
  {625} (\bibinfo {year} {2000})}\BibitemShut {NoStop}%
\bibitem [{\citenamefont {Lin}\ \emph {et~al.}(2005)\citenamefont {Lin},
  \citenamefont {Meron}, \citenamefont {Cui}, \citenamefont {Rice},\ and\
  \citenamefont {Diamant}}]{Lin:2005}%
  \BibitemOpen
  \bibfield  {author} {\bibinfo {author} {\bibfnamefont {B.}~\bibnamefont
  {Lin}}, \bibinfo {author} {\bibfnamefont {M.}~\bibnamefont {Meron}}, \bibinfo
  {author} {\bibfnamefont {B.}~\bibnamefont {Cui}}, \bibinfo {author}
  {\bibfnamefont {S.~A.}\ \bibnamefont {Rice}}, \ and\ \bibinfo {author}
  {\bibfnamefont {H.}~\bibnamefont {Diamant}},\ }\href {\doibase
  10.1103/physrevlett.94.216001} {\bibfield  {journal} {\bibinfo  {journal}
  {Phys. Rev. Lett.}\ }\textbf {\bibinfo {volume} {94}},\ \bibinfo {pages}
  {216001} (\bibinfo {year} {2005})}\BibitemShut {NoStop}%
\bibitem [{\citenamefont {Spitzer}(1970)}]{Spitzer:1970}%
  \BibitemOpen
  \bibfield  {author} {\bibinfo {author} {\bibfnamefont {F.}~\bibnamefont
  {Spitzer}},\ }\href {\doibase 10.1016/0001-8708(70)90034-4} {\bibfield
  {journal} {\bibinfo  {journal} {Adv. Math.}\ }\textbf {\bibinfo {volume}
  {5}},\ \bibinfo {pages} {246} (\bibinfo {year} {1970})}\BibitemShut {NoStop}%
\bibitem [{\citenamefont {Derrida}\ and\ \citenamefont
  {Gerschenfeld}(2009{\natexlab{a}})}]{Derrida:2009}%
  \BibitemOpen
  \bibfield  {author} {\bibinfo {author} {\bibfnamefont {B.}~\bibnamefont
  {Derrida}}\ and\ \bibinfo {author} {\bibfnamefont {A.}~\bibnamefont
  {Gerschenfeld}},\ }\href {\doibase 10.1007/s10955-009-9772-7} {\bibfield
  {journal} {\bibinfo  {journal} {J. Stat. Phys.}\ }\textbf {\bibinfo {volume}
  {136}},\ \bibinfo {pages} {1} (\bibinfo {year}
  {2009}{\natexlab{a}})}\BibitemShut {NoStop}%
\bibitem [{\citenamefont {Imamura}\ \emph {et~al.}(2017)\citenamefont
  {Imamura}, \citenamefont {Mallick},\ and\ \citenamefont
  {Sasamoto}}]{Imamura:2017}%
  \BibitemOpen
  \bibfield  {author} {\bibinfo {author} {\bibfnamefont {T.}~\bibnamefont
  {Imamura}}, \bibinfo {author} {\bibfnamefont {K.}~\bibnamefont {Mallick}}, \
  and\ \bibinfo {author} {\bibfnamefont {T.}~\bibnamefont {Sasamoto}},\ }\href
  {\doibase 10.1103/PhysRevLett.118.160601} {\bibfield  {journal} {\bibinfo
  {journal} {Phys. Rev. Lett.}\ }\textbf {\bibinfo {volume} {118}},\ \bibinfo
  {pages} {160601} (\bibinfo {year} {2017})}\BibitemShut {NoStop}%
\bibitem [{\citenamefont {Imamura}\ \emph {et~al.}(2021)\citenamefont
  {Imamura}, \citenamefont {Mallick},\ and\ \citenamefont
  {Sasamoto}}]{Imamura:2021}%
  \BibitemOpen
  \bibfield  {author} {\bibinfo {author} {\bibfnamefont {T.}~\bibnamefont
  {Imamura}}, \bibinfo {author} {\bibfnamefont {K.}~\bibnamefont {Mallick}}, \
  and\ \bibinfo {author} {\bibfnamefont {T.}~\bibnamefont {Sasamoto}},\ }\href
  {\doibase 10.1007/s00220-021-03954-x} {\bibfield  {journal} {\bibinfo
  {journal} {Commun. Math. Phys.}\ }\textbf {\bibinfo {volume} {384}},\
  \bibinfo {pages} {1409} (\bibinfo {year} {2021})}\BibitemShut {NoStop}%
\bibitem [{\citenamefont {Krapivsky}\ \emph
  {et~al.}(2015{\natexlab{a}})\citenamefont {Krapivsky}, \citenamefont
  {Mallick},\ and\ \citenamefont {Sadhu}}]{Krapivsky:2015a}%
  \BibitemOpen
  \bibfield  {author} {\bibinfo {author} {\bibfnamefont {P.~L.}\ \bibnamefont
  {Krapivsky}}, \bibinfo {author} {\bibfnamefont {K.}~\bibnamefont {Mallick}},
  \ and\ \bibinfo {author} {\bibfnamefont {T.}~\bibnamefont {Sadhu}},\ }\href
  {\doibase 10.1007/s10955-015-1291-0} {\bibfield  {journal} {\bibinfo
  {journal} {J. Stat. Phys.}\ }\textbf {\bibinfo {volume} {160}},\ \bibinfo
  {pages} {885} (\bibinfo {year} {2015}{\natexlab{a}})}\BibitemShut {NoStop}%
\bibitem [{\citenamefont {Poncet}\ \emph {et~al.}(2021)\citenamefont {Poncet},
  \citenamefont {Grabsch}, \citenamefont {Illien},\ and\ \citenamefont
  {B\'enichou}}]{Poncet:2021}%
  \BibitemOpen
  \bibfield  {author} {\bibinfo {author} {\bibfnamefont {A.}~\bibnamefont
  {Poncet}}, \bibinfo {author} {\bibfnamefont {A.}~\bibnamefont {Grabsch}},
  \bibinfo {author} {\bibfnamefont {P.}~\bibnamefont {Illien}}, \ and\ \bibinfo
  {author} {\bibfnamefont {O.}~\bibnamefont {B\'enichou}},\ }\href {\doibase
  10.1103/PhysRevLett.127.220601} {\bibfield  {journal} {\bibinfo  {journal}
  {Phys. Rev. Lett.}\ }\textbf {\bibinfo {volume} {127}},\ \bibinfo {pages}
  {220601} (\bibinfo {year} {2021})}\BibitemShut {NoStop}%
\bibitem [{\citenamefont {Grabsch}\ \emph {et~al.}(2022)\citenamefont
  {Grabsch}, \citenamefont {Poncet}, \citenamefont {Rizkallah}, \citenamefont
  {Illien},\ and\ \citenamefont {B{\'e}nichou}}]{Grabsch:2022}%
  \BibitemOpen
  \bibfield  {author} {\bibinfo {author} {\bibfnamefont {A.}~\bibnamefont
  {Grabsch}}, \bibinfo {author} {\bibfnamefont {A.}~\bibnamefont {Poncet}},
  \bibinfo {author} {\bibfnamefont {P.}~\bibnamefont {Rizkallah}}, \bibinfo
  {author} {\bibfnamefont {P.}~\bibnamefont {Illien}}, \ and\ \bibinfo {author}
  {\bibfnamefont {O.}~\bibnamefont {B{\'e}nichou}},\ }\href {\doibase
  10.1126/sciadv.abm5043} {\bibfield  {journal} {\bibinfo  {journal} {Sci.
  Adv.}\ }\textbf {\bibinfo {volume} {8}},\ \bibinfo {pages} {eabm5043}
  (\bibinfo {year} {2022})}\BibitemShut {NoStop}%
\bibitem [{\citenamefont {Bettelheim}\ \emph
  {et~al.}(2022{\natexlab{a}})\citenamefont {Bettelheim}, \citenamefont
  {Smith},\ and\ \citenamefont {Meerson}}]{Bettelheim:2022}%
  \BibitemOpen
  \bibfield  {author} {\bibinfo {author} {\bibfnamefont {E.}~\bibnamefont
  {Bettelheim}}, \bibinfo {author} {\bibfnamefont {N.~R.}\ \bibnamefont
  {Smith}}, \ and\ \bibinfo {author} {\bibfnamefont {B.}~\bibnamefont
  {Meerson}},\ }\href {\doibase 10.1103/PhysRevLett.128.130602} {\bibfield
  {journal} {\bibinfo  {journal} {Phys. Rev. Lett.}\ }\textbf {\bibinfo
  {volume} {128}},\ \bibinfo {pages} {130602} (\bibinfo {year}
  {2022}{\natexlab{a}})}\BibitemShut {NoStop}%
\bibitem [{\citenamefont {Bettelheim}\ \emph
  {et~al.}(2022{\natexlab{b}})\citenamefont {Bettelheim}, \citenamefont
  {Smith},\ and\ \citenamefont {Meerson}}]{Bettelheim:2022a}%
  \BibitemOpen
  \bibfield  {author} {\bibinfo {author} {\bibfnamefont {E.}~\bibnamefont
  {Bettelheim}}, \bibinfo {author} {\bibfnamefont {N.~R.}\ \bibnamefont
  {Smith}}, \ and\ \bibinfo {author} {\bibfnamefont {B.}~\bibnamefont
  {Meerson}},\ }\href {\doibase 10.1088/1742-5468/ac8a4d} {\bibfield  {journal}
  {\bibinfo  {journal} {J. Stat. Mech: Theory Exp.}\ }\textbf {\bibinfo
  {volume} {2022}},\ \bibinfo {pages} {093103} (\bibinfo {year}
  {2022}{\natexlab{b}})}\BibitemShut {NoStop}%
\bibitem [{\citenamefont {Mallick}\ \emph {et~al.}(2022)\citenamefont
  {Mallick}, \citenamefont {Moriya},\ and\ \citenamefont
  {Sasamoto}}]{Mallick:2022}%
  \BibitemOpen
  \bibfield  {author} {\bibinfo {author} {\bibfnamefont {K.}~\bibnamefont
  {Mallick}}, \bibinfo {author} {\bibfnamefont {H.}~\bibnamefont {Moriya}}, \
  and\ \bibinfo {author} {\bibfnamefont {T.}~\bibnamefont {Sasamoto}},\ }\href
  {\doibase 10.1103/PhysRevLett.129.040601} {\bibfield  {journal} {\bibinfo
  {journal} {Phys. Rev. Lett.}\ }\textbf {\bibinfo {volume} {129}},\ \bibinfo
  {pages} {040601} (\bibinfo {year} {2022})}\BibitemShut {NoStop}%
\bibitem [{\citenamefont {Krajenbrink}\ and\ \citenamefont
  {Le~Doussal}(2023)}]{Krajenbrink:2022}%
  \BibitemOpen
  \bibfield  {author} {\bibinfo {author} {\bibfnamefont {A.}~\bibnamefont
  {Krajenbrink}}\ and\ \bibinfo {author} {\bibfnamefont {P.}~\bibnamefont
  {Le~Doussal}},\ }\href {\doibase 10.1103/PhysRevE.107.014137} {\bibfield
  {journal} {\bibinfo  {journal} {Phys. Rev. E}\ }\textbf {\bibinfo {volume}
  {107}},\ \bibinfo {pages} {014137} (\bibinfo {year} {2023})}\BibitemShut
  {NoStop}%
\bibitem [{\citenamefont {Ablowitz}\ and\ \citenamefont
  {Segur}(1981)}]{Ablowitz:1981}%
  \BibitemOpen
  \bibfield  {author} {\bibinfo {author} {\bibfnamefont {M.~J.}\ \bibnamefont
  {Ablowitz}}\ and\ \bibinfo {author} {\bibfnamefont {H.}~\bibnamefont
  {Segur}},\ }\href@noop {} {\emph {\bibinfo {title} {Solitons and the inverse
  scattering transform}}}\ (\bibinfo  {publisher} {SIAM},\ \bibinfo {year}
  {1981})\BibitemShut {NoStop}%
\bibitem [{\citenamefont {Rizkallah}\ \emph {et~al.}(2023)\citenamefont
  {Rizkallah}, \citenamefont {Grabsch}, \citenamefont {Illien},\ and\
  \citenamefont {Bénichou}}]{Rizkallah:2022}%
  \BibitemOpen
  \bibfield  {author} {\bibinfo {author} {\bibfnamefont {P.}~\bibnamefont
  {Rizkallah}}, \bibinfo {author} {\bibfnamefont {A.}~\bibnamefont {Grabsch}},
  \bibinfo {author} {\bibfnamefont {P.}~\bibnamefont {Illien}}, \ and\ \bibinfo
  {author} {\bibfnamefont {O.}~\bibnamefont {Bénichou}},\ }\href {\doibase
  10.1088/1742-5468/aca8fb} {\bibfield  {journal} {\bibinfo  {journal} {J.
  Stat. Mech: Theory Exp.}\ }\textbf {\bibinfo {volume} {2023}},\ \bibinfo
  {pages} {013202} (\bibinfo {year} {2023})}\BibitemShut {NoStop}%
\bibitem [{\citenamefont {Spitzer}(1974)}]{Spitzer:1974}%
  \BibitemOpen
  \bibfield  {author} {\bibinfo {author} {\bibfnamefont {F.}~\bibnamefont
  {Spitzer}},\ }\href {\doibase 10.1090/S0002-9947-1974-0375533-6} {\bibfield
  {journal} {\bibinfo  {journal} {Trans. Amer. Math. Soc.}\ }\textbf {\bibinfo
  {volume} {198}},\ \bibinfo {pages} {191} (\bibinfo {year}
  {1974})}\BibitemShut {NoStop}%
\bibitem [{\citenamefont {Illien}\ \emph {et~al.}(2013)\citenamefont {Illien},
  \citenamefont {B\'enichou}, \citenamefont {Mej\'{\i}a-Monasterio},
  \citenamefont {Oshanin},\ and\ \citenamefont {Voituriez}}]{Illien:2013}%
  \BibitemOpen
  \bibfield  {author} {\bibinfo {author} {\bibfnamefont {P.}~\bibnamefont
  {Illien}}, \bibinfo {author} {\bibfnamefont {O.}~\bibnamefont {B\'enichou}},
  \bibinfo {author} {\bibfnamefont {C.}~\bibnamefont {Mej\'{\i}a-Monasterio}},
  \bibinfo {author} {\bibfnamefont {G.}~\bibnamefont {Oshanin}}, \ and\
  \bibinfo {author} {\bibfnamefont {R.}~\bibnamefont {Voituriez}},\ }\href
  {\doibase 10.1103/PhysRevLett.111.038102} {\bibfield  {journal} {\bibinfo
  {journal} {Phys. Rev. Lett.}\ }\textbf {\bibinfo {volume} {111}},\ \bibinfo
  {pages} {038102} (\bibinfo {year} {2013})}\BibitemShut {NoStop}%
\bibitem [{\citenamefont {Poncet}\ \emph {et~al.}(2022)\citenamefont {Poncet},
  \citenamefont {Grabsch}, \citenamefont {B\'enichou},\ and\ \citenamefont
  {Illien}}]{Poncet:2022}%
  \BibitemOpen
  \bibfield  {author} {\bibinfo {author} {\bibfnamefont {A.}~\bibnamefont
  {Poncet}}, \bibinfo {author} {\bibfnamefont {A.}~\bibnamefont {Grabsch}},
  \bibinfo {author} {\bibfnamefont {O.}~\bibnamefont {B\'enichou}}, \ and\
  \bibinfo {author} {\bibfnamefont {P.}~\bibnamefont {Illien}},\ }\href
  {\doibase 10.1103/PhysRevE.105.054139} {\bibfield  {journal} {\bibinfo
  {journal} {Phys. Rev. E}\ }\textbf {\bibinfo {volume} {105}},\ \bibinfo
  {pages} {054139} (\bibinfo {year} {2022})}\BibitemShut {NoStop}%
\bibitem [{\citenamefont {Sadhu}\ and\ \citenamefont
  {Derrida}(2015)}]{Sadhu:2015}%
  \BibitemOpen
  \bibfield  {author} {\bibinfo {author} {\bibfnamefont {T.}~\bibnamefont
  {Sadhu}}\ and\ \bibinfo {author} {\bibfnamefont {B.}~\bibnamefont
  {Derrida}},\ }\href {\doibase 10.1088/1742-5468/2015/09/p09008} {\bibfield
  {journal} {\bibinfo  {journal} {J. Stat. Mech: Theory Exp.}\ }\textbf
  {\bibinfo {volume} {2015}},\ \bibinfo {pages} {P09008} (\bibinfo {year}
  {2015})}\BibitemShut {NoStop}%
\bibitem [{\citenamefont {Owen}(1980)}]{Owen:1980}%
  \BibitemOpen
  \bibfield  {author} {\bibinfo {author} {\bibfnamefont {D.~B.}\ \bibnamefont
  {Owen}},\ }\href {\doibase 10.1080/03610918008812164} {\bibfield  {journal}
  {\bibinfo  {journal} {Commun. Stat. Simul. Comput.}\ }\textbf {\bibinfo
  {volume} {9}},\ \bibinfo {pages} {389} (\bibinfo {year} {1980})}\BibitemShut
  {NoStop}%
\bibitem [{\citenamefont {Arabadzhyan}\ and\ \citenamefont
  {Engibaryan}(1987)}]{Arabadzhyan:1987}%
  \BibitemOpen
  \bibfield  {author} {\bibinfo {author} {\bibfnamefont {L.}~\bibnamefont
  {Arabadzhyan}}\ and\ \bibinfo {author} {\bibfnamefont {N.}~\bibnamefont
  {Engibaryan}},\ }\href {\doibase 10.1007/BF01085507} {\bibfield  {journal}
  {\bibinfo  {journal} {J. Sov. Math.}\ }\textbf {\bibinfo {volume} {36}},\
  \bibinfo {pages} {745} (\bibinfo {year} {1987})}\BibitemShut {NoStop}%
\bibitem [{\citenamefont {Bray}\ \emph {et~al.}(2013)\citenamefont {Bray},
  \citenamefont {Majumdar},\ and\ \citenamefont {Schehr}}]{Bray:2013}%
  \BibitemOpen
  \bibfield  {author} {\bibinfo {author} {\bibfnamefont {A.~J.}\ \bibnamefont
  {Bray}}, \bibinfo {author} {\bibfnamefont {S.~N.}\ \bibnamefont {Majumdar}},
  \ and\ \bibinfo {author} {\bibfnamefont {G.}~\bibnamefont {Schehr}},\ }\href
  {\doibase 10.1080/00018732.2013.803819} {\bibfield  {journal} {\bibinfo
  {journal} {Adv. Phys.}\ }\textbf {\bibinfo {volume} {62}},\ \bibinfo {pages}
  {225} (\bibinfo {year} {2013})}\BibitemShut {NoStop}%
\bibitem [{\citenamefont {Krajenbrink}\ and\ \citenamefont
  {Le~Doussal}(2021)}]{Krajenbrink:2021}%
  \BibitemOpen
  \bibfield  {author} {\bibinfo {author} {\bibfnamefont {A.}~\bibnamefont
  {Krajenbrink}}\ and\ \bibinfo {author} {\bibfnamefont {P.}~\bibnamefont
  {Le~Doussal}},\ }\href {\doibase 10.1103/PhysRevLett.127.064101} {\bibfield
  {journal} {\bibinfo  {journal} {Phys. Rev. Lett.}\ }\textbf {\bibinfo
  {volume} {127}},\ \bibinfo {pages} {064101} (\bibinfo {year}
  {2021})}\BibitemShut {NoStop}%
\bibitem [{\citenamefont {Polyanin}\ and\ \citenamefont
  {Manzhirov}(2008)}]{Polyanin:2008}%
  \BibitemOpen
  \bibfield  {author} {\bibinfo {author} {\bibfnamefont {A.~D.}\ \bibnamefont
  {Polyanin}}\ and\ \bibinfo {author} {\bibfnamefont {A.~V.}\ \bibnamefont
  {Manzhirov}},\ }\href@noop {} {\emph {\bibinfo {title} {Handbook of integral
  equations}}}\ (\bibinfo  {publisher} {CRC press},\ \bibinfo {year}
  {2008})\BibitemShut {NoStop}%
\bibitem [{\citenamefont {Krapivsky}\ and\ \citenamefont
  {Meerson}(2012)}]{Krapivsky:2012}%
  \BibitemOpen
  \bibfield  {author} {\bibinfo {author} {\bibfnamefont {P.}~\bibnamefont
  {Krapivsky}}\ and\ \bibinfo {author} {\bibfnamefont {B.}~\bibnamefont
  {Meerson}},\ }\href {\doibase 10.1103/PhysRevE.86.031106} {\bibfield
  {journal} {\bibinfo  {journal} {Phys. Rev. E}\ }\textbf {\bibinfo {volume}
  {86}},\ \bibinfo {pages} {031106} (\bibinfo {year} {2012})}\BibitemShut
  {NoStop}%
\bibitem [{\citenamefont {Derrida}\ and\ \citenamefont
  {Gerschenfeld}(2009{\natexlab{b}})}]{Derrida:2009a}%
  \BibitemOpen
  \bibfield  {author} {\bibinfo {author} {\bibfnamefont {B.}~\bibnamefont
  {Derrida}}\ and\ \bibinfo {author} {\bibfnamefont {A.}~\bibnamefont
  {Gerschenfeld}},\ }\href {\doibase 10.1007/s10955-009-9830-1} {\bibfield
  {journal} {\bibinfo  {journal} {J. Stat. Phys.}\ }\textbf {\bibinfo {volume}
  {137}},\ \bibinfo {pages} {978} (\bibinfo {year}
  {2009}{\natexlab{b}})}\BibitemShut {NoStop}%
\bibitem [{\citenamefont {Spohn}(1983)}]{Spohn:1983}%
  \BibitemOpen
  \bibfield  {author} {\bibinfo {author} {\bibfnamefont {H.}~\bibnamefont
  {Spohn}},\ }\href {\doibase 10.1088/0305-4470/16/18/029} {\bibfield
  {journal} {\bibinfo  {journal} {J. Phys. A}\ }\textbf {\bibinfo {volume}
  {16}},\ \bibinfo {pages} {4275} (\bibinfo {year} {1983})}\BibitemShut
  {NoStop}%
\bibitem [{\citenamefont {Derrida}(2007)}]{Derrida:2007}%
  \BibitemOpen
  \bibfield  {author} {\bibinfo {author} {\bibfnamefont {B.}~\bibnamefont
  {Derrida}},\ }\href {\doibase 10.1088/1742-5468/2007/07/p07023} {\bibfield
  {journal} {\bibinfo  {journal} {J. Stat. Mech.}\ }\textbf {\bibinfo {volume}
  {2007}},\ \bibinfo {pages} {P07023} (\bibinfo {year} {2007})}\BibitemShut
  {NoStop}%
\bibitem [{\citenamefont {Krapivsky}\ \emph
  {et~al.}(2015{\natexlab{b}})\citenamefont {Krapivsky}, \citenamefont
  {Mallick},\ and\ \citenamefont {Sadhu}}]{Krapivsky:2015}%
  \BibitemOpen
  \bibfield  {author} {\bibinfo {author} {\bibfnamefont {P.~L.}\ \bibnamefont
  {Krapivsky}}, \bibinfo {author} {\bibfnamefont {K.}~\bibnamefont {Mallick}},
  \ and\ \bibinfo {author} {\bibfnamefont {T.}~\bibnamefont {Sadhu}},\ }\href
  {\doibase 10.1088/1742-5468/2015/09/P09007} {\bibfield  {journal} {\bibinfo
  {journal} {J. Stat. Mech: Theory Exp.}\ }\textbf {\bibinfo {volume} {2015}},\
  \bibinfo {pages} {P09007} (\bibinfo {year} {2015}{\natexlab{b}})}\BibitemShut
  {NoStop}%
\bibitem [{\citenamefont {Zarfaty}\ and\ \citenamefont
  {Meerson}(2016)}]{Zarfaty:2016}%
  \BibitemOpen
  \bibfield  {author} {\bibinfo {author} {\bibfnamefont {L.}~\bibnamefont
  {Zarfaty}}\ and\ \bibinfo {author} {\bibfnamefont {B.}~\bibnamefont
  {Meerson}},\ }\href {\doibase 10.1088/1742-5468/2016/03/033304} {\bibfield
  {journal} {\bibinfo  {journal} {J. Stat. Mech: Theory Exp.}\ }\textbf
  {\bibinfo {volume} {2016}},\ \bibinfo {pages} {033304} (\bibinfo {year}
  {2016})}\BibitemShut {NoStop}%
\bibitem [{\citenamefont {Krug}\ and\ \citenamefont
  {Garcia}(2000)}]{Krug:2000}%
  \BibitemOpen
  \bibfield  {author} {\bibinfo {author} {\bibfnamefont {J.}~\bibnamefont
  {Krug}}\ and\ \bibinfo {author} {\bibfnamefont {J.}~\bibnamefont {Garcia}},\
  }\href {\doibase 10.1023/A:1018688421856} {\bibfield  {journal} {\bibinfo
  {journal} {J. Stat. Phys.}\ }\textbf {\bibinfo {volume} {99}},\ \bibinfo
  {pages} {31} (\bibinfo {year} {2000})}\BibitemShut {NoStop}%
\bibitem [{\citenamefont {Kundu}\ and\ \citenamefont
  {Cividini}(2016)}]{Kundu:2016}%
  \BibitemOpen
  \bibfield  {author} {\bibinfo {author} {\bibfnamefont {A.}~\bibnamefont
  {Kundu}}\ and\ \bibinfo {author} {\bibfnamefont {J.}~\bibnamefont
  {Cividini}},\ }\href {\doibase 10.1209/0295-5075/115/54003} {\bibfield
  {journal} {\bibinfo  {journal} {Europhys. Lett.}\ }\textbf {\bibinfo {volume}
  {115}},\ \bibinfo {pages} {54003} (\bibinfo {year} {2016})}\BibitemShut
  {NoStop}%
\bibitem [{\citenamefont {Hager}\ \emph {et~al.}(2001)\citenamefont {Hager},
  \citenamefont {Krug}, \citenamefont {Popkov},\ and\ \citenamefont
  {Sch\"utz}}]{Hager:2001}%
  \BibitemOpen
  \bibfield  {author} {\bibinfo {author} {\bibfnamefont {J.~S.}\ \bibnamefont
  {Hager}}, \bibinfo {author} {\bibfnamefont {J.}~\bibnamefont {Krug}},
  \bibinfo {author} {\bibfnamefont {V.}~\bibnamefont {Popkov}}, \ and\ \bibinfo
  {author} {\bibfnamefont {G.~M.}\ \bibnamefont {Sch\"utz}},\ }\href {\doibase
  10.1103/PhysRevE.63.056110} {\bibfield  {journal} {\bibinfo  {journal} {Phys.
  Rev. E}\ }\textbf {\bibinfo {volume} {63}},\ \bibinfo {pages} {056110}
  (\bibinfo {year} {2001})}\BibitemShut {NoStop}%
\bibitem [{\citenamefont {Baek}\ \emph {et~al.}(2017)\citenamefont {Baek},
  \citenamefont {Kafri},\ and\ \citenamefont {Lecomte}}]{Baek:2017}%
  \BibitemOpen
  \bibfield  {author} {\bibinfo {author} {\bibfnamefont {Y.}~\bibnamefont
  {Baek}}, \bibinfo {author} {\bibfnamefont {Y.}~\bibnamefont {Kafri}}, \ and\
  \bibinfo {author} {\bibfnamefont {V.}~\bibnamefont {Lecomte}},\ }\href
  {\doibase 10.1103/PhysRevLett.118.030604} {\bibfield  {journal} {\bibinfo
  {journal} {Phys. Rev. Lett.}\ }\textbf {\bibinfo {volume} {118}},\ \bibinfo
  {pages} {030604} (\bibinfo {year} {2017})}\BibitemShut {NoStop}%
\bibitem [{\citenamefont {Kipnis}\ \emph {et~al.}(1982)\citenamefont {Kipnis},
  \citenamefont {Marchioro},\ and\ \citenamefont {Presutti}}]{Kipnis:1982}%
  \BibitemOpen
  \bibfield  {author} {\bibinfo {author} {\bibfnamefont {C.}~\bibnamefont
  {Kipnis}}, \bibinfo {author} {\bibfnamefont {C.}~\bibnamefont {Marchioro}}, \
  and\ \bibinfo {author} {\bibfnamefont {E.}~\bibnamefont {Presutti}},\ }\href
  {\doibase 10.1007/BF01011740} {\bibfield  {journal} {\bibinfo  {journal} {J.
  Stat. Phys.}\ }\textbf {\bibinfo {volume} {27}},\ \bibinfo {pages} {65}
  (\bibinfo {year} {1982})}\BibitemShut {NoStop}%
\bibitem [{\citenamefont {Hurtado}\ and\ \citenamefont
  {Garrido}(2009)}]{Hurtado:2009}%
  \BibitemOpen
  \bibfield  {author} {\bibinfo {author} {\bibfnamefont {P.~I.}\ \bibnamefont
  {Hurtado}}\ and\ \bibinfo {author} {\bibfnamefont {P.~L.}\ \bibnamefont
  {Garrido}},\ }\href {\doibase 10.1088/1742-5468/2009/02/p02032} {\bibfield
  {journal} {\bibinfo  {journal} {J. Stat. Mech.}\ }\textbf {\bibinfo {volume}
  {2009}},\ \bibinfo {pages} {P02032} (\bibinfo {year} {2009})}\BibitemShut
  {NoStop}%
\bibitem [{\citenamefont {Ferrari}\ and\ \citenamefont
  {Fontes}(1998)}]{Ferrari:1998}%
  \BibitemOpen
  \bibfield  {author} {\bibinfo {author} {\bibfnamefont {P.}~\bibnamefont
  {Ferrari}}\ and\ \bibinfo {author} {\bibfnamefont {L.}~\bibnamefont
  {Fontes}},\ }\href {\doibase 10.1214/EJP.v3-28} {\bibfield  {journal}
  {\bibinfo  {journal} {Electron. J. Probab.}\ }\textbf {\bibinfo {volume}
  {3}},\ \bibinfo {pages} {1 } (\bibinfo {year} {1998})}\BibitemShut {NoStop}%
\bibitem [{\citenamefont {Rajesh}\ and\ \citenamefont
  {Majumdar}(2001)}]{Rajesh:2001}%
  \BibitemOpen
  \bibfield  {author} {\bibinfo {author} {\bibfnamefont {R.}~\bibnamefont
  {Rajesh}}\ and\ \bibinfo {author} {\bibfnamefont {S.~N.}\ \bibnamefont
  {Majumdar}},\ }\href {https://link.aps.org/doi/10.1103/PhysRevE.64.036103}
  {\bibfield  {journal} {\bibinfo  {journal} {Phys. Rev. E}\ }\textbf {\bibinfo
  {volume} {64}},\ \bibinfo {pages} {036103} (\bibinfo {year}
  {2001})}\BibitemShut {NoStop}%
\bibitem [{\citenamefont {Cividini}\ \emph
  {et~al.}(2016{\natexlab{a}})\citenamefont {Cividini}, \citenamefont {Kundu},
  \citenamefont {Majumdar},\ and\ \citenamefont {Mukamel}}]{Cividini:2016}%
  \BibitemOpen
  \bibfield  {author} {\bibinfo {author} {\bibfnamefont {J.}~\bibnamefont
  {Cividini}}, \bibinfo {author} {\bibfnamefont {A.}~\bibnamefont {Kundu}},
  \bibinfo {author} {\bibfnamefont {S.~N.}\ \bibnamefont {Majumdar}}, \ and\
  \bibinfo {author} {\bibfnamefont {D.}~\bibnamefont {Mukamel}},\ }\href
  {\doibase 10.1088/1742-5468/2016/05/053212} {\bibfield  {journal} {\bibinfo
  {journal} {J. Stat. Mech: Theory Exp.}\ }\textbf {\bibinfo {volume} {2016}},\
  \bibinfo {pages} {053212} (\bibinfo {year} {2016}{\natexlab{a}})}\BibitemShut
  {NoStop}%
\bibitem [{\citenamefont {Cividini}\ \emph
  {et~al.}(2016{\natexlab{b}})\citenamefont {Cividini}, \citenamefont {Kundu},
  \citenamefont {Majumdar},\ and\ \citenamefont {Mukamel}}]{Cividini:2016a}%
  \BibitemOpen
  \bibfield  {author} {\bibinfo {author} {\bibfnamefont {J.}~\bibnamefont
  {Cividini}}, \bibinfo {author} {\bibfnamefont {A.}~\bibnamefont {Kundu}},
  \bibinfo {author} {\bibfnamefont {S.~N.}\ \bibnamefont {Majumdar}}, \ and\
  \bibinfo {author} {\bibfnamefont {D.}~\bibnamefont {Mukamel}},\ }\href
  {\doibase 10.1088/1751-8113/49/8/085002} {\bibfield  {journal} {\bibinfo
  {journal} {J. Phys. A}\ }\textbf {\bibinfo {volume} {49}},\ \bibinfo {pages}
  {085002} (\bibinfo {year} {2016}{\natexlab{b}})}\BibitemShut {NoStop}%
\bibitem [{Note1()}]{Note1}%
  \BibitemOpen
  \bibinfo {note} {This model can also be defined by taking a limit of the more
  general Katz–Lebowitz–Spohn (KLS) model~\cite
  {Hager:2001,Baek:2017}}\BibitemShut {NoStop}%
\bibitem [{\citenamefont {Shaw}\ \emph {et~al.}(2003)\citenamefont {Shaw},
  \citenamefont {Zia},\ and\ \citenamefont {Lee}}]{Shaw2003}%
  \BibitemOpen
  \bibfield  {author} {\bibinfo {author} {\bibfnamefont {L.~B.}\ \bibnamefont
  {Shaw}}, \bibinfo {author} {\bibfnamefont {R.~K.}\ \bibnamefont {Zia}}, \
  and\ \bibinfo {author} {\bibfnamefont {K.~H.}\ \bibnamefont {Lee}},\ }\href
  {\doibase 10.1103/PhysRevE.68.021910} {\bibfield  {journal} {\bibinfo
  {journal} {Phys. Rev. E}\ }\textbf {\bibinfo {volume} {68}},\ \bibinfo
  {pages} {021910} (\bibinfo {year} {2003})}\BibitemShut {NoStop}%
\bibitem [{\citenamefont {Bonnemain}\ \emph {et~al.}(2021)\citenamefont
  {Bonnemain}, \citenamefont {Gobron},\ and\ \citenamefont
  {Ullmo}}]{Bonnemain:2021}%
  \BibitemOpen
  \bibfield  {author} {\bibinfo {author} {\bibfnamefont {T.}~\bibnamefont
  {Bonnemain}}, \bibinfo {author} {\bibfnamefont {T.}~\bibnamefont {Gobron}}, \
  and\ \bibinfo {author} {\bibfnamefont {D.}~\bibnamefont {Ullmo}},\ }\href
  {\doibase 10.1063/5.0039742} {\bibfield  {journal} {\bibinfo  {journal} {J.
  Math. Phys.}\ }\textbf {\bibinfo {volume} {62}},\ \bibinfo {pages} {083302}
  (\bibinfo {year} {2021})}\BibitemShut {NoStop}%
\bibitem [{\citenamefont {Krajenbrink}\ and\ \citenamefont
  {Le~Doussal}(2022)}]{Krajenbrink:2022a}%
  \BibitemOpen
  \bibfield  {author} {\bibinfo {author} {\bibfnamefont {A.}~\bibnamefont
  {Krajenbrink}}\ and\ \bibinfo {author} {\bibfnamefont {P.}~\bibnamefont
  {Le~Doussal}},\ }\href {\doibase 10.1103/PhysRevE.105.054142} {\bibfield
  {journal} {\bibinfo  {journal} {Phys. Rev. E}\ }\textbf {\bibinfo {volume}
  {105}},\ \bibinfo {pages} {054142} (\bibinfo {year} {2022})}\BibitemShut
  {NoStop}%
\bibitem [{\citenamefont {Bertini}\ \emph {et~al.}(2001)\citenamefont
  {Bertini}, \citenamefont {De~Sole}, \citenamefont {Gabrielli}, \citenamefont
  {Jona-Lasinio},\ and\ \citenamefont {Landim}}]{Bertini:2001}%
  \BibitemOpen
  \bibfield  {author} {\bibinfo {author} {\bibfnamefont {L.}~\bibnamefont
  {Bertini}}, \bibinfo {author} {\bibfnamefont {A.}~\bibnamefont {De~Sole}},
  \bibinfo {author} {\bibfnamefont {D.}~\bibnamefont {Gabrielli}}, \bibinfo
  {author} {\bibfnamefont {G.}~\bibnamefont {Jona-Lasinio}}, \ and\ \bibinfo
  {author} {\bibfnamefont {C.}~\bibnamefont {Landim}},\ }\href {\doibase
  10.1103/PhysRevLett.87.040601} {\bibfield  {journal} {\bibinfo  {journal}
  {Phys. Rev. Lett.}\ }\textbf {\bibinfo {volume} {87}},\ \bibinfo {pages}
  {040601} (\bibinfo {year} {2001})}\BibitemShut {NoStop}%
\bibitem [{\citenamefont {Bertini}\ \emph {et~al.}(2002)\citenamefont
  {Bertini}, \citenamefont {De~Sole}, \citenamefont {Gabrielli}, \citenamefont
  {Jona-Lasinio},\ and\ \citenamefont {Landim}}]{Bertini:2002}%
  \BibitemOpen
  \bibfield  {author} {\bibinfo {author} {\bibfnamefont {L.}~\bibnamefont
  {Bertini}}, \bibinfo {author} {\bibfnamefont {A.}~\bibnamefont {De~Sole}},
  \bibinfo {author} {\bibfnamefont {D.}~\bibnamefont {Gabrielli}}, \bibinfo
  {author} {\bibfnamefont {G.}~\bibnamefont {Jona-Lasinio}}, \ and\ \bibinfo
  {author} {\bibfnamefont {C.}~\bibnamefont {Landim}},\ }\href {\doibase
  10.1023/A:1014525911391} {\bibfield  {journal} {\bibinfo  {journal} {J. Stat.
  Phys.}\ }\textbf {\bibinfo {volume} {107}},\ \bibinfo {pages} {635} (\bibinfo
  {year} {2002})}\BibitemShut {NoStop}%
\bibitem [{\citenamefont {Bertini}\ \emph {et~al.}(2005)\citenamefont
  {Bertini}, \citenamefont {Gabrielli},\ and\ \citenamefont
  {Lebowitz}}]{Bertini:2005}%
  \BibitemOpen
  \bibfield  {author} {\bibinfo {author} {\bibfnamefont {L.}~\bibnamefont
  {Bertini}}, \bibinfo {author} {\bibfnamefont {D.}~\bibnamefont {Gabrielli}},
  \ and\ \bibinfo {author} {\bibfnamefont {J.~L.}\ \bibnamefont {Lebowitz}},\
  }\href {\doibase 10.1007/s10955-005-5527-2} {\bibfield  {journal} {\bibinfo
  {journal} {J. Stat. Phys.}\ }\textbf {\bibinfo {volume} {121}},\ \bibinfo
  {pages} {843} (\bibinfo {year} {2005})}\BibitemShut {NoStop}%
\bibitem [{\citenamefont {Bertini}\ \emph {et~al.}(2009)\citenamefont
  {Bertini}, \citenamefont {De~Sole}, \citenamefont {Gabrielli}, \citenamefont
  {Jona-Lasinio},\ and\ \citenamefont {Landim}}]{Bertini:2009}%
  \BibitemOpen
  \bibfield  {author} {\bibinfo {author} {\bibfnamefont {L.}~\bibnamefont
  {Bertini}}, \bibinfo {author} {\bibfnamefont {A.}~\bibnamefont {De~Sole}},
  \bibinfo {author} {\bibfnamefont {D.}~\bibnamefont {Gabrielli}}, \bibinfo
  {author} {\bibfnamefont {G.}~\bibnamefont {Jona-Lasinio}}, \ and\ \bibinfo
  {author} {\bibfnamefont {C.}~\bibnamefont {Landim}},\ }\href {\doibase
  10.1007/s10955-008-9670-4} {\bibfield  {journal} {\bibinfo  {journal} {J.
  Stat. Phys.}\ }\textbf {\bibinfo {volume} {135}},\ \bibinfo {pages} {857}
  (\bibinfo {year} {2009})}\BibitemShut {NoStop}%
\bibitem [{\citenamefont {Bertini}\ \emph {et~al.}(2015)\citenamefont
  {Bertini}, \citenamefont {De~Sole}, \citenamefont {Gabrielli}, \citenamefont
  {Jona-Lasinio},\ and\ \citenamefont {Landim}}]{Bertini:2015}%
  \BibitemOpen
  \bibfield  {author} {\bibinfo {author} {\bibfnamefont {L.}~\bibnamefont
  {Bertini}}, \bibinfo {author} {\bibfnamefont {A.}~\bibnamefont {De~Sole}},
  \bibinfo {author} {\bibfnamefont {D.}~\bibnamefont {Gabrielli}}, \bibinfo
  {author} {\bibfnamefont {G.}~\bibnamefont {Jona-Lasinio}}, \ and\ \bibinfo
  {author} {\bibfnamefont {C.}~\bibnamefont {Landim}},\ }\href {\doibase
  10.1103/RevModPhys.87.593} {\bibfield  {journal} {\bibinfo  {journal} {Rev.
  Mod. Phys.}\ }\textbf {\bibinfo {volume} {87}},\ \bibinfo {pages} {593}
  (\bibinfo {year} {2015})}\BibitemShut {NoStop}%
\end{thebibliography}

%

\end{document}